\documentclass[fleqn,twoside]{article}
\usepackage{amsmath,amssymb,espcrc2,color,cite}
\usepackage[usenames,dvipsnames]{xcolor}

\usepackage{graphicx,soul}
\usepackage[figuresright]{rotating}

\newcommand{\url}[1]{{\tt #1}}
\newcommand{\lsim}
{\;\raisebox{-.3em}{$\stackrel{\displaystyle <}{\sim}$}\;}
\newcommand{\gsim}
{\;\raisebox{-.3em}{$\stackrel{\displaystyle >}{\sim}$}\;}

\newcommand{\DRbar}{\ensuremath{\smash{\overline{\mathrm{DR}}}}}
\newcommand{\gmt}{\ensuremath{(g-2)_\mu}}
\newcommand{\br}{{\rm BR}}
\newcommand{\bsg}{BR($b \to s \gamma$)}

\newcommand{\btn}{BR($B_u \to \tau \nu_\tau$)}

\newcommand{\bmm}{\ensuremath{\br(B_s \to \mu^+\mu^-)}}

\newcommand{\bsdmm}{\ensuremath{\br(B_{s, d} \to \mu^+\mu^-)}}

\newcommand{\ssi}{\ensuremath{\sigma^{\rm SI}_p}}

\newcommand{\ALRe}{\ensuremath{A_{\rm LR}^e}}
\newcommand{\sweff}{\sin^2\theta_{\mathrm{eff}}}
\newcommand{\MW}{\ensuremath{M_W}}
\newcommand{\MZ}{\ensuremath{M_Z}}
\newcommand{\Mh}{\ensuremath{M_h}}
\newcommand{\MH}{\ensuremath{M_H}}
\newcommand{\MA}{\ensuremath{M_A}}

\newcommand{\mt}{m_t}

\newcommand{\gl}{\ensuremath{{\tilde g}}}
\newcommand{\mgl}{\ensuremath{m_{\tilde g}}}
\newcommand{\sq}{\tilde q}
\newcommand{\asq}{\bar{\tilde q}}
\newcommand{\msq}{\ensuremath{m_{\tilde q}}}

\newcommand{\stau}[1]{\ensuremath{\tilde \tau_{#1}}}

\newcommand{\SuR}{\ensuremath{\tilde{u}_R}}
\newcommand{\SdR}{\ensuremath{\tilde{d}_R}}

\newcommand{\SucR}{\ensuremath{\tilde{u}_R/\tilde{c}_R}}

\newcommand{\cha}[1]{\tilde \chi^\pm_{#1}}

\newcommand{\mcha}[1]{\ensuremath{m_{\tilde \chi^\pm_{#1}}}}
\newcommand{\neu}[1]{\tilde \chi^0_{#1}}
\newcommand{\mneu}[1]{\ensuremath{m_{\tilde \chi^0_{#1}}}}

\newcommand{\mste}{m_{\tilde t_1}}

\newcommand{\staue}{\tilde \tau_1}
\newcommand{\mstaue}{m_{\staue}}

\newcommand{\tb}{\ensuremath{\tan\beta}}

\newcommand{\tev}{\ensuremath{\,\, \mathrm{TeV}}}
\newcommand{\gev}{\ensuremath{\,\, \mathrm{GeV}}}

\newcommand{\ifb}{\ensuremath{{\rm fb}^{-1}}}
\def\order#1{\ensuremath{{\cal O}(#1)}}

\def\reffi#1{\mbox{Fig.~\ref{#1}}}
\def\reffis#1{\mbox{Figs.~\ref{#1}}}
\def\refta#1{\mbox{Table~\ref{#1}}}

\def\refse#1{\mbox{Sect.~\ref{#1}}}

\definecolor{orange}{rgb}{1,0.5,0}

\definecolor{Gray}{named}{Gray}

\newcommand{\ETslash}{\, \ensuremath{/ \hspace{-.7em} E_T}}

\graphicspath{{figs/}}

\title{\vspace{-4.5cm}
\bf \LARGE Likelihood Analysis of Supersymmetric SU(5) GUTs \\ \vspace{0.5em}}

\author{
{\bf E.~Bagnaschi}\address[DESY]
   {DESY, Notkestra{\ss}e 85, D--22607 Hamburg, Germany},
{\bf J.C.~Costa}{\address[Imperial]
   {High\,Energy\,Physics\,Group,\,Blackett\,Laboratory,\,Imperial\,College,\,Prince\,Consort\,Road,\,London\,SW7\,2AZ,\,UK},
{\bf K.~Sakurai}\address[Durham]{Institute for Particle Physics Phenomenology, Department of Physics, 
University of Durham, Science Laboratories, South Road, Durham, DH1 3LE, UK}\hbox{$^{\rm ,}$}\address[Warsaw]{Institute of Theoretical Physics, Faculty of Physics, University of Warsaw, ul.~Pasteura 5, PL--02--093 Warsaw, Poland},
{\bf M.~Borsato}\address[USdC]{Universidade de Santiago de Compostela, 
E-15706 Santiago de Compostela, Spain},
{\bf O.~Buchmueller}\addressmark[Imperial],
\bf R.~Cavanaugh}\address[FNAL]
   {Fermi National Accelerator Laboratory, P.O. Box 500, 
    Batavia, Illinois 60510, USA}\hbox{$^{\rm ,}$}\address[UIC]
   {Physics Department, University of Illinois at Chicago, Chicago, 
    Illinois 60607-7059, USA},
{\bf V.~Chobanova}\addressmark[USdC],
{\bf M.~Citron}\addressmark[Imperial],
{\bf A.~De~Roeck}\address[CERNEP]
   {Experimental Physics Department, CERN, CH--1211 Geneva 23, Switzerland; \\  Antwerp University, B--2610 Wilrijk, Belgium},
 {\bf M.J.~Dolan}\address[SLAC]
{ARC Centre of Excellence for Particle Physics at the Terascale, School of Physics, University of Melbourne, 3010, Australia},
{\bf J.R.~Ellis}\address[KCL]{Theoretical Particle Physics
  and Cosmology Group, Department of Physics, King's College London, London~WC2R~2LS, UK; \\ Theoretical Physics Department, CERN, CH--1211 Geneva 23, Switzerland},
{\bf H.~Fl\"acher}\address[Bristol]
   {H.H.~Wills Physics Laboratory, University of Bristol, Tyndall Avenue, Bristol BS8 1TL, UK},
{\bf S.~Heinemeyer}\address[Madrid]
{Campus of International Excellence UAM+CSIC, Cantoblanco, E--28049 Madrid, Spain;\\
  Instituto de F\'{\i}sica Te{\'o}rica UAM-CSIC, C/ Nicolas Cabrera 13-15, E--28049 Madrid, Spain; \\
   Instituto de F\'{\i}sica de Cantabria (CSIC-UC), Avda.\ de Los Castros s/n, 
    E--39005 Santander, Spain},
{\bf G.~Isidori}\address[Zurich]
{Physik-Institut, Universit\"at Z\"urich, CH-8057 Z\"urich, Switzerland},
{\bf M.~Lucio}\addressmark[USdC],
{\bf D.~Mart\'inez~Santos}\addressmark[USdC],
{\bf K.A.~Olive}\address[Minnesota] 
{William I.\ Fine Theoretical Physics Institute, School of Physics and
 Astronomy, University of Minnesota, Minneapolis, Minnesota 55455, USA}, 
{\bf A.~Richards}\addressmark[Imperial],
{\bf K.J.~de~Vries}\addressmark[Imperial],
{\bf G.~Weiglein}\addressmark[DESY]
}

\begin{document}
\begin{abstract}
\vspace{0.25cm}

We perform a likelihood analysis of the constraints from 
accelerator experiments and astrophysical observations
on {supersymmetric (SUSY)} models with SU(5) boundary conditions on soft {SUSY}-breaking
parameters at the GUT scale. The 
parameter space of the models studied has 7 parameters: a universal gaugino mass $m_{1/2}$, 
distinct masses for the scalar partners of matter fermions in five- and ten-dimensional representations
of SU(5), $m_5$ and $m_{10}$, and for the $\mathbf{5}$ and $\mathbf{\bar 5}$ Higgs representations
{$m_{H_u}$} and {$m_{H_d}$}, a universal trilinear soft {SUSY}-breaking parameter $A_0$, and 
the ratio of Higgs vevs $\tan \beta$. 
In addition to previous constraints from direct sparticle searches, low-energy
and flavour observables, we incorporate constraints based on preliminary
results from 13~TeV LHC searches
for jets + $\ETslash$ events and long-lived particles, {as well as} the latest PandaX-II and LUX searches
for {direct Dark Matter detection}. In addition to previously-identified mechanisms for 
bringing the supersymmetric relic density into the range allowed by cosmology, we identify a novel
${\tilde u_R}/{\tilde c_R} - \neu1$ coannihilation mechanism that appears in the supersymmetric SU(5) GUT model
{and discuss the role of ${\tilde \nu_\tau}$ coannihilation}.
We find complementarity between the prospects for direct Dark Matter detection
and SUSY searches at the LHC.

\vspace{0.25cm}
\begin{center}
{\tt KCL-PH-TH/2016-57, CERN-PH-TH/2016-217, DESY 16-156, IFT-UAM/CSIC-16-105 \\
FTPI-MINN-16/29, UMN-TH-3609/16, FERMILAB-PUB-16-453-CMS, IPPP/16/97}
\end{center}
\end{abstract}
\thispagestyle{empty}
\newpage

\vspace{4.0cm}

\maketitle



\section{Introduction}
\label{sec:intro}

In the absence so far of any experimental indications of supersymmetry 
(SUSY)~\cite{ATLAS20,1507.05525,ATLAS:2016kts,CMS20,CMS:2016xva}, nor any
clear theoretical guidance how SUSY may be broken,
the building of models and the exploration of phenomenological constraints on them~\cite{oldmc,mc9,mc10,mc11,MC12-DM,mcweb}
have adopted a range of assumptions. One point of view has been to consider the
simple parametrization of soft SUSY breaking in which the
gaugino and scalar masses, as well as the trilinear soft SUSY-breaking
parameters, are all constrained to be universal at the SUSY GUT scale (the
CMSSM~\cite{DN,funnel,CMSSM,eo6,ehow+,oldmc,mc9,otherCMSSM}). An alternative point of view has been to discard all universality assumptions,
and treat the soft SUSY-breaking parameters as all independent 
phenomenological quantities (the pMSSM~\cite{pMSSM10,mc11,otherpMSSM}), imposing diagonal mass matrices and the
{minimal flavour violation (MFV)} criterion.
Intermediate between these extremes, models with one or two non-universal soft SUSY-breaking
contributions to Higgs masses (the NUHM1~\cite{NUHM1,eosknuhm,elos,eelnos,oldmc,mc9} and NUHM2~\cite{NUHM2,Sneutrino,eosknuhm,elos,eelnos,mc10}) have also been considered.

It is interesting to explore also models that are less simplified than the CMSSM,
but not as agnostic as the pMSSM, in that they incorporate a limited number of simplifying
assumptions. GUTs motivate the assumption that the gaugino masses are universal, and
constraints on flavour-changing neutral interactions suggest that the soft SUSY-breaking masses
for scalars with identical quantum numbers are also universal. However,
there is no compelling phenomenological reason why
the soft SUSY-breaking masses for scalars with different quantum numbers should be
universal.

Specific GUTs may also provide some guidance in this respect. For example, in an SO(10)
GUT the scalar masses of all particles in a given generation belonging to a single $\mathbf{16}$
representation of SO(10) would be universal, as would those for the $\mathbf{5}$ and $\mathbf{\bar 5}$
SU(5) Higgs representations that belong to a single $\mathbf{10}$ of SO(10) and break
electroweak symmetry, as in the NUHM1. 
In contrast, the SU(5) framework is less restrictive, allowing different
masses for scalars in $\mathbf{\bar 5}$ and $\mathbf{10}$ representations~\cite{SU5}, 
and also for the $\mathbf{5}$ and 
$\mathbf{\bar 5}$ Higgs representations. 
Thus it is a 1-parameter extension of the NUHM2.
In this paper we explore the theoretical, phenomenological, experimental
and cosmological constraints on this SU(5)-based SUSY GUT model.

This {relaxation of} universality is relevant for the evaluation of several different constraints
from both the LHC and elsewhere. For example, the most powerful LHC constraints on the CMSSM, NUHM1
and NUHM2 are those from the classic $\ETslash$ searches \cite{ATLAS20,CMS20}. 
These constrain principally the right-handed
squarks, whose decays are dominated by the ${\tilde q_R} \to q \,\neu1$ channel that maximizes the $\ETslash$ signature. 
On the other hand, the decay chains of left-handed squarks are more complicated, typically involving the $\cha1$,
resulting in a dilution of the $\ETslash$ signature and more importance for final states including leptons. 
In a SUSY SU(5) GUT, the left-handed squarks and the right-handed up-type squarks appear in
$\mathbf{10}$ representations whereas the right-handed down-type squarks appear in
$\mathbf{\bar 5}$ representations, with independent soft SUSY-breaking masses.
Hence the impacts of the LHC $\ETslash$ and other constraints need to be re-evaluated.

The possible difference between the soft SUSY-breaking contributions to the 
masses of the 
squarks appearing in a $\mathbf{10}$ of SU(5), 
i.e., up-type squarks and left-handed down-type squarks, and those 
appearing in a $\mathbf{\bar 5}$ of SU(5),
i.e., right-handed down-type squarks,
offers a new avenue for compressing the stop spectrum. Also, as we shall see, with $m_5 \ne m_{10}$
there is the possibility that $m_{{\tilde u_R}, {\tilde c_R}}$ 
are much smaller than the other squark masses, leading to another
type of compressed spectrum~\footnote{This possibility has also been noted in
a supersymmetric SO(10) GUT framework~\cite{Baer:2008jn}.}.

In principle, the constraints from flavour observables may also act differently when $m_5 \ne m_{10}$.
For example, the soft SUSY-breaking masses of the left- and right-handed charge +2/3
quarks are independent, and flavour observables such as \bsg\ and \bmm\ depend on both of them,
in general.

Another experimental constraint whose interpretation may be affected by the non-universality
of scalar masses is \gmt. {\it A priori}, a SUSY explanation of the discrepancy between
the Standard Model (SM) prediction and the experimental measurement of \gmt\ requires relatively
light smuons, either right- and/or left-handed, which are in $\mathbf{10}$ and
$\mathbf{\bar 5}$ representations, respectively. 
It is interesting to investigate to what extent
the tension between a
SUSY interpretation of \gmt\ and the LHC constraints on squarks 
that is present in more constrained SUSY models
could be alleviated
by the extra degree of freedom afforded by the $\mathbf{\bar 5} - \mathbf{10}$ disconnect in SU(5). 

Finally, we recall that in large parts of the regions of the CMSSM, NUHM1 and NUHM2
parameter spaces favoured at the 68\% CL the relic $\neu1$ density is brought into the
range allowed by Planck \cite{Planck15} and other data via coannihilation with the stau and other sleptons
\cite{stauco,Citron:2012fg}.
In an SU(5) GUT, the left- and right-handed sleptons are in different representations,
$\mathbf{\bar 5}$ and $\mathbf{10}$, respectively. Hence they have different masses, in general,
providing more flexibility in the realization of coannihilation. 
{Specifically}, as mentioned above, the
freedom to have $m_5 \ne m_{10}$ allows the possibility that the right-handed up- and
charm-flavour squarks, ${\tilde u_R}$ and ${\tilde c_R}$, are much lighter than the other squarks,
opening up the novel possibility of ${\tilde u_R}/{\tilde c_R} - \neu1$
coannihilation, as we discuss below.

Our analysis of the available experimental constraints largely follows those in our previous
studies of other variants of the MSSM~\cite{oldmc,mc9,mc10,mc11,MC12-DM,mcweb}, the main new feature being that we incorporate
the constraints based on the preliminary results
from LHC searches for jets + $\ETslash$ events with $\sim 13$/fb 
of data at 13 TeV~\cite{CMS:2016xva}.
For this purpose, we recast available results for simplified models with
the mass hierarchies $\mgl \gg \msq$ and vice versa.
We also include
the preliminary constraints from LHC searches in 13-TeV data for the heavy MSSM
Higgs bosons and long-lived charged particles,
and incorporate in combination the recent PandaX \cite{pandax} and LUX \cite{lux16} data.

The SUSY SU(5) GUT model we study is set up in Section~2, and our
implementations of constraints and analysis procedure are summarized in Section~3.
Section~4 describes how we characterize different Dark Matter {(DM)} mechanisms, including
the novel ${\tilde u_R}/{\tilde c_R} - \neu1$ coannihilation mechanism, {${\tilde \nu_\tau}$
coannihilation} and a
hybrid possibility. Section~5 {contains} our results in {several} model parameter planes,
and Section~6 describes various one-dimensional likelihood functions including those
for {several} sparticle masses, \gmt\ and various other observables. 
Higgs boson branching ratios {(BRs)} are presented in Section~7, followed by
  a comparison of the SU(5) with the NUHM2 results in Section~8.
The possibility of a
long-lived $\staue$ is discussed in Section~9, and the prospects for direct DM
detection are discussed in Section~10. Finally, Section~11 presents a summary
and some conclusions.


\section{Supersymmetric SU(5) GUT Model}

We assume a universal, SU(5)-invariant gaugino mass parameter $m_{1/2}$, which is
input at the GUT scale, as are the other SUSY-breaking parameters listed below.

We assume the conventional multiplet assignments of matter fields in the minimal superymmetric GUT:
\begin{equation}
(q_L, u^c_L, e^c_L)_i \; \in \; \mathbf{10}_i, \; \; (\ell_L, d^c_L)_i \; \in \; \mathbf{\bar 5}_i \, ,
\label{assignments}
\end{equation}
where the subscript $i = 1, 2, 3$ is a generation index. The only relevant Yukawa couplings
are those of the third generation, particularly that of the $t$ quark (and possibly the $b$ quark 
and the $\tau$ lepton) that may play an important role
in generating electroweak symmetry breaking. In our discussion of flavour constraints,
we assume {the MFV scenario in which} generation mixing is described by the
Cabibbo-Kobayashi-Maskawa {(CKM)} model.
This is motivated by phenomenological constraints on low-energy flavour-changing neutral
interactions, as is our assumption that the soft SUSY-breaking scalar masses for
the different $\mathbf{10}_i$ and $\mathbf{\bar 5}_i$ representations are
universal in generation space, and are denoted by
$m_{10}$ and $m_{5}$, respectively. In contrast to the CMSSM, NUHM1 and NUHM2, we
allow $m_5 \ne m_{10}$. We assume a universal soft trilinear SUSY-breaking parameter~$A_0$.

We assume the existence of two Higgs doublets $H_u$ and $H_d$ in $\mathbf{5}$ and $\mathbf{\bar 5}$ 
representations that break electroweak symmetry and give masses to the charge +2/3 and charge -1/3 
and -1 matter fields, respectively. It is well known that this assumption gives a (reasonably)
successful relation between the masses of the $b$ quark and the $\tau$ lepton~\cite{mbmtau}, but not for
the lighter charge -1/3 quarks and charged leptons. We assume that whatever physics resolves
this issue is irrelevant for our analysis, as would be the case, for instance, if corrections to the
naive SU(5) mass relations were generated by higher-dimensional superpotential terms~\cite{EG}.
In the absence of any phenomenological constraints, we allow the soft SUSY-breaking
contributions to the $H_u$ and $H_d$ masses, $m_{H_u}$ and $m_{H_d}$, to be different
from each other, as in the NUHM2, as well as from $m_5$ and $m_{10}$. As in the CMSSM, NUHM1
and NUHM2, we allow the ratio of Higgs vacuum expectation values, $\tan \beta$, to be a free
parameter.

In addition to these electroweak Higgs representations, we require one or more Higgs
representations to break the SU(5) GUT symmetry. The minimal possibility is a single
$\mathbf{24}$ representation $\Sigma$, but we do not commit ourselves to this minimal scenario.
It is well known that this scenario has problems with rapid proton decay~\footnote{We note that this problem
becomes less severe for supersymmetry-breaking scales beyond a TeV \cite{eelnos}.}
 and GUT
threshold effects on gauge coupling unification. We assume that these issues are 
resolved by the appearance of additional fields at or around the GUT scale that are otherwise
irrelevant for TeV-scale phenomenology. 
The effective low-energy Higgsino mixing coupling
$\mu$ is a combination of an input bilinear $H_u H_d$ coupling and possible trilinear and
higher-order couplings to GUT-scale Higgs multiplets such as $H_u \Sigma H_d$. We assume
that these combine to yield $\mu = \order{1} \tev$ and positive, without entering into the
possibility of some dynamical mechanism, and commenting {below} only briefly on the case $\mu < 0$.


\section{Implementations of Constraints and Analysis Procedure}


Our treatments in this paper of many of the relevant constraints follow very closely the implementations
in our previous analyses of other supersymmetric models~\cite{oldmc,mc9,mc10,mc11,MC12-DM}.
For the convenience of the reader, we summarise the constraints in Table~\ref{tab:constraints}.
In the following subsections we review our implementations, highlighting new constraints
and instances where we implement constraints differently from our previous work. 

\subsection{Electroweak and Flavour Constraints}

We treat as Gaussian constraints all electroweak precision observables, all $B$-physics and $K$-physics
observables except for \bsdmm.
The $\chi^2$ contribution from \bsdmm, combined here in the quantity
$R_{\mu\mu}$~\cite{mc9},  is calculated using a combination of the
CMS~\cite{CMSBsmm} and LHCb~\cite{LHCbBsmm} results described
in~\cite{CMSLHCbBsmm} with the more recent result from ATLAS~\cite{ATLASBsmm}.
{We extract the corresponding $\chi^2$ contribution in 
Table~\ref{tab:constraints} by applying to the 2-dimensional likelihood
provided by the combination of these experiments the minimal flavour violation (MFV) assumption
that applies in the SU(5) model}.
We calculate the elements of the CKM matrix using only experimental
observables that are not included in our set of flavour constraints.

We have updated our implementations of all the flavour constraints, and now use the current world average
value of $\mt$~\cite{WAmt}. These and all other constraints whose implementations have been changed are indicated by arrows and
boldface in Table~\ref{tab:constraints}.

\begin{table*}[htb!]
\renewcommand{\arraystretch}{1.15}
\begin{center}
\small{
\begin{tabular}{|c|c|c|} \hline
Observable & Source & Constraint \\
& Th./Ex.  & \\
\hline \hline
$\to \quad \mathbf{\mt}$ [GeV]         & \cite{WAmt} &{ $\mathbf{173.34\pm0.76}$} \\
\hline
$\Delta\alpha_{\rm had}^{(5)}(\MZ)$ 
                    & {\cite{PDG2015}}   & { $0.02771 \pm 0.00011$} \\
\hline
$\MZ$ [GeV]         & \cite{lepewwg,gfitter2013}             & $91.1875\pm0.0021$  \\ 
\hline\hline
$ \Gamma_{Z}$ [GeV] &\cite{Svenetal}
                                 /\cite{lepewwg,gfitter2013} & $2.4952\pm0.0023\pm0.001_{\rm SUSY}$   \\ 
\hline
$\sigma_{\rm had}^{0}$ [nb] &\cite{Svenetal} 
                                /\cite{lepewwg,gfitter2013}  & $41.540\pm0.037$    \\
\hline
$R_l$ &\cite{Svenetal}   
      /\cite{lepewwg,gfitter2013} &$20.767\pm0.025$    \\ 
\hline
$ A_{\rm FB}(\ell)$ &\cite{Svenetal}   
                   /\cite{lepewwg,gfitter2013} &$0.01714\pm0.00095$ \\ 
\hline
$ A_{\ell}(P_\tau)$ &\cite{Svenetal}   
                  /\cite{lepewwg,gfitter2013} & 0.1465 $\pm$ 0.0032 \\ 
\hline
$ R_{\rm b}$ &\cite{Svenetal}   
            /\cite{lepewwg,gfitter2013} & 0.21629 $\pm$ 0.00066 \\ 
\hline
$ R_{\rm c}$ &\cite{Svenetal}   
            /\cite{lepewwg,gfitter2013} & 0.1721 $\pm$ 0.0030 \\ 
\hline
$ A_{\rm FB}({b})$ &\cite{Svenetal}   
                  /\cite{lepewwg,gfitter2013} & 0.0992 $\pm$ 0.0016 \\ 
\hline
$ A_{\rm FB}({c})$ &\cite{Svenetal}   
                  /\cite{lepewwg,gfitter2013} & 0.0707 $\pm$ 0.0035 \\ 
\hline
$ A_{b}$  &\cite{Svenetal}   
          /\cite{lepewwg,gfitter2013} & 0.923 $\pm$ 0.020 \\ 
\hline
$ A_{c}$ &\cite{Svenetal}   
         /\cite{lepewwg,gfitter2013} & 0.670 $\pm$ 0.027 \\ 
\hline
${\ALRe}$ &\cite{Svenetal}   
                     /\cite{lepewwg,gfitter2013} & 0.1513 $\pm$ 0.0021 \\ 
\hline
$ \sin^2 \theta_{\rm w}^{\ell}(Q_{\rm fb})$ 
        &\cite{Svenetal}   
        /\cite{lepewwg,gfitter2013} & 0.2324 $\pm$ 0.0012 \\ 
\hline
$\MW$ [GeV]
     &\cite{Svenetal}
     /\cite{lepewwg,gfitter2013} & {$80.385 \pm 0.015\pm0.010_{\rm SUSY}$ } \\
\hline\hline
$ a_{\mu}^{\rm EXP} - a_{\mu}^{\rm SM}$
     &\cite{g-2}
     /\cite{newBNL}
     &$(30.2 \pm 8.8 \pm 2.0_{\rm SUSY})\times10^{-10}$ \\
\hline
$\to \quad \mathbf{\Mh}$ [GeV]
     & \cite{FH,FeynHiggs}
     / \cite{Aad:2015zhl} & { $\mathbf{125.09 \pm 0.24 \pm 1.5_{\rm SUSY}}$} \\
\hline\hline
{$\to \quad$ \bf{BR}${_{{b \to s \gamma}}^{\rm EXP/SM}}$}
     &{\cite{Misiak}/\cite{HFAG}}
     &{$\mathbf{1.021 \pm 0.066_{\rm EXP}}$ } \\
     & 
     &\phantom{1.021}{$\mathbf{\pm 0.070_{\rm TH, SM} \pm 0.050_{\rm TH,SUSY}}$}  \\
\hline
$\to \quad  \mathbf{R_{\mathbf{\mu\mu}}}$
     & \cite{BobethBmm}/\cite{CMSLHCbBsmm,ATLASBsmm}   & {\bf 2D likelihood}, {\bf {MFV}} \\

\hline
{$\to \quad$ \bf{BR}${_{B \to \tau\nu}^{\rm EXP/SM}}$}
     & {\cite{HFAG,Kronenbitter:2015kls}}
     &  $\mathbf{1.02 \pm 0.19_{\rm EXP} \pm 0.13_{\rm SM}}$ \\
\hline
{$\to \quad \mathbf{BR}_{B \to X_s \ell \ell}^{\rm EXP/SM}$}
     & \cite{Xsllth}/\cite{HFAG}
     & $\mathbf{0.99 \pm 0.29_{\rm EXP} \pm 0.06_{\rm SM}}$ \\
\hline
{$\to \quad \mathbf{BR}_{K \to \mu \nu}^{\rm EXP/SM}$}
     & \cite{SuFla,Marciano:2004uf} /\cite{PDG2015}
     & $\mathbf{0.9998 \pm 0.0017_{\rm EXP} \pm 0.0090_{\rm TH}}$ \\
\hline
{$\to \quad \mathbf{BR}_{K \to \pi \nu \bar{\nu}}^{\rm EXP/SM}$}
     & \cite{BurasKpinn15}/\cite{Artamonov:2008qb}
     & $\mathbf{2.2 \pm 1.39_{\rm EXP} \pm 0.20_{\rm TH}}$ \\
\hline
{$\to \quad \mathbf{\Delta M}_{B_s}^{\rm EXP/SM}$}
     & {\cite{Buras:2000qz, SuFla} /\cite{HFAG}}
     & {$\mathbf{1.016 \pm  0.074_{\rm SM}}$ }\\
\hline
{$\to \quad {\frac{\mathbf{\Delta M}_{B_s}^{\rm EXP/SM}}
           {\mathbf{\Delta M}_{B_d}^{\rm EXP/SM}}}$}
     & \cite{Buras:2000qz,SuFla} /{\cite{HFAG}}
     & {$\mathbf{0.84 \pm 0.12_{\rm SM}}$ }\\
\hline
{$\to \quad \mathbf{\Delta \epsilon}_K^{\rm EXP/SM}$}
     & \cite{Buras:2000qz,SuFla} /{\cite{PDG2015}}
     & $\mathbf{1.14 \pm 0.10_{\rm EXP+TH}}$ \\
\hline \hline
$\to \quad \mathbf{\Omega_{\mathbf\rm CDM} h^2}$
     &\cite{MicroMegas,SSARD}/\cite{Planck15}
         &{$\mathbf{0.1186 \pm 0.0020}_{\rm EXP} \mathbf{\pm 0.0024_{\rm TH}}$} \\
\hline
$\to \mathbf{\ssi}$ & \cite{pandax,lux16} & $\mathbf{(\mneu{1}, \ssi)}$ {\bf plane} \\
\hline\hline
$\to \quad$ {\bf{Heavy stable charged particles}} & {\cite{CMS:2016ybj}} & {\bf Fast simulation based on} \cite{CMS:2016ybj, Khachatryan:2015lla} \\
\hline
$\to \quad \mathbf{\sq \to q \neu1}, \mathbf{\tilde g \to f \bar f \neu1}$ & {\cite{CMS:2016xva}} &
$\mathbf{\sigma \cdot {\rm BR}}$ {\bf limits in the}
$\mathbf{(\msq, m_{\tilde\chi_1^0})}, \mathbf{(\mgl, m_{\tilde\chi_1^0})}$ \bf{planes} \\ \hline
$\to \mathbf{H/A \to \tau^+ \tau^-}$ & \cite{CMSHA,HBtautau,ATLAS:2016fpj} & 
{\bf 2D likelihood, $\mathbf{\sigma \cdot {\rm BR}}$ limit}  \\
\hline 
\end{tabular}
\caption{\it List of experimental constraints used in this work, including experimental and (where applicable)
theoretical errors: supersymmetric theory uncertainties are indicated separately. Instances where our
implementations differ from those in Table~1 in~\protect\cite{mc11} are
indicated by arrows and boldface.}  \label{tab:constraints}}  

\end{center}

\end{table*}


\subsection{Higgs Constraints} 
\label{sec:higgsconstraints}

We use the combination of ATLAS and CMS
measurements of the mass of the Higgs boson: 
$\Mh = 125.09 \pm 0.24 \gev$~\cite{Aad:2015zhl}. We employ the {\tt FeynHiggs 2.11.2} code~\cite{FH,FeynHiggs}
to evaluate
the constraint this imposes on the parameter space, assuming a one-$\sigma$ theoretical
uncertainty of $1.5 \gev$~\footnote{We use a modified version
of {\tt FeynHiggs 2.11.2} that includes two-loop QCD corrections in the
evaluation of the \DRbar\ running top mass and an improved evalution
of the top mass in the \DRbar-on-shell conversion for the scalar tops.}.

The $\chi^2$ contributions of {85} Higgs search channels from the LHC and the Tevatron
are evaluated using {\tt HiggsSignals}, see~\cite{HiggsSignals}, 
where a complete list of references can be found. 
The $\chi^2$ contributions from the limits from searches for the heavy
neutral MSSM Higgs bosons in
the $H/A \to \tau^+\tau^-$ channels are evaluated using the code 
{\tt HiggsBounds}~\cite{HiggsBounds,HBtautau}, which incorporates the
results of CMS searches~\cite{CMSHA,HBtautau} with $\sim 25~\ifb$ of
8\,TeV data. The contributions from the two
possible production modes, $gg \to H/A$ and $b \bar b \to H/A$, are
combined in
a consistent manner, depending on the MSSM parameters. 
The results from {\tt HiggsBounds} have been compared
with the published CMS analysis, and are in very good agreement~\cite{HBtautau}. 
The corresponding $\chi^2$ contribution is labelled as ``2D likelihood'' in 
Table~\ref{tab:constraints}.
For the corresponding constraint with $~13$ fb$^{-1}$ of 13 TeV data,
we implement an approximate treatment of the 
$\chi^2$ contribution using
the preliminary result of ATLAS \cite{ATLAS:2016fpj},
as we describe in more detail below.
Limits from other Higgs boson searches are not relevant for the
investigation in this paper and are therefore not included.


\subsection{LHC \boldmath{$\ETslash$} constraints at 13 TeV}

ATLAS and CMS have recently announced preliminary results from $\ETslash$ searches
with $\sim 13$/fb of data at 13~TeV, using simplified models for gluino and squark pair production \cite{ATLAS:2016kts, CMS:2016xva}.
These searches assume $\mgl \ll \msq$ and $\msq \ll \mgl$, respectively, and 100\%
{BRs} for the decays $\gl \to f {\bar f} \neu1$ ($f = q, b, t$) and $\sq \to q \neu1$, respectively,
which maximize the possible corresponding~~$\ETslash$ signatures. Neither of these 
assumptions is valid in the SUSY SU(5) GUT model: as we will see in more detail later,
the $\mgl$ and $\msq$ masses are quite similar in much of the favoured region of parameter
space~\footnote{An exception is provided by the ${\tilde u_R}$ and ${\tilde c_R}$, which may
be much lighter than the gluino and other squarks in some regions of
parameter space. We will discuss 
this possibility in detail below.}, and in general other decay modes 
dilute the~~$\ETslash$ signature, although larger-multiplicity final states may compensate through an increase in transverse energy $H_T$~\cite{Cohen:2016nzv}. These other decay modes populate other search channels
including leptons, which we do not consider in this paper as they were of
limited importance in our previous analyses of the CMSSM, NUHM1 and NUHM2, having impact
only for relatively large squark masses and small $m_{1/2}$. 

Fig.~\ref{fig:sigmafactors} displays the ratios of the $\gl \gl$ cross section (left panel)
and the $\sq \sq + \sq \asq$
cross section (right panel) that we find in ranges of $\msq$ and $\mgl$
that are representative of those favoured in our analysis before implementing the LHC 13-TeV
$\ETslash$ constraint, relative to the cross sections found in the simplified models with
$\mgl \ll \msq$ and $\msq \ll \mgl$, respectively. We have used {\tt NLL-fast-3.1} \cite{Beenakker:2015rna} to obtain 
the cross section at NLO + NLL level.
{In both plots a large area at higher squark masses is visible, as
  well as a thin strip at $\sim 500 \gev$. The latter corresponds to
  lighter ${\tilde u_R}$ and ${\tilde c_R}$ discussed below.}
We see that the $\gl \gl$ cross section (left panel) is generally {\it smaller} than in the corresponding simplified
model by a factor $> 2$ due to the destructive interference between the $s$-channel gluon exchange diagram and 
the $t$-channel squark exchange diagram in $q \bar q \to \gl \gl$, thus weakening the LHC constraints as discussed below. 
On the other hand, the 
$\sq \sq + \sq \asq$ cross section (right panel) is generally {\it a factor $\gtrsim 10$ larger}
than in the simplified model, except in the  ${\tilde u_R}/{\tilde c_R} - \neu1$ coannihilation strip
at small $m_{\tilde u_R}, m_{\tilde c_R}, \mneu1 \sim 500 \gev$ and $m_{1/2} \sim 2500 \gev$, to which we return later.
The enhancement of the squark cross-section is due to the fact that in the 
squark-neutralino simplified model
there is no production mode with total baryon number 
$B = 2/3$, $qq \to \tilde q \tilde q$, because
gluinos are assumed to be absent.
  On the other hand, in our model $\mgl \sim \min(\msq)$, and $qq \to \tilde q
  \tilde q$ (with $t$-channel $\gl$ exchange)
becomes the dominant squark production mode
in the large $\msq$ region, due to the valence quark-parton
dominance in the proton in the large $x$ regime.

\begin{figure*}[htb!]
\begin{center}
\hspace{6mm}
\resizebox{7.5cm}{!}{\includegraphics{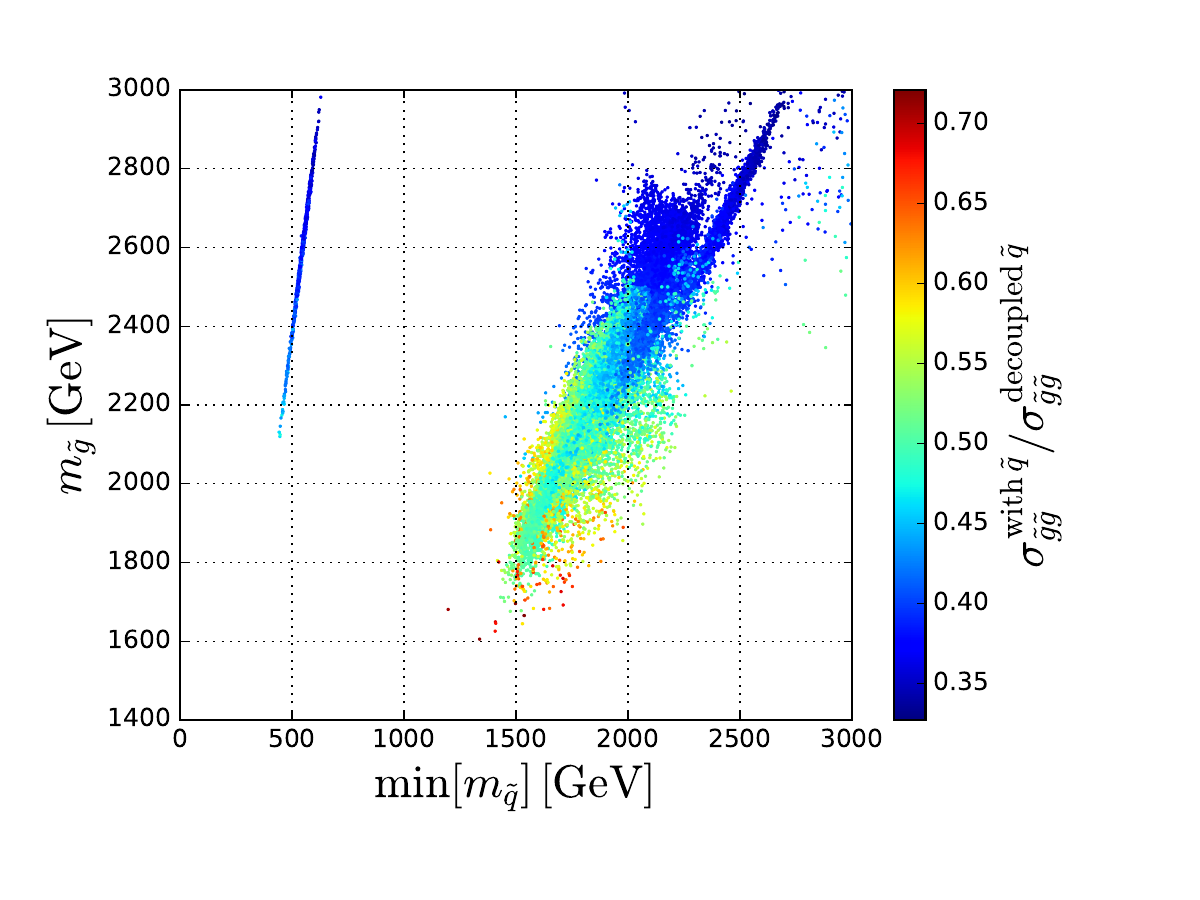}}  \hspace{0.5mm}
\resizebox{7.5cm}{!}{\includegraphics{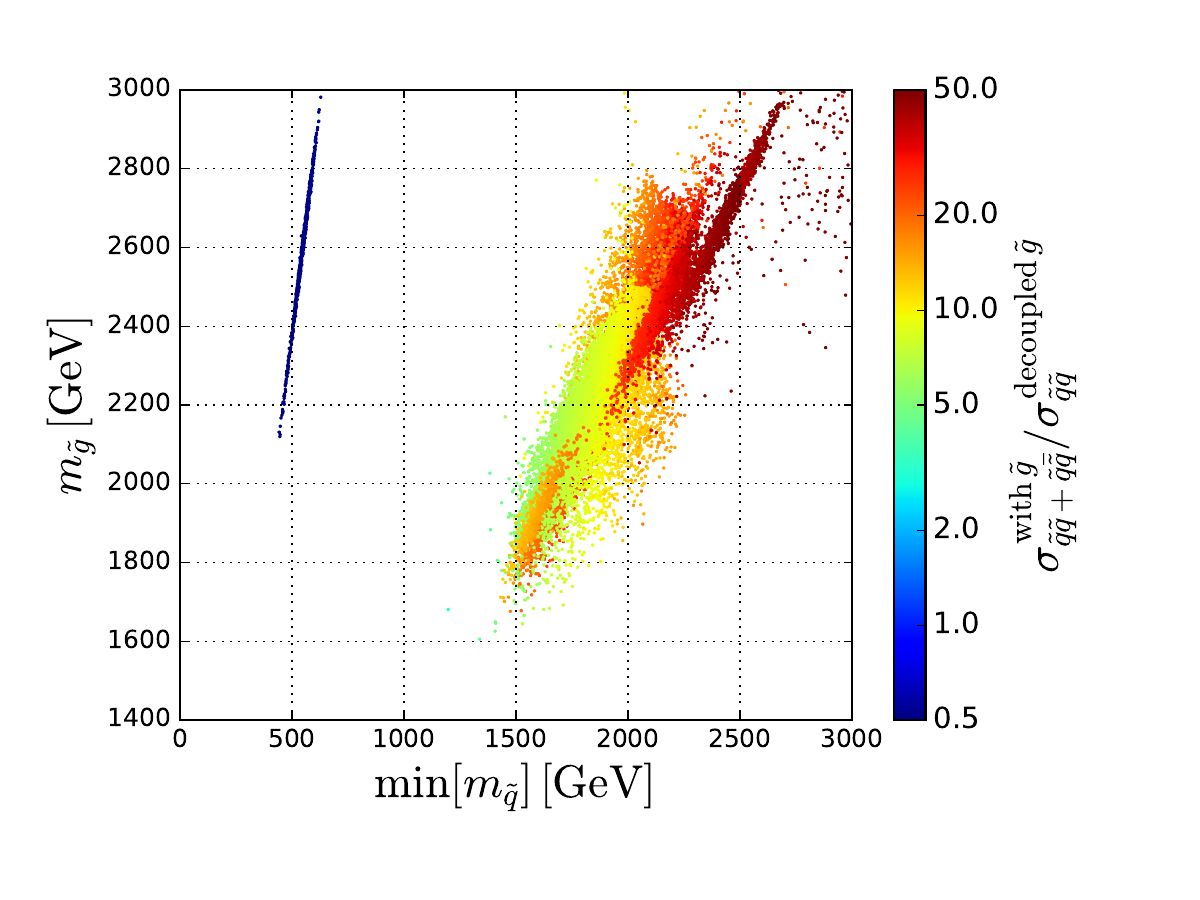}} \\
\end{center}
\vspace{-0.5cm}
\caption{\it 
Left panel: the ratio of the $\gl \gl$ cross section that we find in the
range of $\msq$ and $\mgl$ favoured in our analysis before implementing
the LHC 13-TeV $\ETslash$ constraint, relative to the cross section
found in the simplified model with $\mgl \ll \msq$. Right panel: the
corresponding ratio of the $\sq \sq + \sq \asq$  cross section, relative
to the cross section for $\sq \asq$ found in the simplified model with
$\msq \ll \mgl$. 
}
\label{fig:sigmafactors}
\end{figure*}

Fig.~\ref{fig:gluinodecays} displays the CMS 95\% confidence limits
in the $(\mgl, \mneu1)$ plane from a
hadronic jets plus  $\ETslash$ search~\cite{CMS:2016xva} within a simplified model
assuming that the decay mode $\gl \to q {\bar q} \neu1$ occurs with
100\% BR (solid black lines). 
These limits are compared with the best-fit points 
(green stars) 
and the regions in the fits that are preferred at $\Delta\chi^2 = 2.30$ and $\Delta\chi^2 = 5.99$ 
(red and blue contours, respectively). Here and in the following 
analogous parameter planes, we use the $\Delta\chi^2 = 2.30$ and 
$\Delta\chi^2 = 5.99$ contours as proxies for the 
boundaries of the 68\% and 95\% CL regions in the fit. 

\begin{figure*}[htb!]
\vspace{1.5cm}
\begin{center}
\resizebox{7.5cm}{!}{\includegraphics{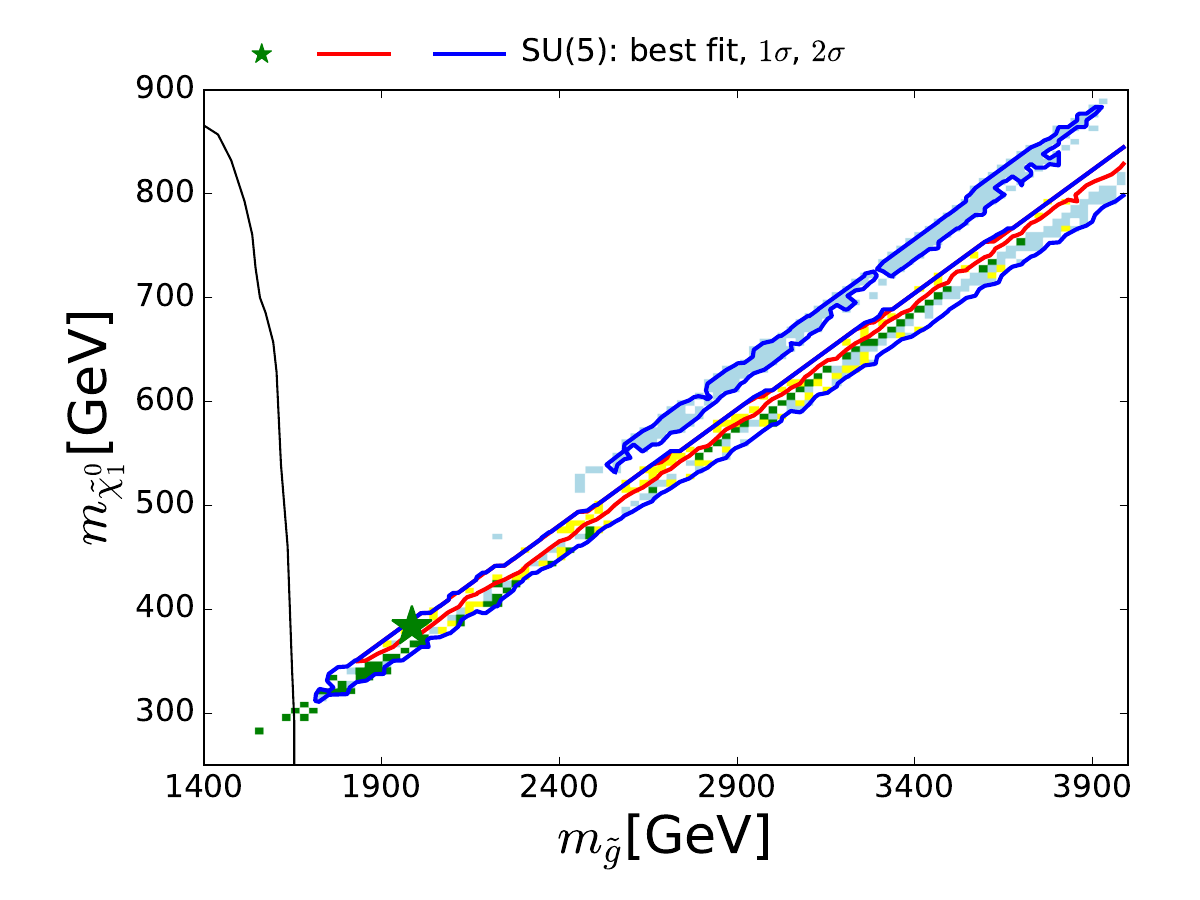}} 
\resizebox{7.5cm}{!}{\includegraphics{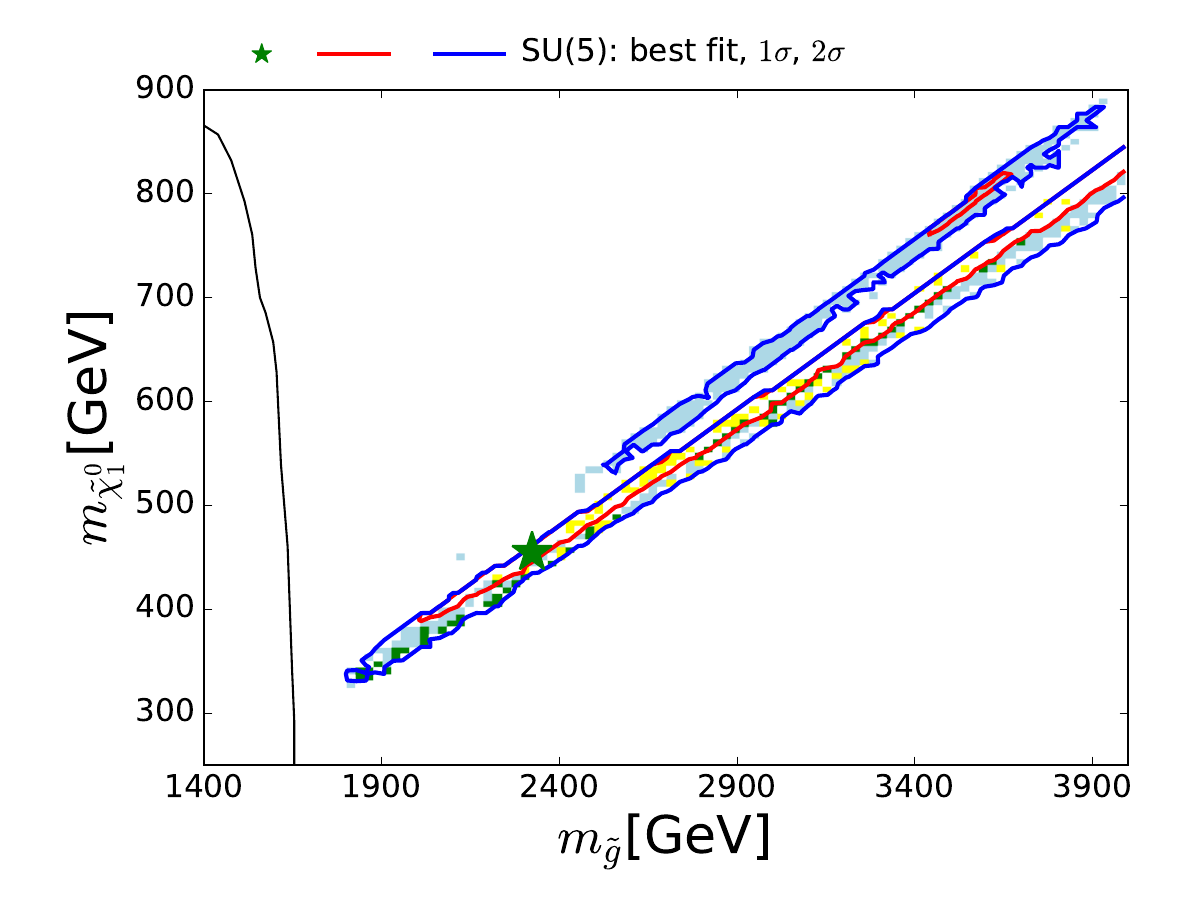}}  \\ 
\resizebox{12cm}{!}{\includegraphics{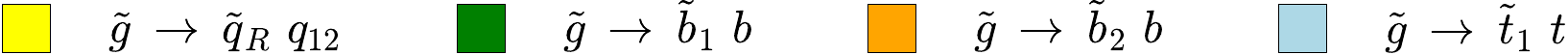}} \\
\end{center}
\vspace{-0.5cm}
\caption{\it 
The solid lines show the CMS 95\% CL exclusion in the $(\mgl, \mneu1)$ plane \cite{CMS:2016xva},
assuming a simplified model with heavy squarks and 100\% BR for $\gl \to q {\bar q} \neu1$.
The left (right) panel shows the best-fit point (green star), 68 and 95\% CL contours (red and blue lines, respectively) for
$(\mgl, \mneu1)$ obtained without (with) the CMS 13-TeV constraint.
The dominant ($> 50$\%) ${\tilde g}$ decays into first- and
second-generation quarks and squarks $\tilde q_{L,R}$ and third-generation
quarks and squarks ${\tilde t/\tilde b}_{1,2}$ found in the SUSY SU(5)
model are colour-coded as indicated.}
\label{fig:gluinodecays}
\end{figure*}

In addition, within the 95\% CL
region in Fig.~\ref{fig:gluinodecays} we have indicated
the dominant ($> 50$\%) ${\tilde g}$ decays found in our analysis.
We note that many model points do not have any decay mode with BR
$> 50$\% within the 95\% CL region
and that, for those that do, the dominant decays are two-body $\gl \to \sq {\bar q}$
modes that were not considered in~\cite{CMS:2016xva}. 
Because of this and the fact that 
the $\gl \gl$ cross section is always smaller than
in the gluino simplified model by a factor $> 2$ (see the left panel of Fig.~\ref{fig:sigmafactors}),
the LHC 13-TeV $\ETslash$ constraint from the gluino simplified model  {has only} negligible impact.
Our LHC 13-TeV $\ETslash$ constraint on the gluino mass 
actually comes indirectly from the squark mass constraint 
estimated using the squark simplified model discussed below,
since {the squark and gluino masses are related via renormalization group evolution in the SU(5) model}. 
The left panel in 
Fig.~\ref{fig:gluinodecays} was obtained before implementing the LHC 13-TeV
$\ETslash$ 95\% confidence limit on gluino and squark pair-production, while in the right panel 
this constraint is included.
We note that the simplified model
exclusion in this analysis extended to $\mgl \lesssim 1650 \gev$, below the gluino mass at the pre-LHC 13~TeV best-fit point,
and barely reaching the 68\% CL contour (solid red line). 

Fig.~\ref{fig:squarkdecays} contains an analogous 
set of planes for CMS $\ETslash$ searches for squarks,
where the CMS limit assuming a simplified model with heavy gluino and 100\%
BRs for $\sq \to q \neu1$ is displayed
(black lines): {the solid lines assume that all the squarks of the first two generations
are degenerate, the dashed lines assume two degenerate squarks, and the dotted lines assume just one squark.}
The planes in the {upper} panels display 
$\mneu1$ and the masses of the
first- and second-generation right-handed up-type squarks {(here
  commonly denoted as $\tilde u_{R}$)},
while the planes in the {lower} panels are for
the down-type squarks {(here commonly denoted as $\tilde d_{R}$)}.
The main decay modes of the \SuR\ ({upper}) and the \SdR\ ({lower})
are indicated
over much of the preferred parameter 
space, and we note that the dominant ($> 50$\%) decay modes of both
right-handed up- and down-type squarks are indeed into the corresponding
quark flavour + $\neu1$ for nearly the whole 68\% CL regions, as assumed
in the squark simplified-model search. This is, however, 
not the case for the left-handed up- and down-type squarks
(not shown), whose dominant decays are into $\cha1$ and electroweak doublet partner quark flavours.
Furthermore, within the displayed 95\% CL regions there are also large
areas
where decays into gluinos, not considered in the simplified model, are
dominant.

\begin{figure*}[htb!]
\begin{center}
\resizebox{7.5cm}{!}{\includegraphics{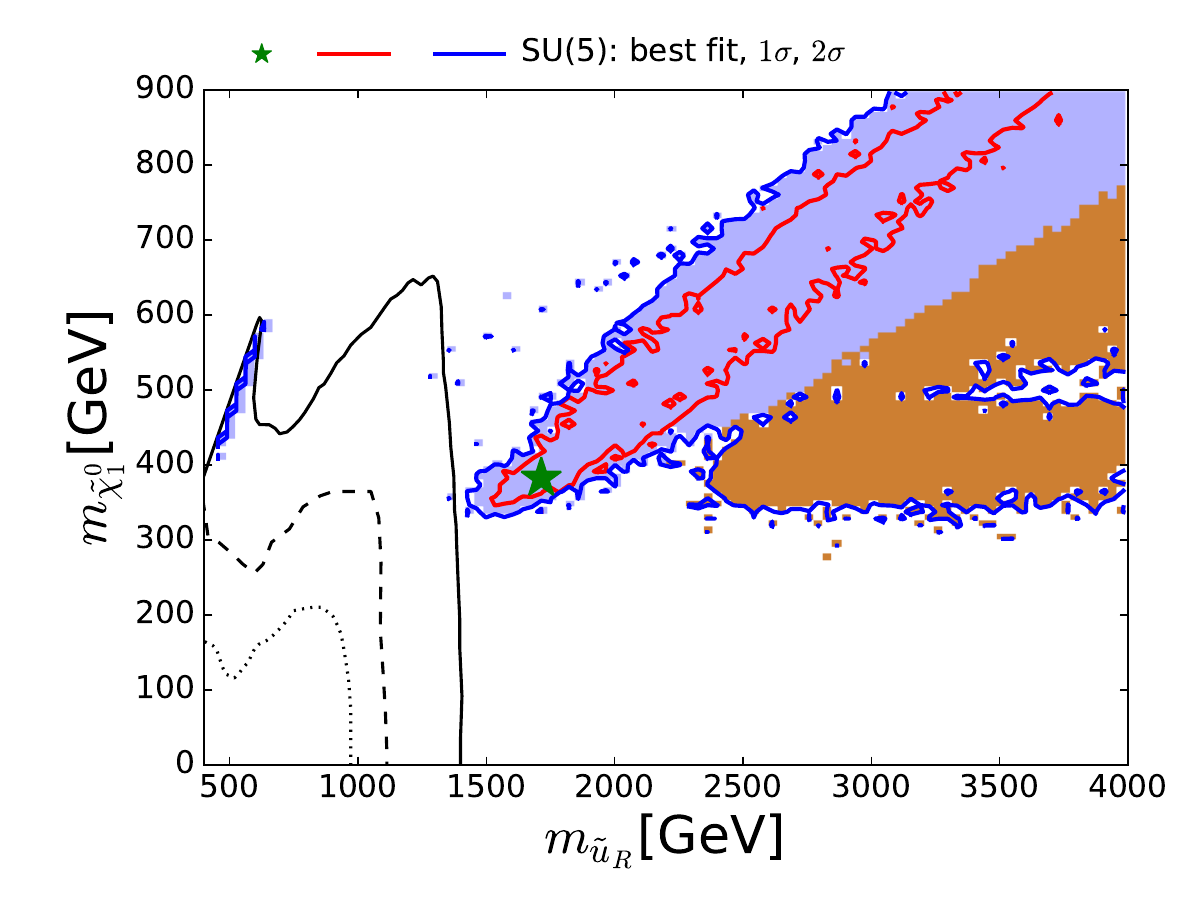}}  \hspace{5mm}
\resizebox{7.5cm}{!}{\includegraphics{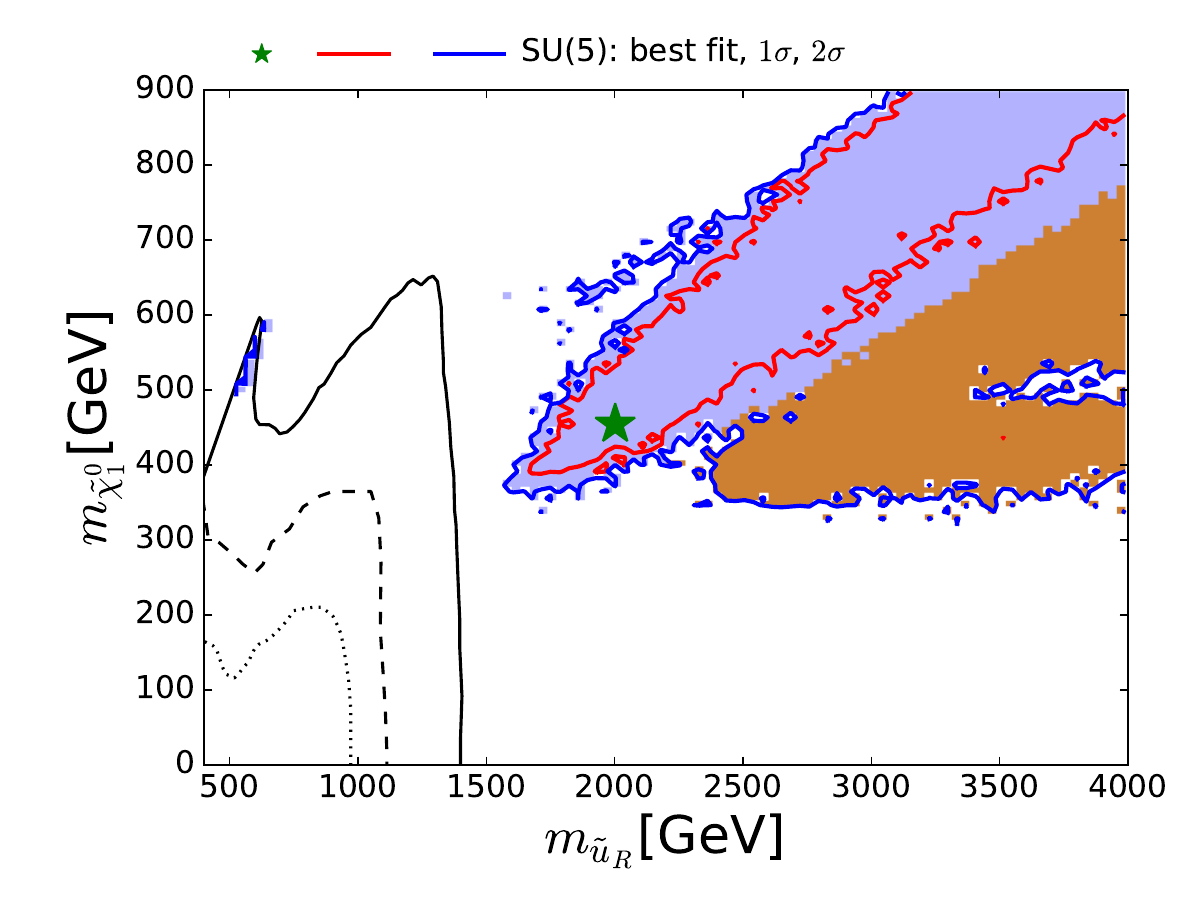}}  \\
\resizebox{6cm}{!}{\includegraphics{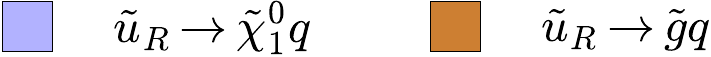}} \\[3em]
\resizebox{7.5cm}{!}{\includegraphics{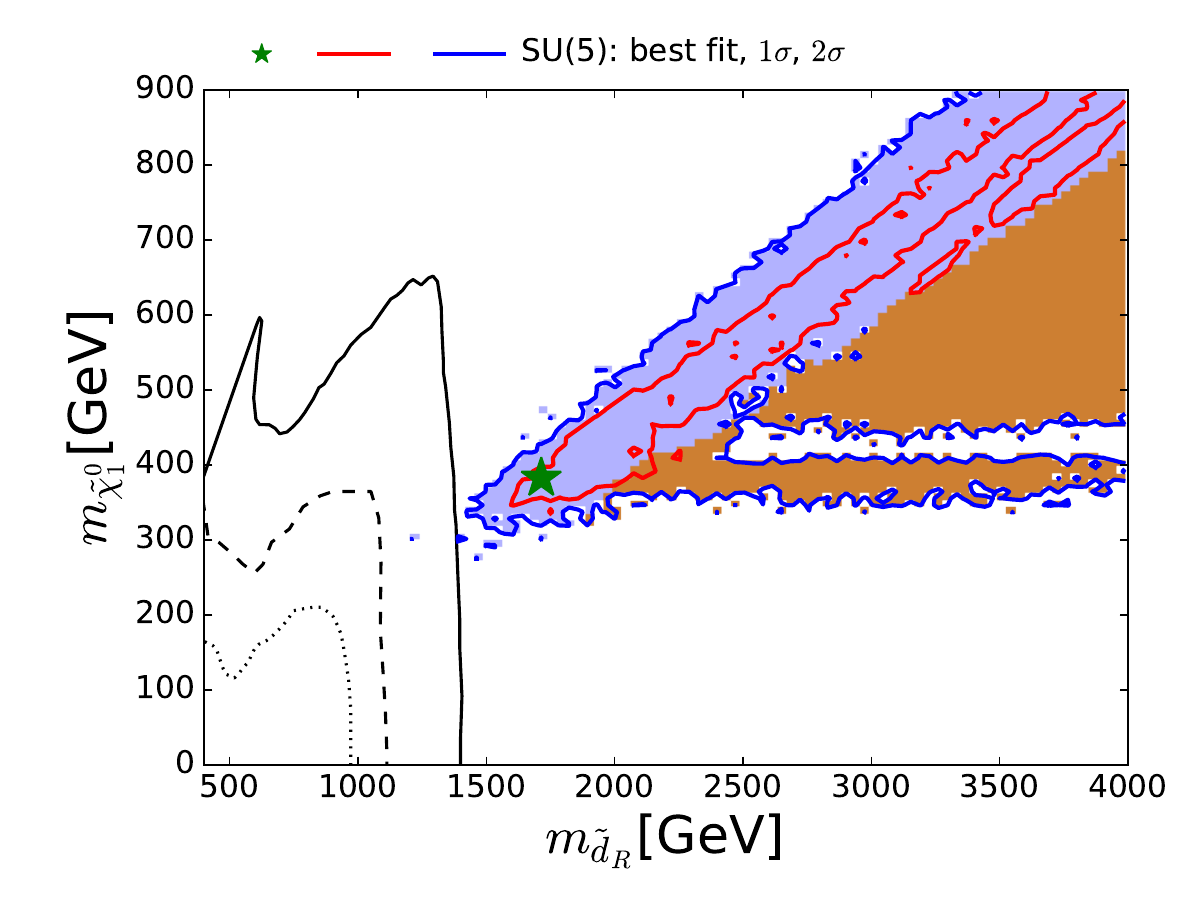}}
\hspace{5mm}
\resizebox{7.5cm}{!}{\includegraphics{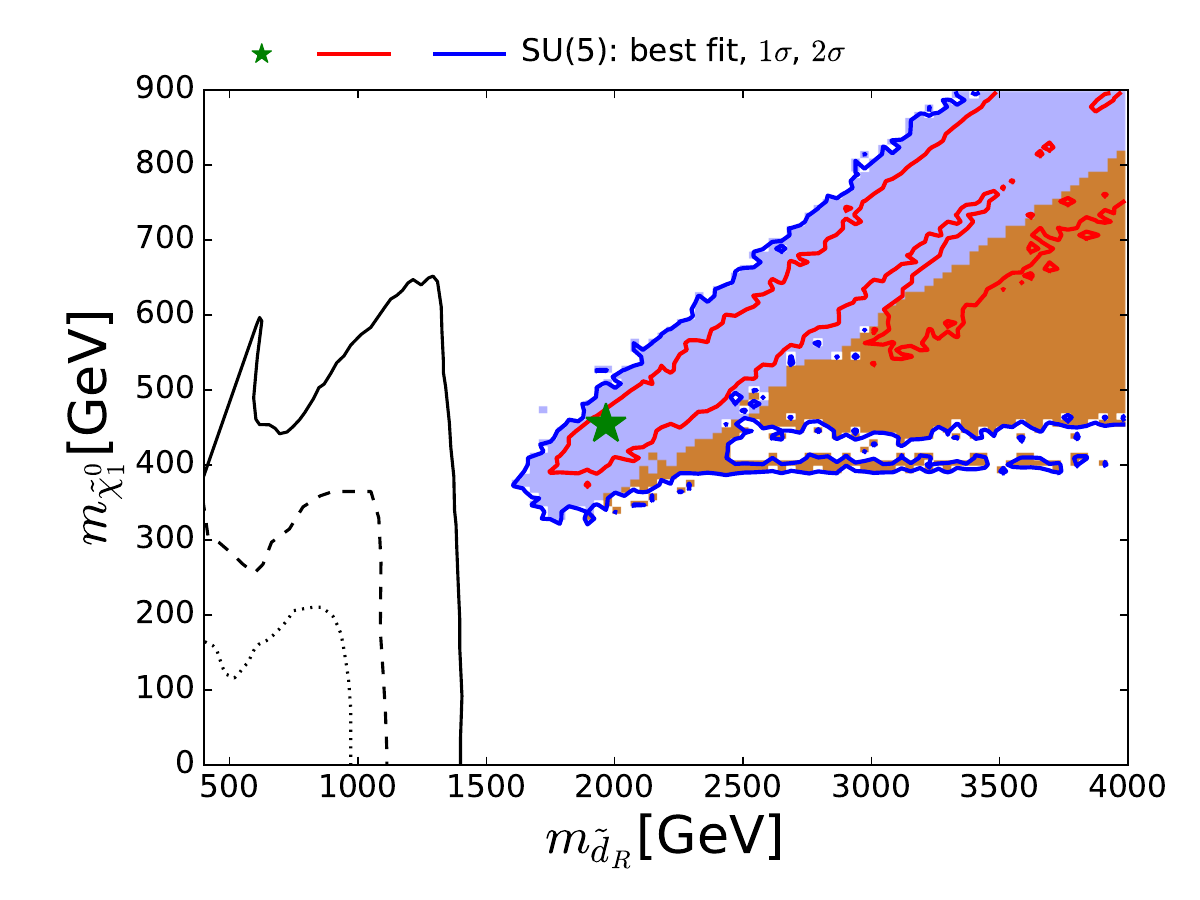}} \\
\resizebox{6cm}{!}{\includegraphics{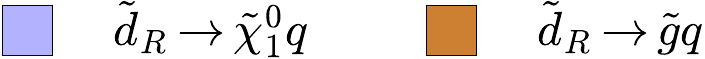}} \\
\end{center}
\caption{\it 
The black lines show the CMS 95\% CL exclusion in the $(\msq, \mneu1)$
plane \cite{CMS:2016xva}, assuming a simplified model with heavy gluinos
and 100\% BR for $\sq \to q  \neu1$: {the solid lines assume that all
  the squarks of the first two generations are degenerate, the dashed
  lines assume two degenerate squarks, and the dotted lines assume just
  one squark.} {All} panels show the best-fit point (green star), 68
and 95\% CL contours (red and blue lines, respectively) for $\mneu1$ and
the masses of the first- and second-generation right-handed up-type
squarks $\tilde u_{R}$ ({upper} panels) and the down-type squarks $\tilde
d_{R}$ ({lower} panels). {In both cases}, the left panels were obtained without the CMS
13-TeV constraint, and the right panels include it. The dominant ($>
50$\%) $\sq$ decays found in the SUSY SU(5) model are colour-coded as
indicated. }
\label{fig:squarkdecays}
\end{figure*}

Because the ${\tilde q_R} \to q \, \neu1$ decays are important, and also because the
$\sq \sq + \sq \asq$ cross section in our sample is much larger than that found at large $\msq$
for $\sq \asq$ in the simplified model with $\msq \ll \mgl$, as seen in the right panel of
Fig.~\ref{fig:sigmafactors}, we have implemented
a recast of this search in our global analysis~\footnote{The $\SucR - \neu1$ coannihilation
strip visible in the upper {panels} of Fig.~\ref{fig:squarkdecays} at $m_{\tilde u_R} \simeq \mneu1 \sim 500 \gev$
is the subject of a later dedicated discussion.}, and the comparison
between the left panels (without this contribution) and the right panels 
(with this contribution)
in Fig.~\ref{fig:squarkdecays} shows the importance of this constraint.

Our implementation of the LHC 13-TeV $\ETslash$ constraint is based on~\cite{CMS:2016xva}.
In this analysis, the CMS Collaboration provides a map of the 95\% CL cross-section upper limit as a function of
$\msq$ and $m_{\tilde \chi_1^0}$ assuming $pp \to \tilde q \asq$
and 100\% BR for $\tilde q \to q \tilde \chi_1^0$.
This is indeed the dominant 
production and decay mode in most parts of the 68\% CL regions of the
considered model, 
as can be seen in Figs.~\ref{fig:sigmafactors} and \ref{fig:squarkdecays}.
For each point we compare our calculation of
$(\sigma_{\sq {\asq}} + \sigma_{\sq\sq})\;{\rm BR}^2_{\tilde q \to q \tilde \chi_1^0}$ with the CMS
95\% CL upper limit on the cross section: $\sigma_{\rm UL}(m_{\tilde q, \tilde \chi_1^0})$.
We model the $\chi^2$ penalty as
\begin{equation}
\chi^2_{\tilde q}({\ETslash}) = 5.99 \cdot \Big[
\frac{(\sigma_{\sq {\asq}} + \sigma_{\sq\sq})\;{\rm BR}^2_{\tilde q \to q \tilde \chi_1^0}}{\sigma_{\rm UL}(m_{\tilde q, \tilde \chi_1^0})}
\Big]^2 \, ,
\label{eq:chi2model}
\end{equation}
so that the CMS 95\% CL upper limit corresponds to $\chi^2({\ETslash}) = 5.99$
and $\chi^2$ scales as the square of the number of signal events, $N_{\rm sig}$, which gives the right scaling.
{We have checked that our implementation (\ref{eq:chi2model}) reproduces the $\pm 1 \,\sigma$ band
in the 2-dimensional exclusion limit provided by CMS~\cite{CMS:2016xva}, 
with a discrepancy that is much smaller than the width of the $\pm 1 \,\sigma$ band.}

{The aforementioned CMS analysis~\cite{CMS:2016xva} also looks at three simplified gluino models assuming
100\% BR for $\tilde g \to f \bar f \tilde \chi_1^0$ with $f = q, b, t$, respectively,
and provides corresponding cross-section upper limit maps as a function of $\mgl$ and $m_{\tilde \chi_1^0}$.
We implement these constraints by defining $\chi^2_{\tilde g \to f \bar f \tilde \chi_1^0}({\ETslash})$
by analogy with Eq.~\eqref{eq:chi2model}.

We also consider the $pp \to \tilde q \tilde g$ process, treating it as follows.
This process is only relevant when $\msq \sim \mgl$.
In this regime, if $\msq > \mgl$ ($\mgl > \msq$), $\tilde q$ ($\tilde g$) tends to decay into $\tilde g$ ($\tilde q$), radiating soft jets.
If these soft jets are ignored, we are left with the $\gl \gl$ ($\sq \sq$) system.
In this approximation, the impact of $pp \to \tilde q \tilde g$ can therefore be estimated by adding an 
extra contribution 
$\sigma_{\tilde q \tilde g} BR_{\tilde q \to q \tilde g}$ ($\sigma_{\tilde q \tilde g} BR_{\tilde g \to q \tilde q}$)
to $\sigma_{\gl \gl}$ ($\sigma_{\sq \sq} + \sigma_{\sq \asq}$).
In general, SUSY searches are designed to look for high $p_T$ objects,
and one loses a small amount of sensitivity by ignoring soft jets.
We therefore believe that our implementation of the $pp \to \tilde q \tilde g$ process is conservative.  

Finally, we estimate the total $\chi^2$ penalty from the LHC 13-TeV $\ETslash$ constraint to be
$\chi^2(\ETslash) = \chi^2_{\tilde q}(\ETslash) + \sum_{f=q,b,t} \chi^2_{\tilde g \to f \bar f \tilde \chi_1^0}(\ETslash)$.\footnote{{One could be concerned that summing up the $\chi^2$ contributions from different simplified model limits 
would overestimate the exclusion limit, since these signal regions are not necessarily independent.
This would be indeed the case if the same event sample were confronted with multiple overlapping signal regions.
In our case, however, the signal sample is divided into statistically independent sub-samples, 
corresponding to the simplified model topologies $\tilde g \to b \bar b \tilde \chi_1^0, \to t \bar t \tilde \chi_1^0$, etc., 
and these sub-samples are confronted with the corresponding simplified model limits only once.
In such a case the $\chi^2(\ETslash)$ estimate (\ref{eq:chi2model}) provides 
a conservative limit when there is no significant excess in the data.
}}
}


\subsection{Constraints on long-lived charged particles}

We also include in our analysis LHC constraints from searches for 
heavy long-lived charged particles (HLCP)
that are, in general, relevant to coannihilation regions where the 
mass difference between the lightest SUSY particle (LSP) and the
next-to-lightest SUSY particle (NLSP)
may be small and the NLSP may therefore be long-lived. As we discuss
below, important roles
are played in our analysis by $\staue$, $\cha1$ and {\SucR} coannihilation,
but only in the $\staue$ case is the NLSP - LSP mass difference small enough to
offer the possibility of a long-lived charged particle.
We implement in our global analysis the preliminary CMS 13-TeV result \cite{CMS:2016ybj} 
using tracking and time-of-flight measurements,
based on the recipe and the efficiency map as a function of the pseudo-rapidity 
and velocity of the HLCP given in \cite{Khachatryan:2015lla}.
We use {\tt Pythia 8} \cite{Sjostrand:2007gs} and {\tt Atom} \cite{Atom} to generate and analyse the events,
and assume that the efficiencies for detecting slow-moving $\staue$s are similar at 8 and 13 TeV. 
\footnote{A similar recasting method was used in \cite{Evans:2016zau}. See
also \cite{Heisig:2015yla} for another 
approach using simplified model topologies.}
The efficiency contains a lifetime-dependent factor $\propto {\rm exp}(- d m/ p \tau)$, where $d$ is a distance $d \simeq 10$~m
that depends on the pseudorapidity, and $m, p$ and $\tau$ are the mass, momentum and
lifetime of the long-lived particle. This factor drops rapidly for particles with
lifetimes $\lesssim 10$~ps, corresponding to $m_{\staue} - \mneu1 \gtrsim 1.6 \gev$.


\subsection{Constraints on heavy neutral Higgs bosons from Run II}

Concerning the production of heavy neutral Higgs bosons, in addition to the
$8$ TeV constraints on $H/A \to \tau^+ \tau^-$ provided by  
{\tt HiggsBounds}, we also take into account the preliminary exclusion
limits obtained by ATLAS from searches for generic spin-0 bosons $\phi$
in the $\tau\tau$ final state with an integrated luminosity of 
13.3 fb$^{-1}$ at 13~TeV that were presented at the ICHEP 2016 conference and
described in~\cite{ATLAS:2016fpj} {(see also the CMS results in~\cite{CMSHA13})}.  
Upper bounds on $\sigma \times {\rm BR}({\phi} \to \tau \tau)$ are reported
for each ${M_\phi}$ separately
for the gluon fusion production channel  
and for production in association with a $b {\bar b}$ pair assuming
there is no contamination between the modes, {assuming a single resonance}.
We compute the cross sections and the BRs {in the MSSM}
using {\tt FeynHiggs}, adding the contributions for $\phi = H$
and $\phi = A$, {using the average of
the two masses, which are degenerate within the experimental resolution}. This result is
compared with the upper limit from the corresponding channel
neglecting contamination. This approach leads
to a conservative limit since we underestimate the signal yield 
in each channel by neglecting the contamination (the events from the other production mode).
As in Eq.~(\ref{eq:chi2model}), the $\chi^2$ penalties are modelled as
\begin{equation}
\chi^2({Y_i}) = 4 \cdot \Big( \frac{ \sigma_{X_i} \cdot {\rm BR}_{\tau^+ \tau^-} }{ \sigma_{Y_i}^{\rm UL}(\MA) } \Big)^2 \,,
\end{equation}
where $X_i = (gg \to {H/A}$, $pp \to b \bar b {H/A})$ is the production mode, $Y_i = (ggF, bb{\phi})$ is the corresponding search channel
and $\sigma^{\rm UL}(\MA)$ is the 95\% CL upper limit evaluated at $\MA
{(\approx \MH)}$ by ATLAS \cite{ATLAS:2016fpj}.
Finally we take the stronger $\chi^2$ rather than combining them, 
in order to be on the
conservative side~\footnote{{ A more conservative approach would be to
choose the strongest search channel based on the expected sensitivity rather than that observed.
However, the expected limit} {shown in~\cite{ATLAS:2016fpj} is similar to that observed,
so we do not expect that this approach would lead to a significant change. 
}}:
$\chi^2({H/A \to \tau^+ \tau^-}) = \max( \chi^2({ggF}), \chi^2({bb{\phi}}) )$.


\subsection{Other constraints}

The most important other constraint update is that on spin-independent DM scattering.
We incorporate in our global fit the recent result published by the PandaX-II experiment \cite{pandax}, 
which we
combine with the new result from the LUX Collaboration \cite{lux16}, as discussed in more detail in
Section~8.


For the electroweak observables we use {{\tt FeynWZ}~\cite{Svenetal}}, 
and for the flavour constraints
we use {{\tt SuFla}~\cite{SuFla}. 
For the Higgs observables, we use 
{\tt FeynHiggs 2.11.2}~\cite{FH,FeynHiggs} {(including the updates
  discussed in \refse{sec:higgsconstraints})}, 
{\tt HiggsBounds 4.3.1}~\cite{HiggsBounds}
and {\tt HiggsSignals 1.4.0}~\cite{HiggsSignals}. 
We calculate the sparticle spectrum using 
{\tt SoftSusy 3.3.10}~\cite{SoftSusy}
and sparticle decays using {\tt SDECAY~1.3b}~\cite{Sdecay} and 
{\tt StauDecay 0.1}\cite{Citron:2012fg}. 
The DM density and scattering rate are calculated using 
{\tt  micrOMEGAs 3.2}~\cite{MicroMegas} and {\tt SSARD}~\cite{SSARD}, 
respectively. Finally, we use {\tt SLHALib~2.2}~\cite{SLHA} 
to interface the different codes.


\subsection{Sampling procedure}

As discussed in the previous Section, the SUSY SU(5) GUT model we study
has 7 parameters: $m_{1/2}$, $m_5$, $m_{10}$, $m_{H_u}$, $m_{H_d}$, $A_0$ and $\tb$.
The ranges of these parameters that we scan in our analysis are listed in Table~\ref{tab:ranges}.
The quoted negative values actually correspond to negative values of $m^2_5, m^2_{10}, m^2_{H_u}$
and $m^2_{H_d}$: for convenience, we use the notation ${\rm sign}(m^2) \times \sqrt{|m^2|} \to m$.
The negative values of $m_5$ and $m_{10}$ that are included in the scans
may be
compatible with early-Universe cosmology \cite{eglos}, and yield acceptable tachyon-free spectra.
In the portions of the scans with negative values of $m_{H_u}$ and $m_{H_d}$, although
the effect of the top quark Yukawa coupling in the renormalization group equations
is important, it may not be the mechanism responsible for generating
electroweak symmetry breaking, since $m_{H_u}$ and $m_{H_d}$ are negative already at the input scale.

\begin{table*}[htb!]
\begin{center}
\begin{tabular}{|c|c|c|} \hline
Parameter   &  Range      & Number of  \\ 
            &             & segments   \\ 
\hline         
$m_{1/2}$        &  ( 0  , 4)      & {2} \\
$m_{5}$        &  ( - 2.6  , 8)      & 2 \\
$m_{10}$        &  ( - 1.3 , 4)      & 3 \\
$m_{H_u}$        &  ( -7  , 7)      & 3 \\
$m_{H_d}$        &  ( -7  , 7)      & 3 \\
$A_0$        	&  ( -8  , 8)      & 1 \\
\tb         &  ( 2  , 68)      & 1 \\
\hline \hline
Total number of boxes &   & 108   \\
\hline
\end{tabular}
\caption{\it Ranges of the SUSY SU(5) GUT parameters sampled, together with the numbers of
segments into which each range was divided, and the corresponding total number of sample boxes.
The mass parameters are expressed in TeV units.} 
\label{tab:ranges}
\end{center}
\end{table*}

We sample this parameter space using {\tt MultiNest v2.18}~\cite{multinest}, dividing the 7-dimensional
parameter space into 108 boxes, as also described in Table~\ref{tab:ranges}. This has two advantages:
it enables us to run {\tt MasterCode} on many nodes in parallel, and it enables us to probe more
efficiently for local features in the likelihood function. For each box, 
we choose a prior such that 80\% of the sample has a flat distribution within the nominal range, 
while 20\% of the sample is in normally-distributed tails outside the box. Our resultant
total sample overlaps smoothly between boxes, 
avoiding any spurious features at
the box boundaries. The total number of points in our sample is $\sim 125 \times 10^6$, 
of which $\sim 8 \times 10^6$ have $\Delta \chi^2 < 10$.


\section{Dark Matter Mechanisms}

The relic density of the LSP, assumed here to be the
lightest neutralino, $\neu1$, {which is stable in supersymmetric SU(5) because of $R$-parity},
may be brought into the narrow range allowed by the Planck 
satellite and other measurements \cite{Planck15} via a combination of different mechanisms. It was
emphasized previously \cite{MC12-DM} in studies of the CMSSM, NUHM1 and NUHM2 that simple
annihilations of pairs of LSPs into conventional particles would not have been
sufficient to bring the relic $\neu1$ density down into the Planck range for values of
$\mneu1$ compatible with the LHC search limits
and other constraints on these models.
Instead, there has to be some extra mechanism for suppressing the LSP density.
Examples include enhanced, rapid annihilation through
direct-channel resonances such as $Z, h, H/A$. Another possibility is coannihilation with some other,
almost-degenerate sparticle species~\cite{DN,coannihilation}: candidates for the coannihilating species identified in
previous studies include the $\staue, {\tilde \mu}, {\tilde e}, {\tilde \nu}, {\tilde t_1}$ and $\cha1$.

We introduced in~\cite{MC12-DM} measures on the sparticle mass parameters that quantify the
mass degeneracies relevant to the
above-mentioned coannihilation and rapid annihilation processes, of which the following are relevant to our analysis of
the SUSY SU(5) GUT model~\footnote{We note that the focus-point
mechanism~\cite{FMM} does not play a role in the SU(5) model.}:
\begin{align}
{\staue} {\rm ~coann.~(pink):} \hspace{06mm} 
\left(\frac{\mstaue}{\mneu{1}} - 1 \right)   & \,<\,  0.15 \, , \nonumber \\[.3em]
\cha{1} {\rm ~coann.~(green):} \hspace{02mm} 
\left( \frac{\mcha{1}}{\mneu{1}} - 1 \right)  & \,<\, 0.1 \, , \nonumber \\[.3em]
A/H {\rm ~funnel~(pale~blue):} \hspace{02mm} 
\left|\frac{\MA}{\mneu{1}} - 2 \right|  & \,<\,  0.4 \, . 
\label{shadings}
\end{align}
We also indicate above the colour codes used in subsequent figures to identify regions
where each of these degeneracy conditions applies.
We have verified in a previous study~\cite{MC12-DM} that
CMSSM, NUHM1 and NUHM2 points that satisfy the DM density
constraint fulfill one or more of the mass-degeneracy conditions, and that they identify 
correctly the mechanisms that yield the largest fractions of final states,
which are usually $\gtrsim 50$\%~\cite{mc10,KdV}.

{In much of the region satisfying the $\staue$ degeneracy criterion above,
the ${\tilde \nu_\tau}$ has a similar mass, and can contribute to coannihilation~\cite{Sneutrino}.
We highlight the parts of the sample where sneutrino coannihilation is important by introducing
a shading for regions where the ${\tilde \nu_\tau}$ is the next-to-lightest sparticle (NLSP),
and obeys the degeneracy condition
\begin{align}
{\tilde \nu_\tau}^{\rm NLSP} {\rm ~coann.~(orange):} 
\left(\frac{m_{\tilde \nu_\tau}}{\mneu{1}} - 1 \right)   & \,<\,  0.1 \, .
\label{tausneutrino}
\end{align}
We discuss later the importance of this supplementary DM mechanism.}

As we discuss in this paper, a novel possibility in the SU(5) SUSY GUT is coannihilation
with right-handed up-type squarks, ${\tilde u_R}$ and ${\tilde c_R}$, which may be much lighter
than the other squarks in this model, as a consequence of the freedom to have $m_5 \ne m_{10}$.
We quantify the relevant mass degeneracy criterion by
\begin{align}
{\SucR} {\rm ~coann.~(yellow):} \; 
\left( \frac{m_{\SucR}}{\mneu{1}} - 1 \right)  & \,<\, 0.2 \, .  
\label{newshading}
\end{align}
As we shall see in the subsequent figures, this novel
degeneracy condition can play an important role when $m_5 \gg m_{10}$.
The existence of this new coannihilation region was verified using {\tt SSARD}~\cite{SSARD},
an independent code for calculating the supersymmetric spectrum and relic density.

We also distinguish in this analysis `hybrid' regions where the $\staue$
coannihilation and $H/A$ funnel mechanisms may be relevant} simultaneously:
\begin{eqnarray}
& \staue~{\rm coann.} + H/A~{\rm funnel:} & {\rm (purple)} \, ,
\label{hybrids}
\end{eqnarray}
also with the indicated colour code. 

\section{Results}

\subsection{Parameter Planes}

We display in Fig.~\ref{fig:m5m10m12} features of the global $\chi^2$ function
for the SUSY SU(5) GUT model in the $(m_5, m_{1/2})$ plane 
(left panel) and the $(m_{10}, m_{1/2})$ plane (right panel), profiled over the
other model parameters\footnote{We have used  {\tt Matplotlib}~\cite{matplotlib} and {\tt PySLHA}~\cite{pyslha} to plot the results of our analysis.}.
Here and in subsequent parameter planes, the best-fit point is shown as a green star, 
the 68\% CL regions are surrounded by red contours, and the 95\% CL regions are
surrounded by blue contours 
(as mentioned above, 
we use the $\Delta\chi^2 = 2.30$ and 
$\Delta\chi^2 = 5.99$ contours as proxies for the 
boundaries of the 68\% and 95\% CL regions in the fit).
The regions inside the 95\% CL contours are shaded
according to the dominant DM mechanisms discussed in the previous Section,
see the criteria (\ref{shadings}, \ref{newshading}, \ref{hybrids}). In the (relatively limited) unshaded 
regions there is no single dominant DM mechanism.

\begin{figure*}[htb!]
\begin{center}
\resizebox{7.5cm}{!}{\includegraphics{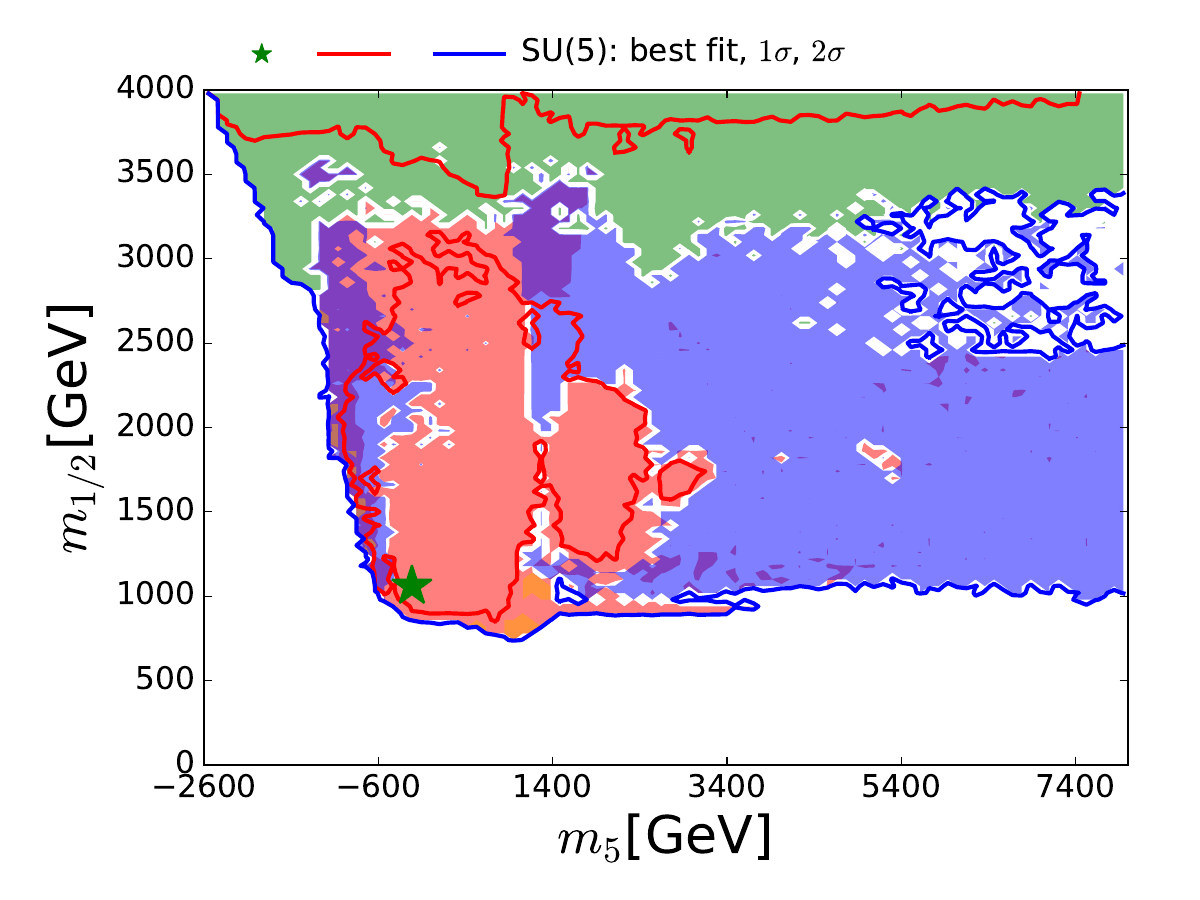}}  \hspace{5mm}
\resizebox{7.5cm}{!}{\includegraphics{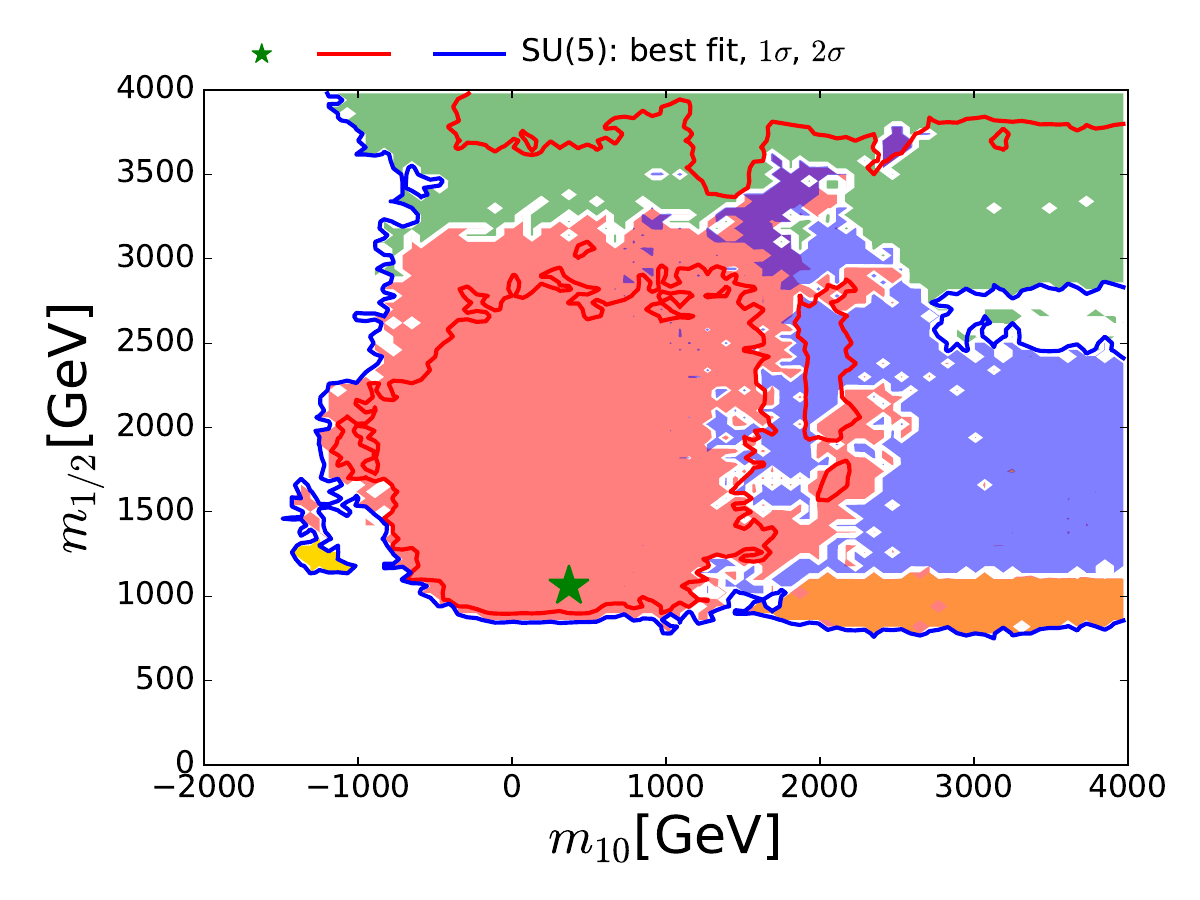}} \\
\vspace{0.5cm}
\resizebox{11cm}{!}{\includegraphics{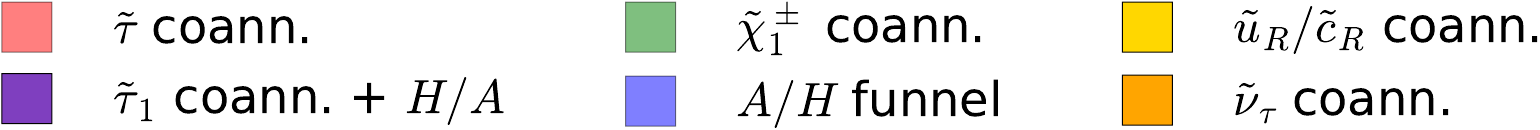}} \\

\end{center}
\vspace{-0.5cm}
\caption{\it 
The $(m_5, m_{1/2})$ plane (left panel) and the $(m_{10}, m_{1/2})$ plane (right panel) in the SUSY SU(5)
GUT model. The best-fit point is shown as a green star, the red contour surrounds the 68\% CL region, and
the blue contour surrounds the 95\% CL region. The coloured shadings represent the dominant DM
mechanisms, as indicated in the lower panel and described in the text.
}
\label{fig:m5m10m12}
\end{figure*}

As we see in Fig.~\ref{fig:m5m10m12},
the best-fit point is at relatively small values of $m_5, m_{10}$
and $m_{1/2}$, close to the lower limit on $m_{1/2}$, whereas the 68\% CL region
extends to much larger values of $m_5, m_{10}$ and $m_{1/2}$. The values
of the model parameters at the best-fit point are listed in Table~\ref{tab:bestfit}~\footnote{The SLHA files
for the best-fit point and other supplementary material can be found in~\cite{mcweb}.}.
The upper row of {numbers} are the results from the current fit including the latest LHC 13-TeV and 
PandaX-II/LUX constraints, and the {numbers}
in parentheses in the bottom row  were obtained using instead the previous LHC 8-TeV and XENON100 constraints,
but the same implementations of the other constraints. 
The most significant effect of the new LHC data has been to increase the best-fit value of $m_{1/2}$
by $\sim 160 \gev$: the changes in the other fit parameters are not significant, in view of the uncertainties. As we
discuss in more detail later, the favoured fit regions are driven by the \gmt\ constraint towards the
boundary of the region excluded by the $\ETslash$ constraint. 
Away from this boundary, the global $\chi^2$ function
is quite flat.

The best-fit point and much of the 68\% CL region lie within the pink shaded region where
$\staue - \neu1$ coannihilation is the dominant DM mechanism. At larger
values of $m_5$ and $m_{10}$ we encounter a blue shaded region where rapid
annihilation via direct-channel $H/A$ poles is dominant. We also see darker shaded
hybrid regions where $\staue$ and $H/A$ annihilation are
important simultaneously. At larger values of $m_{1/2} \gtrsim 3000 \gev$, in the green shaded
regions, the dominant DM mechanism is $\cha1 - \neu1$ coannihilation. {There is also a band
in the $(m_{10}, m_{1/2})$ plane with $m_{10} \gtrsim 1500 \gev$ and
$m_{1/2} \sim 1000 \gev$, allowed at the 95\% CL, where ${\tilde \nu_\tau}^{\rm NLSP}$ coannihilation is important.}

\begin{table*}[htb!]
\begin{center}
\begin{tabular}{|c|c|c|c|c|c|c|} \hline
$m_{1/2}$   &  $m_5$  & $m_{10}$ & $m_{H_u}$ &  $m_{H_d}$ &   $A_0$    & \tb  \\ 
\hline         
1050 & -220 & 380 & -5210 & -4870 & -5680 & 12 \\
(890) & (-80) & (310) & (-4080) & (-4420) & (5020) & (11) \\
\hline
\end{tabular}
\caption{\it Parameters of the best-fit point in the SUSY SU(5) GUT model, with mass parameters given in GeV units.
The {numbers} in parentheses in the bottom row are for a fit that does not include the LHC 13-GeV constraints
and the recent PandaX-II and LUX constraints on DM scattering.
Note that we use the same convention for
the sign of $A_0$ as in~\cite{oldmc,mc9,mc10,mc11,MC12-DM}, which is opposite to the
convention used in, e.g., {\tt SoftSUSY}, and that
we use the notation ${\rm sign}(m^2) \times \sqrt{|m^2|} \to m$
for {$m_5, m_{10}, m_{H_u}$ and $m_{H_d}$}.
}
\label{tab:bestfit}
\end{center}
\end{table*}

We also note the appearance within the 95\% CL region at $m_{1/2} \sim 1000
\gev$, and
$m_{10} \sim - 1000 \gev$ of the novel ${\tilde u_R}/{\tilde c_R} - \neu1$
coannihilation region (shaded yellow). To understand the origin of this novelty, consider the one-loop
renormalization-group equations for the states in the $\mathbf{10}$ representations of
SU(5), namely $(q_L, u^c_L, e^c_L)_i$, above the highest MSSM particle
mass {(all masses are
  understood to be scalar fermion masses,
 and we suppress subscripts \mbox{}$_L$)}:
\begin{eqnarray}
16 \pi^2 \frac{\partial m^2_{q_i}}{\partial t} & = & \delta_{i3} (X_t + X_b) - \frac{32}{3} g_3^2 |M_3|^2  \nonumber \\
& & \hspace{-0.8cm} - 6 g_2^2 |M_2|^2 - {\frac{2}{15}}g_1^2 |M_1|^2 + \frac{1}{5} g_1^2 S \, , \label{q} \\
16 \pi^2 \frac{\partial m^2_{u_i^c}}{\partial t} & = & {2} \delta_{i3} X_t - \frac{32}{3} g_3^2 |M_3|^2 \nonumber \\
& &  - \frac{32}{15} g_1^2 |M_1|^2 - \frac{4}{5} g_1^2 S \, , \label{uc} \\
16 \pi^2 \frac{\partial m^2_{e_i^c}}{\partial t} & = & {2} \delta_{i3} X_\tau - \frac{24}{5} g_1^2 |M_1|^2 + \frac{6}{5} g_1^2 S \, , \label{ec}
\end{eqnarray}
where $t \equiv \ln(Q/Q_0)$ with $Q$ the renormalization scale and $Q_0$ some reference scale,
\begin{eqnarray}
X_t & \equiv & 2|y_t|^2 (m^2_{H_u} + m^2_{q_3} + m^2_{t^c}) + 2|A_t|^2 \, , \label{Xt} \\
X_b & \equiv & 2|y_b|^2 (m^2_{H_d} + m^2_{q_3} + m^2_{b^c}) + 2|A_b|^2 \, , \label{Xb} \\
X_\tau & \equiv & 2|y_\tau|^2 (m^2_{H_d} + m^2_{l_3} + m^2_{\tau^c}) + 2|A_\tau|^2 \, , \label{Xtau}
\end{eqnarray}
and
\begin{eqnarray}
S & \equiv & (m^2_{H_u} - m^2_{H_d}) \nonumber \\
& + & {\rm Tr} \left({m^2_{q} - m^2_{l}} - 2 m^2_{u^c} + m^2_{d^c} + m^2_{e^c} \right) \, ,
\label{S}
\end{eqnarray}
{where the trace in $S$ sums over the generations.}
The ${\tilde u_R}/{\tilde c_R} - \neu1$ coannihilation mechanism becomes important in a region
of the SUSY SU(5) GUT parameter space where $m_5^2$ is very large and positive {($\sim 27$ TeV$^2$)},
$m_{10}^2$ is small and negative {($\sim -1.4$ TeV$^2$)}, $m_{H_u}^2$ is very large and negative 
{($\sim -23$ TeV$^2$)}, and $m_{H_d}^2$ is very large and positive {($\sim 50$ TeV$^2$)}. 
In this region, therefore, $X_t$ is very large and negative {($\sim -35$ TeV$^2$)}, $X_b$ and 
$X_\tau$ are suppressed because of small Yukawa couplings 
($\tb$ is not large in this region), and $S$ is also very large and 
negative {($\sim -73$ TeV$^2$)}, since $m_{H_u}^2 - m_{H_d}^2$ is large and 
negative and ${\rm Tr} (m^2_q - m^2_l - 2 m^2_{u^c} + m^2_{d^c} + m^2_{e^c} )$ 
vanishes at the GUT scale. 
Inspection shows that the $X_t$ terms in (\ref{q}) and (\ref{uc})
drive the stop and sbottom masses upwards, and the $S$ terms in (\ref{q}) and (\ref{ec})
drive the left-handed squark and right-handed slectron masses upwards. On the other hand,
the $S$ term in (\ref{uc}) drives the right-handed squark masses downwards. 
Since there are
no counteracting $X$ terms for the ${\tilde u_R}$ and ${\tilde c_R}$, these have lower masses
than the other sfermions, opening the way to a ${\tilde u_R}/{\tilde c_R} - \neu1$ 
coannihilation region.\footnote{ {An SLHA file corresponding to the ${\tilde u_R}/{\tilde c_R} - \neu1$ 
coannihilation region can be found in~\cite{mcweb}.}}

\begin{figure*}[htb!]
\begin{center}
\resizebox{10cm}{!}{\includegraphics{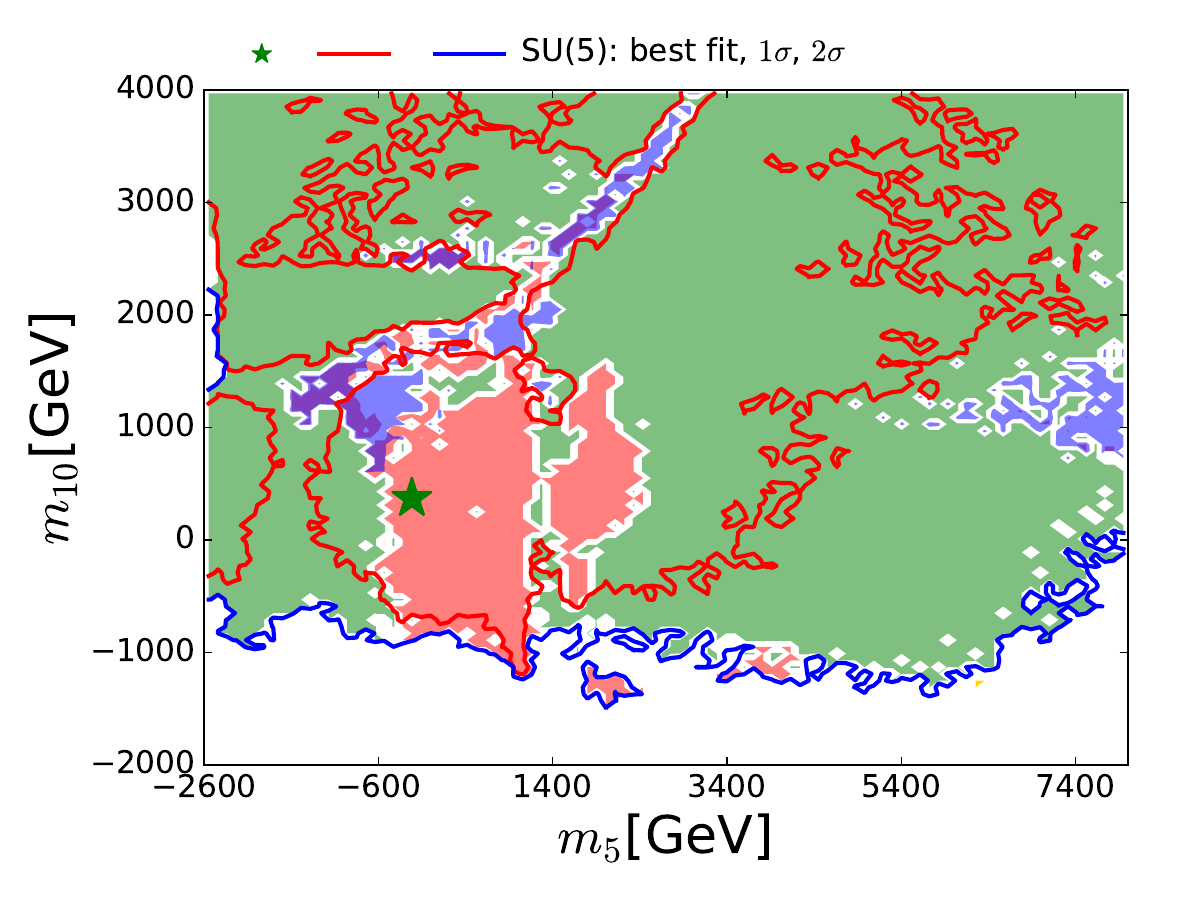}}  \\
\vspace{0.5cm}
\resizebox{11cm}{!}{\includegraphics{dm_legend}} \\

\end{center}
\vspace{-0.5cm}
\caption{\it 
The $(m_5, m_{10})$ plane in the SUSY SU(5)
GUT model. The line colours and shadings are the same as in
Fig.~\protect\ref{fig:m5m10m12}.
}
\label{fig:m5m10}
\end{figure*}

As discussed in more detail later, we used the {\tt Atom}~\cite{Atom} simulation code for a dedicated verification that points
in this region escape all the relevant LHC constraints. 
These points avoid exclusion by the LHC constraints 
through a combination of a strong mass degeneracy,
$m_{{\tilde u_R}/{\tilde c_R}} - \mneu1 \lesssim 50 \gev$,
leading to strong suppression of the standard ~$\ETslash$ signature,
and the reduction of the production rate compared to the simplified model
that assumes mass degeneracy of all 8 light flavour squarks (see Fig.~\ref{fig:sigmafactors}). 
These effects are clearly visible in Fig.~18 of~\cite{1507.05525}.

Fig.~\ref{fig:m5m10} displays the corresponding information in the $(m_5, m_{10})$ plane
of the SUSY SU(5) GUT model. As already reported in Table~\ref{tab:bestfit},
here we see directly that the best-fit point
has very small (and slightly negative) $m_5$, and that $m_{10}$ is somewhat larger, exploiting the possibility
that $m_5 \ne m_{10}$ that is offered in this model. We also see again that the 68\%
CL region extends to values of $m_5$ and $m_{10}$ beyond the $\staue$ coannihilation region. 
We also note that in most of the rest of this plane $\cha1 - \neu1$ coannihilation is dominant, with
only scattered regions where rapid $H/A$ annihilation is important, even in combination
with $\staue$ coannihilation.

Projections of our results in the $(\tb, m_{1/2}), (\tb, m_5)$ and $(\tb, m_{10})$
planes are shown in Fig.~\ref{fig:tbm12m5m10}. We see that values of $\tb \gtrsim 4$
are allowed at the 95\% CL, that the range $\tb \in (8, 57)$ is favoured at
the 68\% CL, and that there is no phenomenological upper limit on $\tb$ at the 95\% CL\footnote{The 
RGE evolution of the Yukawa couplings blows up for $\tan \beta \gsim 60$.}. 
The best-fit point has $\tb = 13$,
as also reported in Table~\ref{tab:bestfit}.

\begin{figure*}[htb!]
\begin{center}
\resizebox{10cm}{!}{\includegraphics{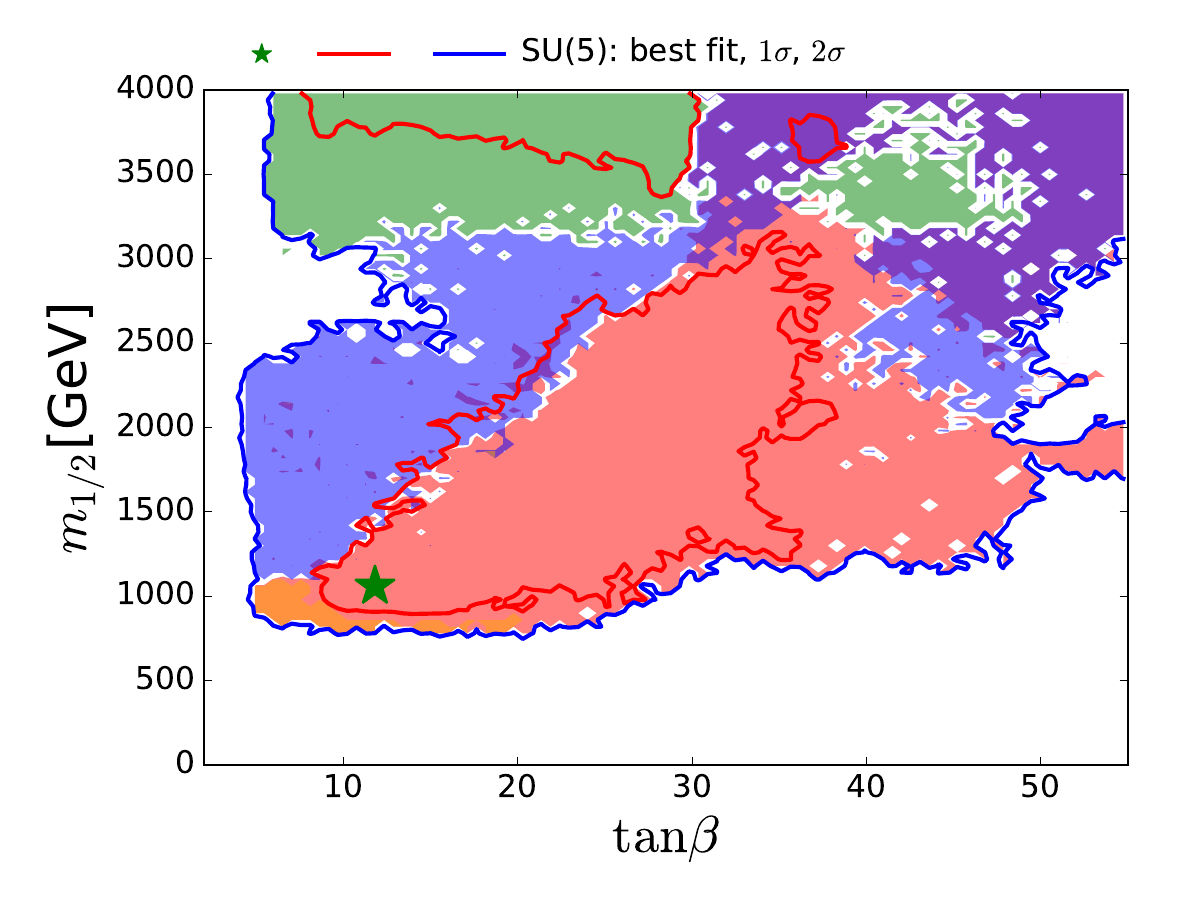}} \\
\vspace{0.5cm}
\resizebox{7.5cm}{!}{\includegraphics{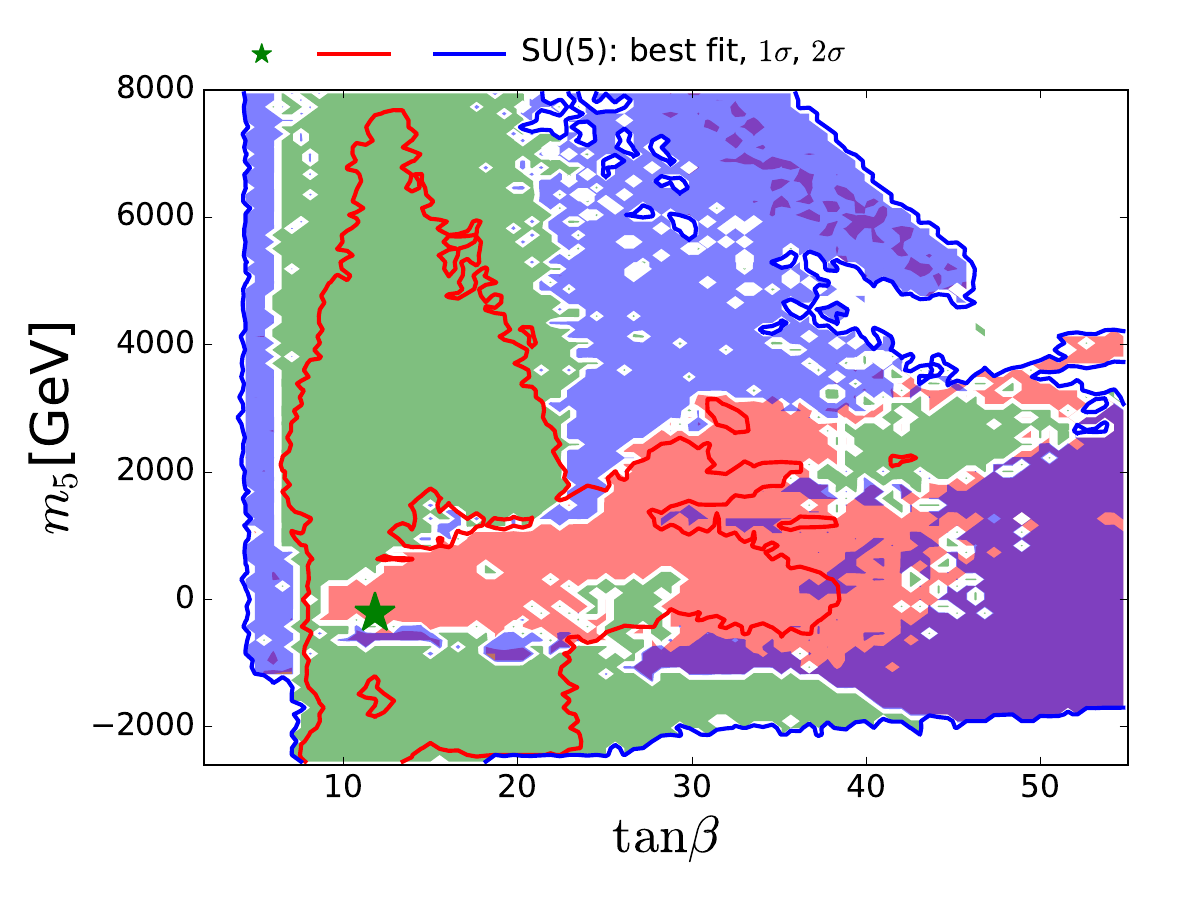}}
\resizebox{7.5cm}{!}{\includegraphics{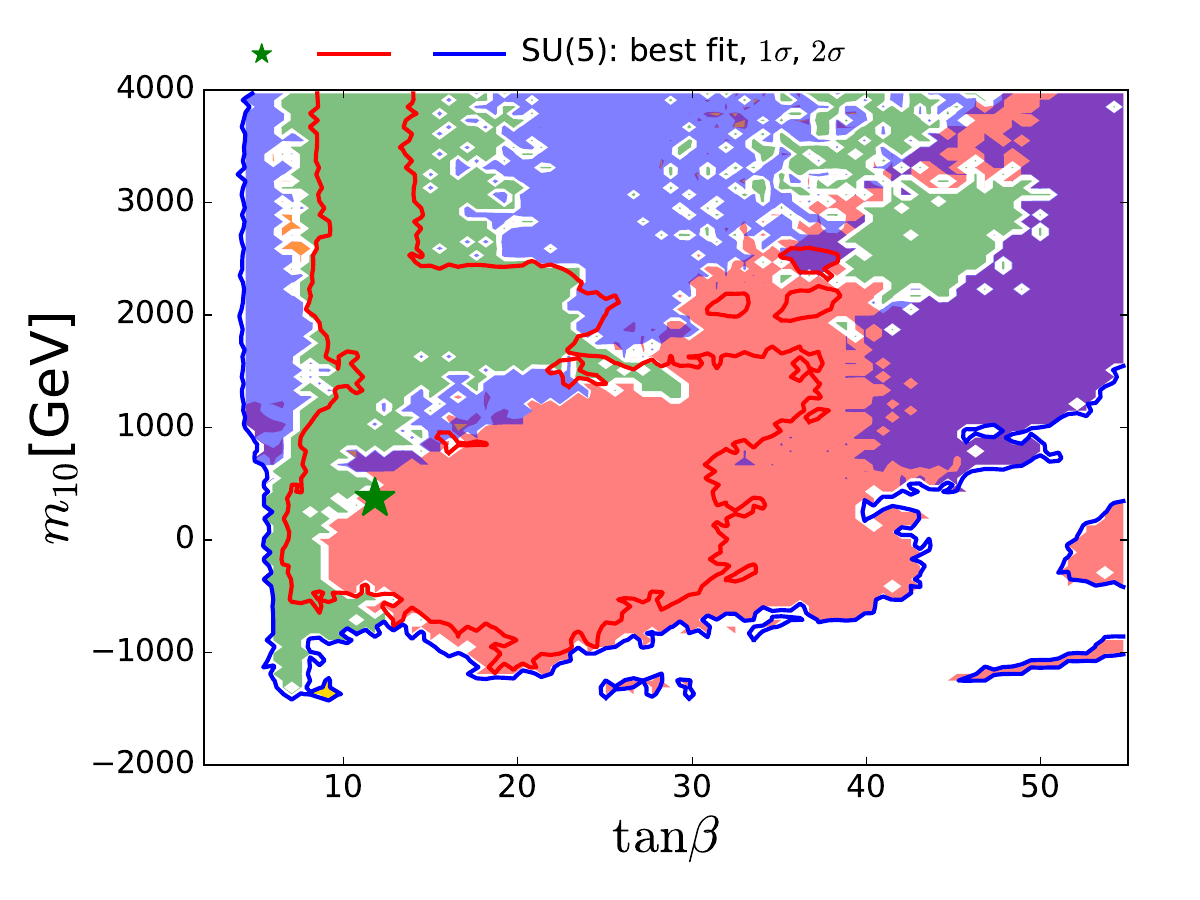}}  \\
\vspace{0.5cm}
\resizebox{11cm}{!}{\includegraphics{dm_legend}} \\

\end{center}
\vspace{-0.5cm}
\caption{\it 
The $(\tb, m_{1/2})$ plane (upper panel), the $(\tb, m_{5})$ plane (lower left panel) and
the $(\tb, m_{10})$ plane (lower right panel) in the SUSY SU(5)
GUT model. The line colours and shadings are the same as in Fig.~\protect\ref{fig:m5m10m12}.
}
\label{fig:tbm12m5m10}
\end{figure*}

The pink $\staue - \neu1$ coannihilation region is very prominent in the $(\tb, m_{1/2})$ projection
shown in the upper panel of Fig.~\ref{fig:tbm12m5m10}, as is the blue rapid $H/A$
annihilation region and the purple $\staue - \neu1$ coannihilation + $H/A$ funnel
hybrid region at large $\tb$ and $m_{1/2}$.
While the $H/A$ funnel appears in the CMSSM only when $\tan \beta \gsim 45$ for $\mu > 0$ \cite{DN,funnel}, in the SU(5) SUSY GUT model, it is found at significantly lower $\tan \beta$,
due to the separation of $m_{H_u}$ and $m_{H_d}$ from $m_{5}$ and $m_{10}$, effectively
making $m_A$ (and $\mu$) free parameters as in the NUHM2. {There is also a region
in the $(\tb, m_{1/2})$ plane with $\tb \lesssim 10$ and
$m_{1/2} \sim 1000 \gev$ where ${\tilde \nu_\tau}^{\rm NLSP}$ coannihilation is important.}


The $\staue - \neu1$ coannihilation region
and the purple $\staue - \neu1$ coannihilation + $H/A$ funnel
hybrid region are prominent for $|m_5| \lesssim 3000 \gev$ in the $(\tb, m_5)$ and $(\tb, m_{10})$
planes shown in the lower part of Fig.~\ref{fig:tbm12m5m10}, with $\cha1 - \neu1$
coannihilation dominant at smaller values of $\tb$, in particular. 
The ${\tilde u_R}/{\tilde c_R} - \neu1$ coannihilation region
appears in a small island for $\tb \sim 8$ and $m_{10} \sim -1200 \gev$ in the $(\tb, m_{10})$ 
plane shown in the lower right panel of Fig.~\ref{fig:tbm12m5m10}.

We display in Fig.~\ref{fig:Mhtbm12m5m10} projections of our results for $\Mh$
versus $m_{1/2}$ (upper left), $\tb$ (upper right), $m_5$ (lower left) and $m_{10}$
(lower right). The predicted values of $\Mh$ are well centred within the expected 
{\tt FeynHiggs} uncertainty range around the value
measured at the LHC, $\Mh = 125.09 \pm 0.24 \gev$ \cite{Aad:2015zhl}. Moreover, the Dark 
Matter mechanisms do not exhibit any preference for values of $\Mh$ above or below
the nominal central value. Thus, there is no apparent tension between this LHC
measurement and the other constraints on the SUSY SU(5) GUT model, with the notable exception of $\gmt$.

\begin{figure*}[htb!]
\begin{center}
\resizebox{7.5cm}{!}{\includegraphics{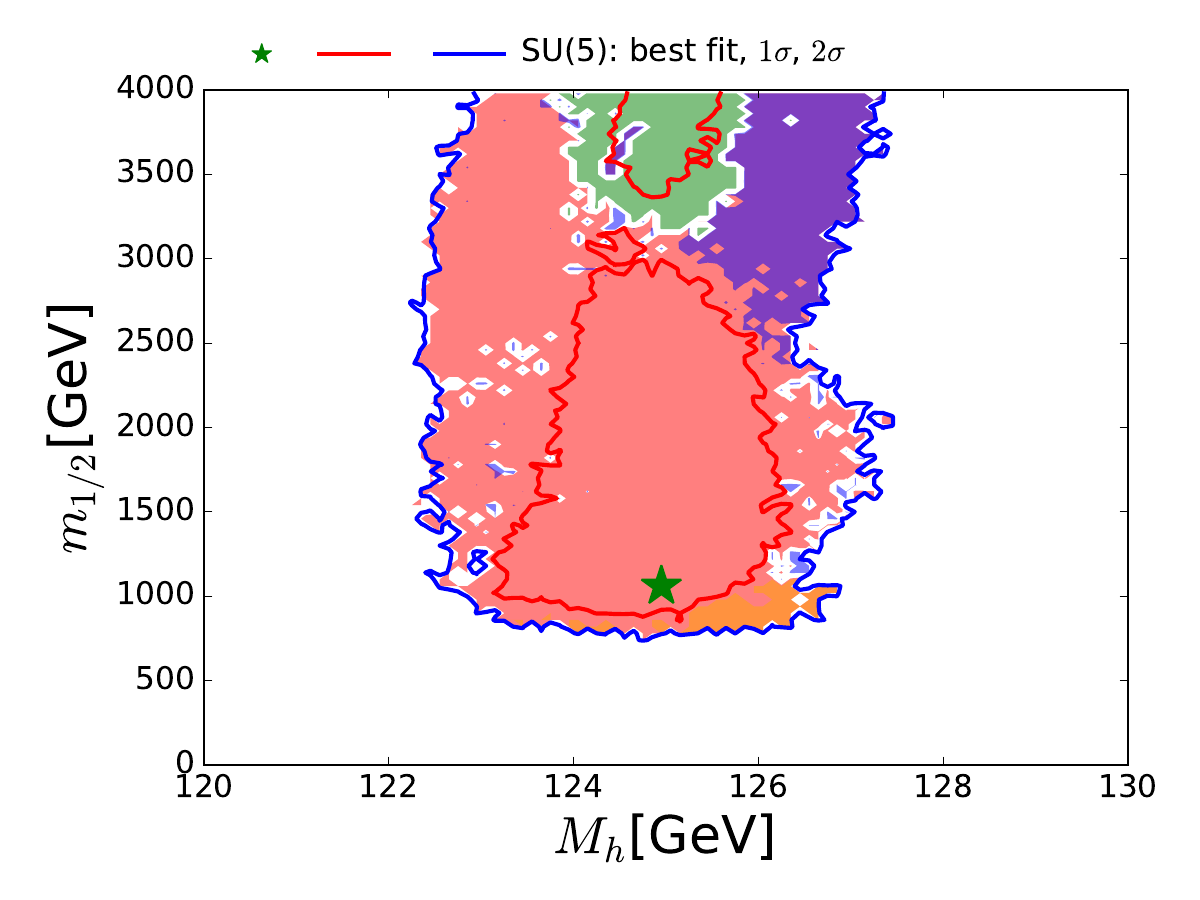}} 
\resizebox{7.5cm}{!}{\includegraphics{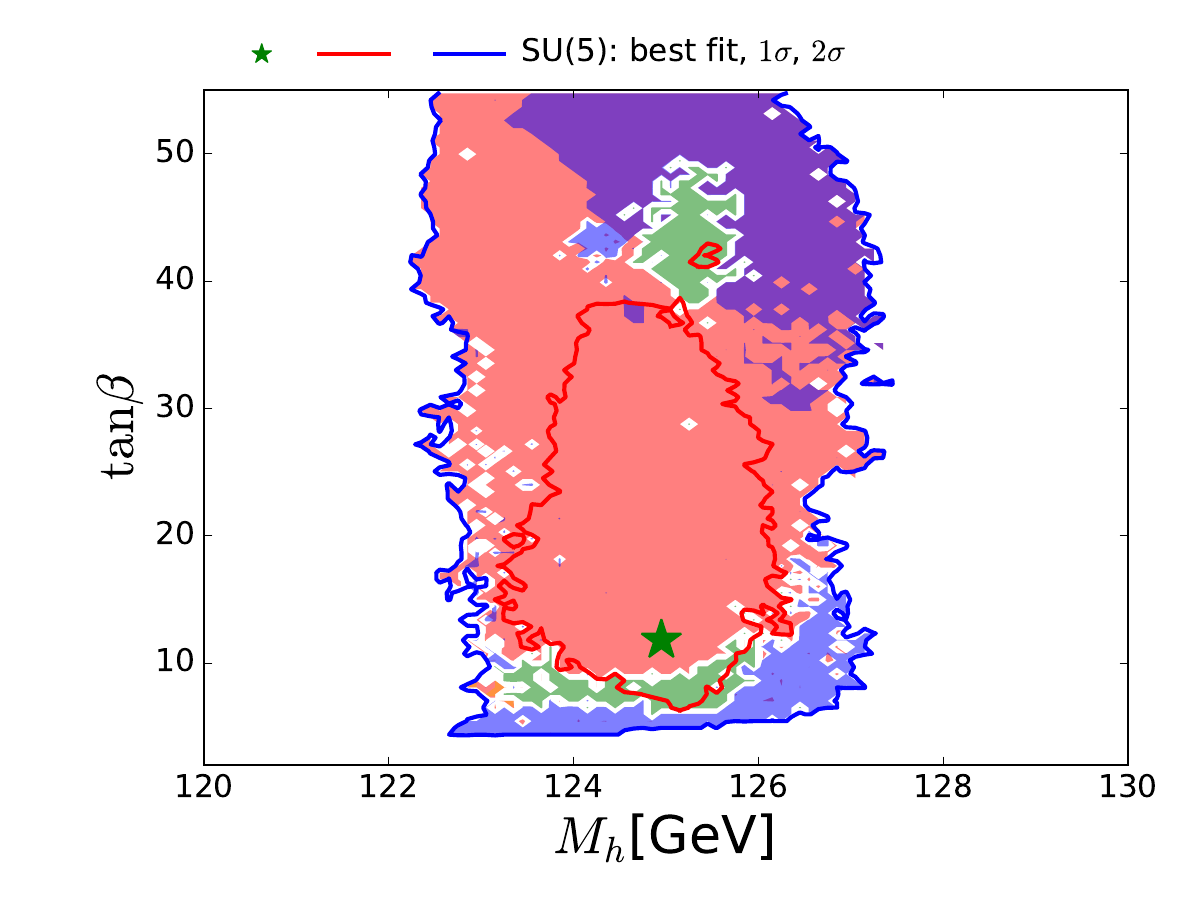}} \\
\vspace{0.5cm}
\resizebox{7.5cm}{!}{\includegraphics{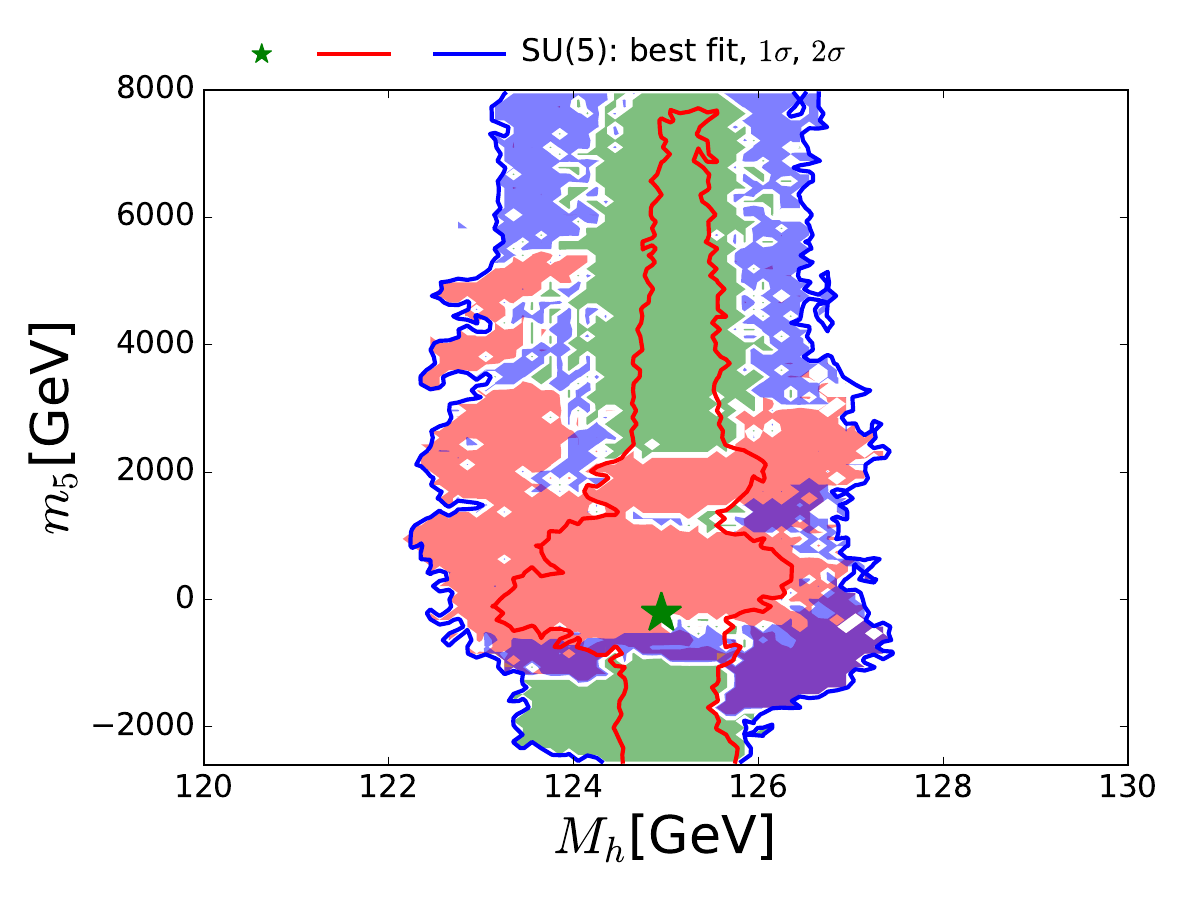}}
\resizebox{7.5cm}{!}{\includegraphics{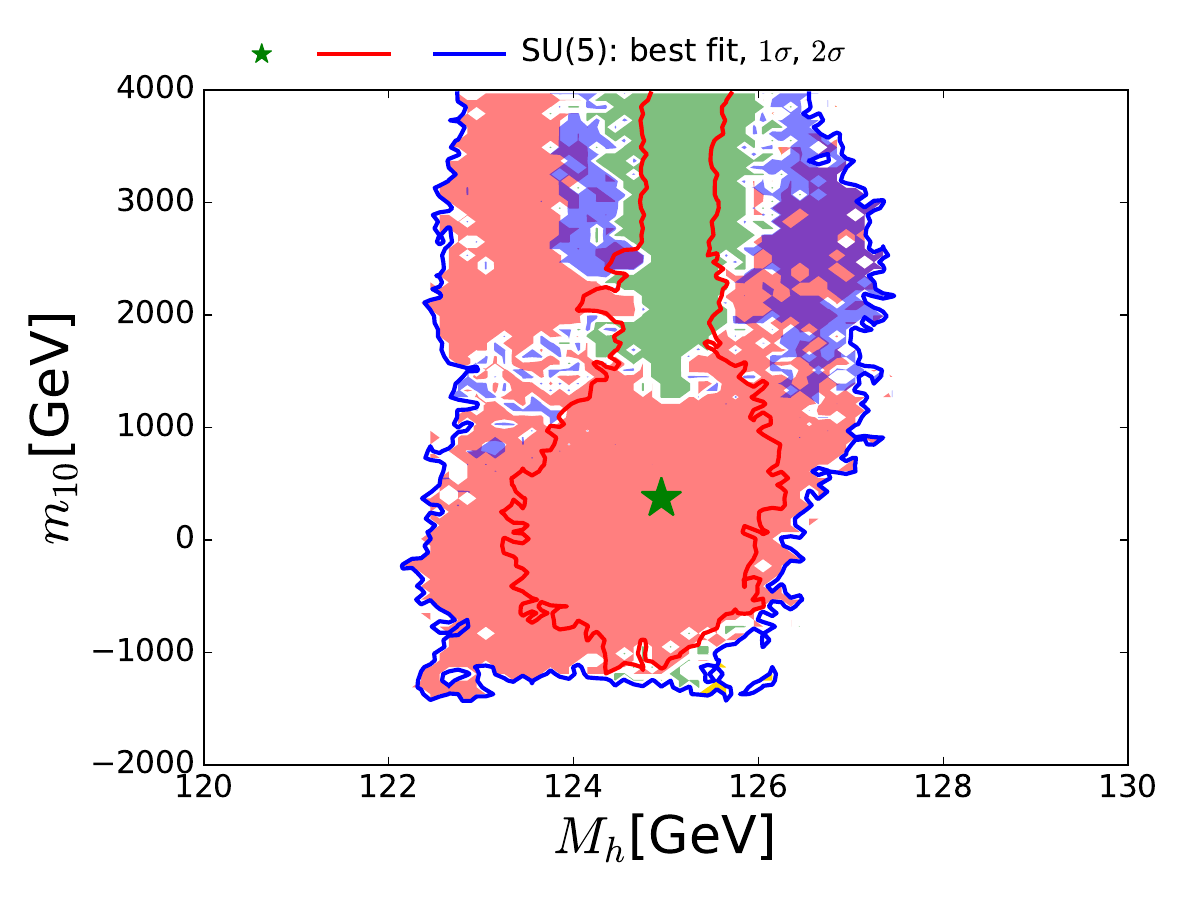}}  \\
\vspace{0.5cm}
\resizebox{11cm}{!}{\includegraphics{dm_legend}} \\

\end{center}
\vspace{-0.5cm}
\caption{\it 
The $(\Mh, m_{1/2})$ plane (upper left panel), the $(\tb, \Mh)$ plane (upper right panel),
the $(\Mh, m_5)$ plane (lower left panel) and the $(\Mh, m_{10})$ plane (lower right panel) in the SUSY SU(5)
GUT model. The line colours and shadings are the same as in Fig.~\protect\ref{fig:m5m10m12}.
}
\label{fig:Mhtbm12m5m10}
\end{figure*}

As is well known, the calculation of $\Mh$ in the MSSM is particularly
sensitive
to the value of the trilinear soft SUSY-breaking parameter $A_0$
as well as the stop squark masses. The latter depend in the
SUSY SU(5) GUT model on $m_{10}$ and $m_{1/2}$, but are insensitive
to $m_5$.

The $(m_{H_u}, m_{H_d})$ plane is shown in Fig.~\ref{fig:mHumHd}.
We see that the best-fit point lies in the quadrant where both $m_{H_u}$
and $m_{H_d}$ are negative, and that the 68\% CL region extends also
to the quadrant where $m_{H_d}$ is negative and $m_{H_u}$ is positive,
as does the $\staue - \neu1$ coannihilation region. On the other hand,
the $\cha1 - \neu1$ coannihilation region lies in the upper quadrants where
$m_{H_d} > 0$. There is also an intermediate region, characterized by the H/A funnel mechanism and its hybridization
with $\staue$ coannihilation, part of which is also allowed at the 68\% CL. {There is also a region
with  $M_{H_u} \sim 4000 \gev, M_{H_d} \sim -3000 \gev$ where ${\tilde \nu_\tau}^{\rm NLSP}$ 
coannihilation is important.}

\begin{figure*}[htb!]
\begin{center}
\resizebox{10cm}{!}{\includegraphics{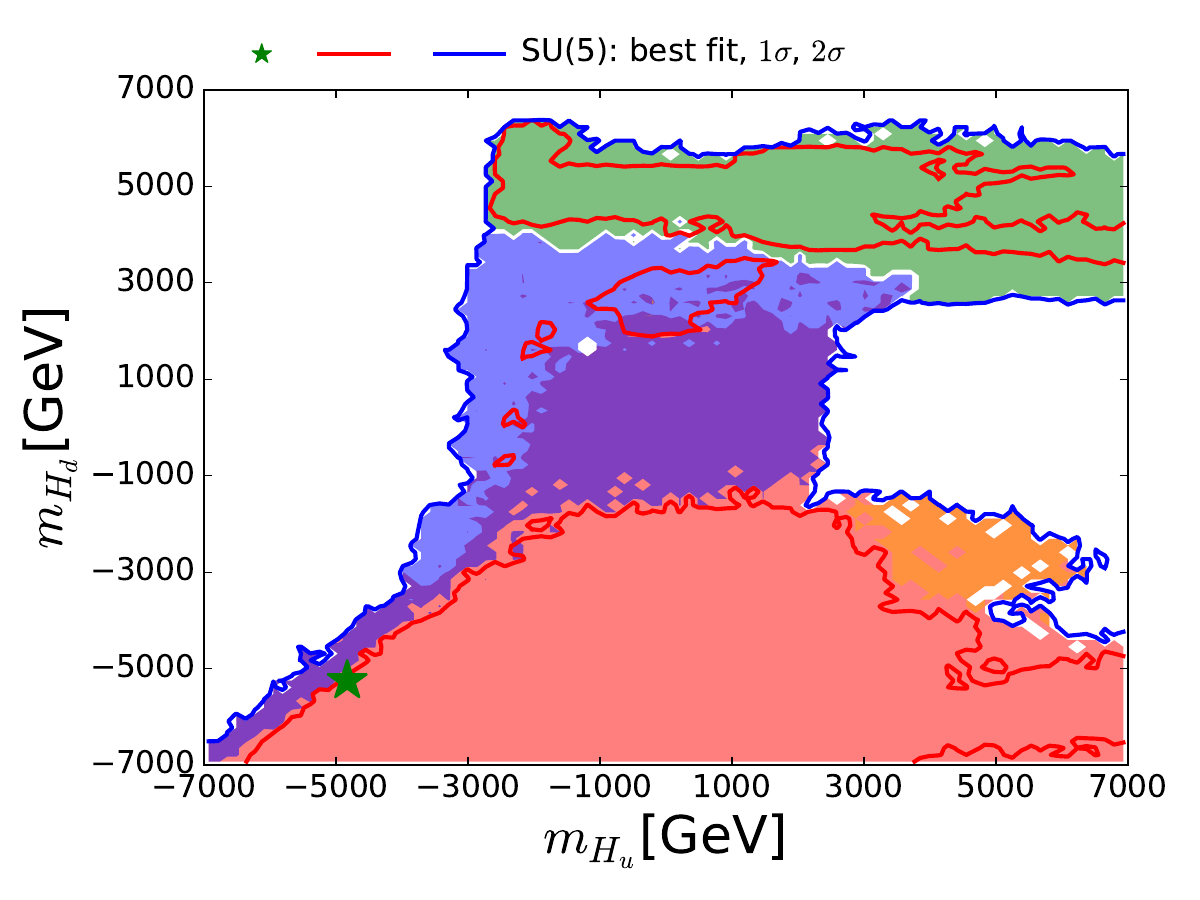}}  \\
\vspace{0.5cm}
\resizebox{11cm}{!}{\includegraphics{dm_legend}} \\

\end{center}
\vspace{-0.5cm}
\caption{\it 
The $(m_{H_u}, m_{H_d})$ plane in the SUSY SU(5)
GUT model. The line colours and shadings are the same as in Fig.~\protect\ref{fig:m5m10m12}.
}
\label{fig:mHumHd}
\end{figure*}

The left panel in Fig.~\ref{fig:mAtanb} displays the $(\MA, \tb)$ plane in the supersymmetric
SU(5) GUT model. {We see that $\MA \gtrsim 800 (1000) \gev$ at the $\Delta \chi^2 = 5.99~(2.30)$ level, corresponding to the 95 (68) \% CL,
which is largely due to the interplay of the indirect constraints on $(\MA, \tb)$ {such
as $\Mh$ (see also~\cite{CMSHA13}) as well as}
the direct constraints from the LHC heavy MSSM Higgs searches. Even for large $\tb$,
where these constraints impose the strongest lower limit on $\MA$, it is much weaker than
our global limit, which is $\MA \gtrsim 2800 (> 4000) \gev$ at the 95 (68) \% CL~\footnote{{We
note, however, that this constraint might weaken with a larger parameter sample.}}. 
We observe the same behavior in the right panel of Fig.~\ref{fig:mAtanb}, where
the one dimensional likelihood profile for $\MA$ is shown.
Indeed, the lightest pseudoscalar mass allowed at $\Delta \chi^2 = 4$ is $\sim920$ GeV.
The best-fit point in the global fit has $(\MA, \tb) \simeq (1600 \gev, 13)$: this is considerably beyond the present
and projected LHC reach, though poorly determined.}

\begin{figure*}[t]
\resizebox{8cm}{!}{\includegraphics{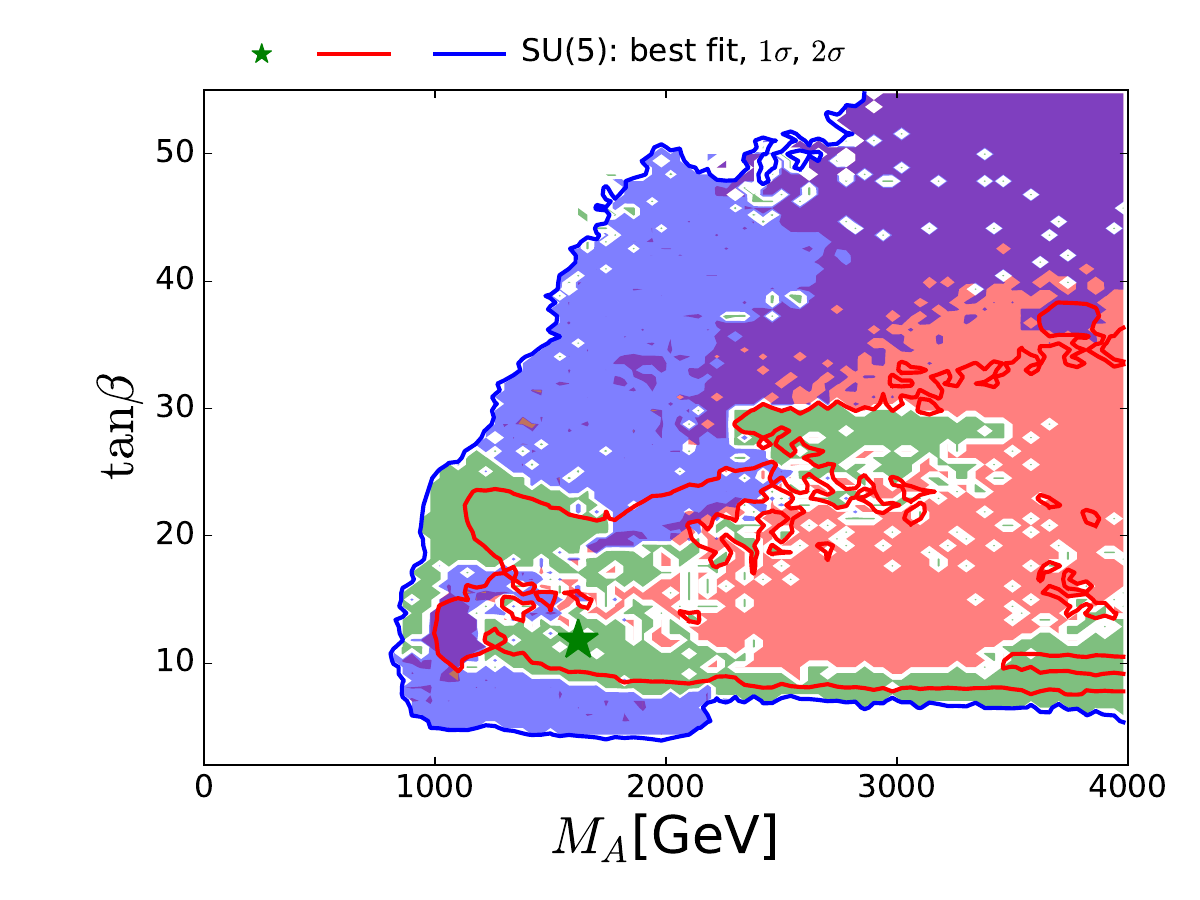}}
\resizebox{8cm}{!}{\includegraphics{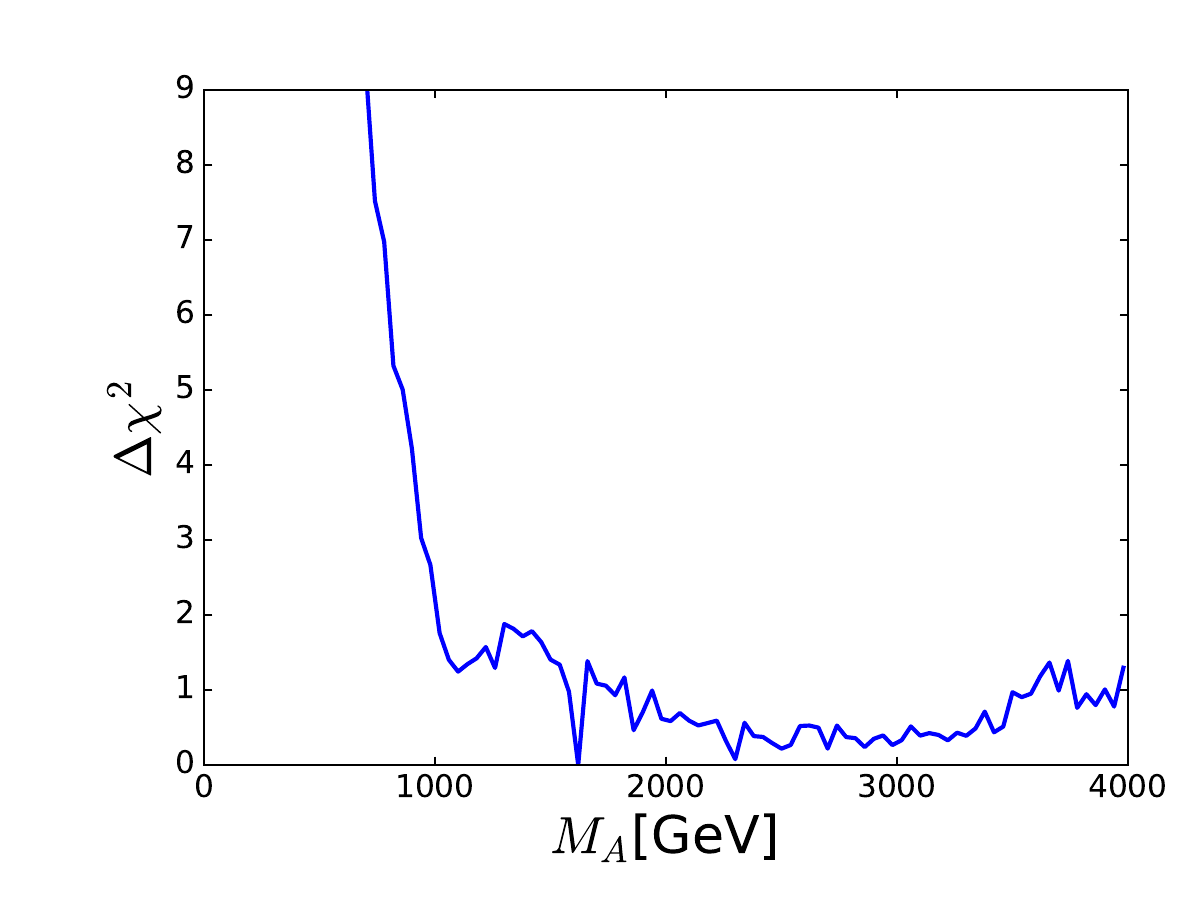}}  \\
\vspace{0.5cm}
\hspace{0.6cm}\resizebox{8cm}{!}{\includegraphics{dm_legend}} \\

\vspace{-0.5cm}
\caption{\it 
On the left, the $(\MA, \tan\beta)$ plane in the SUSY SU(5)
GUT model. The line colours and shadings are the same as in
Fig.~\protect\ref{fig:m5m10m12}. On the right, the $\chi^2$ likelihood function for the pseudoscalar mass.
}
\label{fig:mAtanb}
\end{figure*}


\section{One-Dimensional Likelihood Functions}

We now discuss the one-dimensional $\Delta \chi^2$ functions for various observable quantities.

Fig.~\ref{fig:masschi2} displays those for \mgl\ (top left), $m_{\tilde q_L}$ (top right),
$m_{\tilde d_R}$ (centre left), $m_{\tilde u_R}$ (centre right), 
$m_{\tilde t_1}$ (bottom left) and $m_{\tilde \tau_1}$ (bottom right). The solid blue line
in each panel corresponds to the current analysis of the supersymmetric SU(5) model including LHC Run 2 data at 13 TeV, {the dashed blue line shows the result of an SU(5) fit in which the LHC 13-TeV results
are not included, and the solid grey line corresponds to `fake' NUHM2-like results obtained
by selecting a subset of the SU(5) sample with $m_5/m_{10} \in [0.9, 1.1]$, which we discuss in more detail 
later~\footnote{{The $\Delta \chi^2$ functions for the NUHM2 subsample are calculated relative to its
minimum $\chi^2$, which is {$\sim 0.4$} higher than the minimum $\chi^2$ for the full SU(5) sample.}}

The current SU(5) fit exhibits minima of $\chi^2$ at masses $\lesssim 2.5
\tev$: $\mgl \simeq 2600 \gev$, common squark mass $\msq \simeq 2200 \gev$, 
$m_{\SuR}$, $m_{\SdR}$, $\mste \simeq 2200 \gev$ 
and $\mstaue \simeq 540 \gev$, followed by a rise at higher
mass towards a plateau with $\Delta \chi^2 {\lsim} 2$. The minimum
is relatively 
sharp for \mgl, \msq\ and $m_{\tilde \tau_1}$, whereas it is broader for
$m_{\tilde t_1}$. The exact values are listed in Table~\ref{tab:spectrum}
and depicted in Fig.~\ref{fig:bestfit}. In this figure we also indicate decay branching ratios (BRs) exceeding 20\%
by dashed lines, which are thicker for more important BRs.
Fig.~\ref{fig:spectrum} displays the 1-dimensional 68 and 95\% CL ranges for the
Higgs and sparticle masses in the supersymmetric SU(5) model as darker
and lighter coloured bands, with the best-fit values shown as blue lines.

\begin{figure*}[htb!]
\begin{center}
\resizebox{7.3cm}{!}{\includegraphics{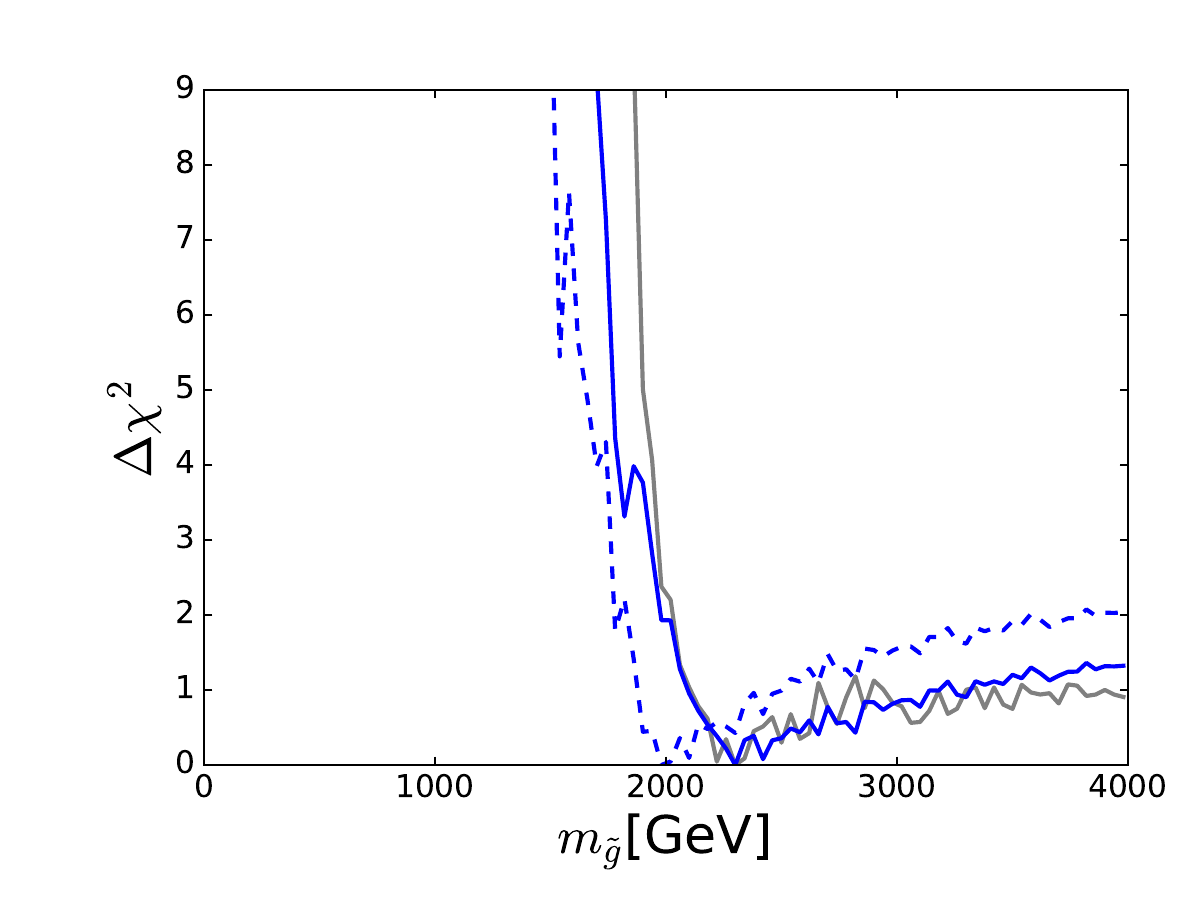}}
\resizebox{7.3cm}{!}{\includegraphics{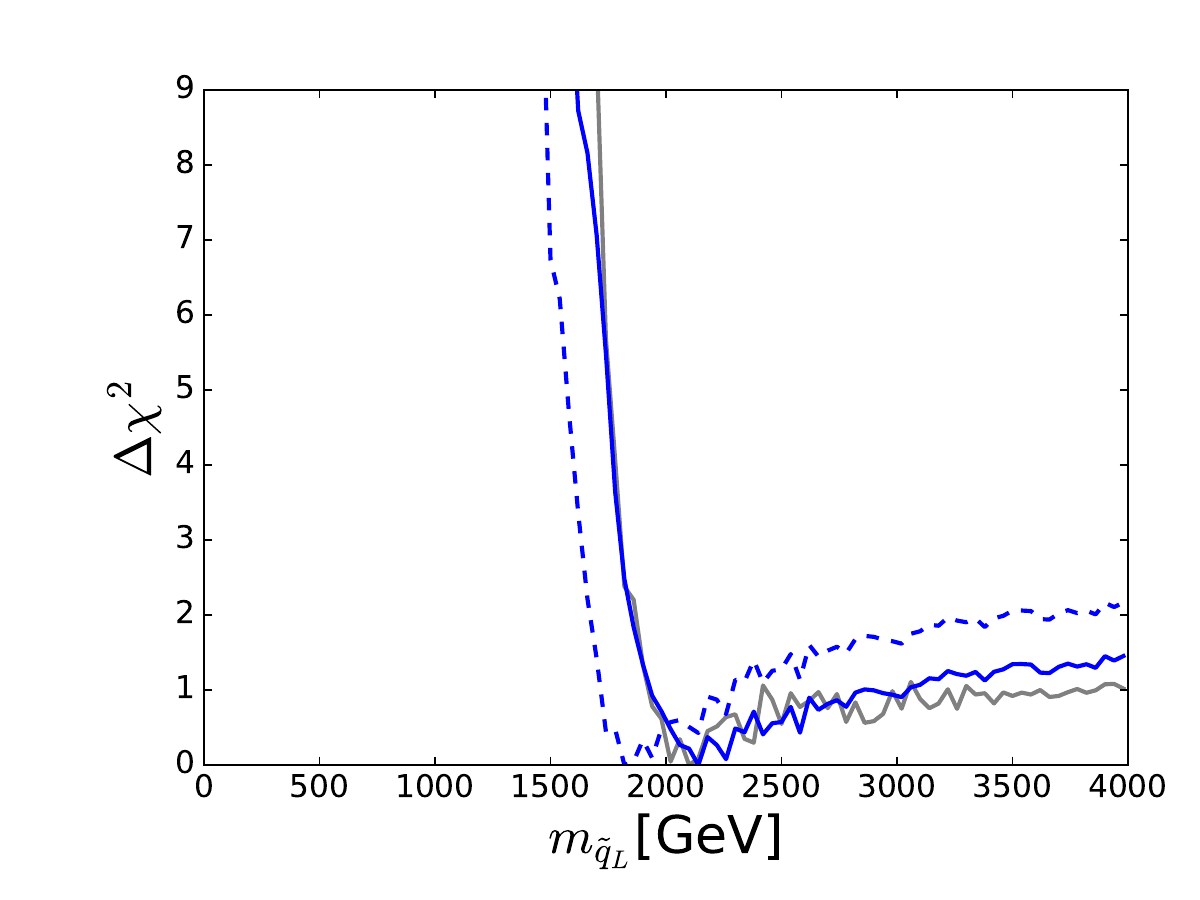}} \\
\resizebox{7.3cm}{!}{\includegraphics{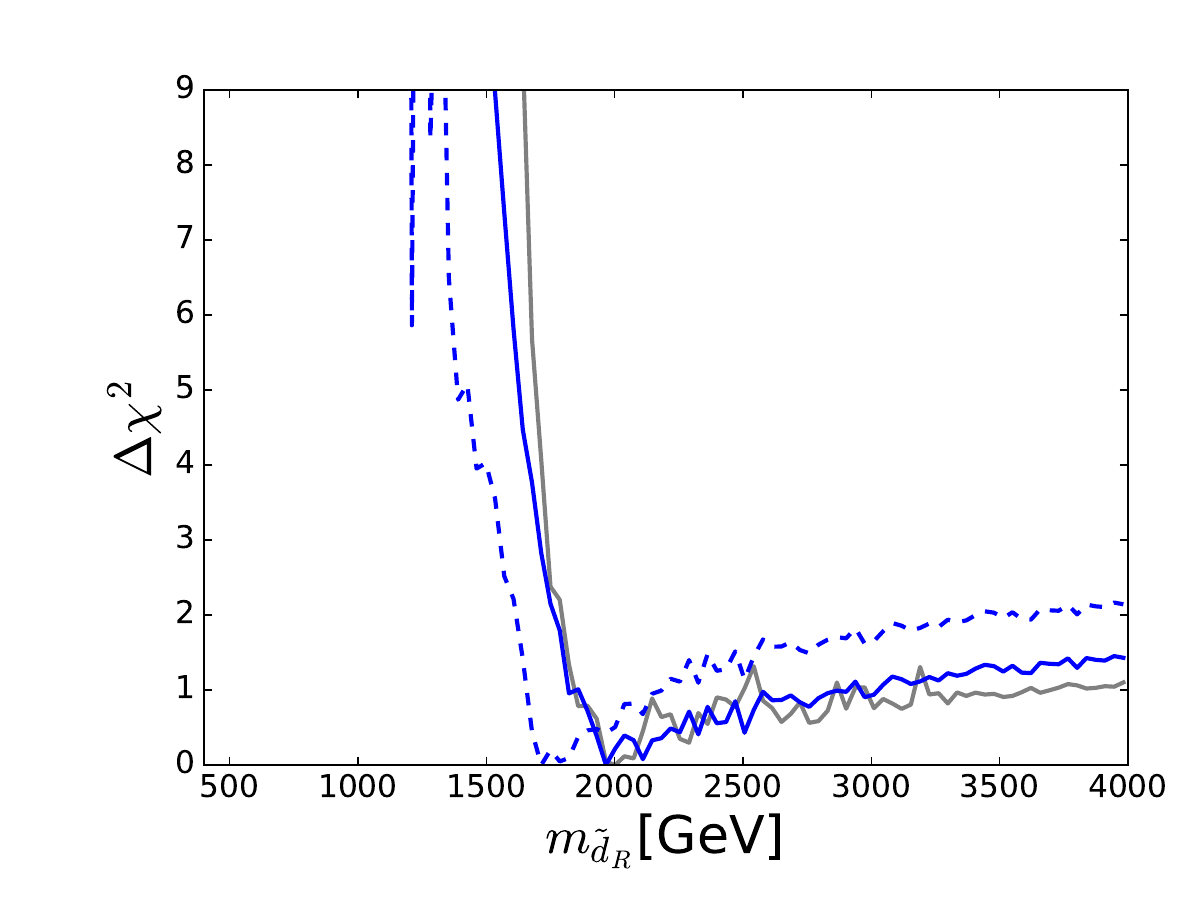}}
\resizebox{7.3cm}{!}{\includegraphics{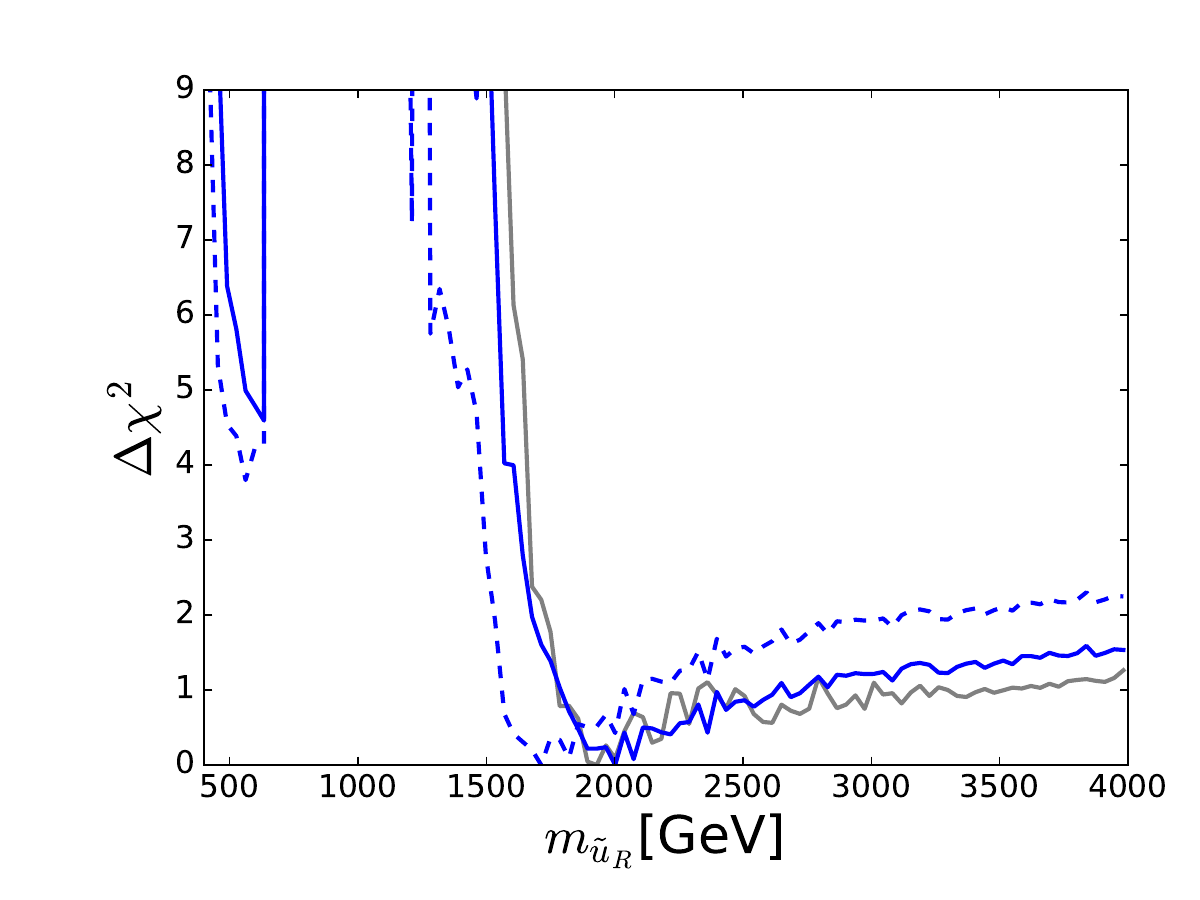}} \\
\resizebox{7.3cm}{!}{\includegraphics{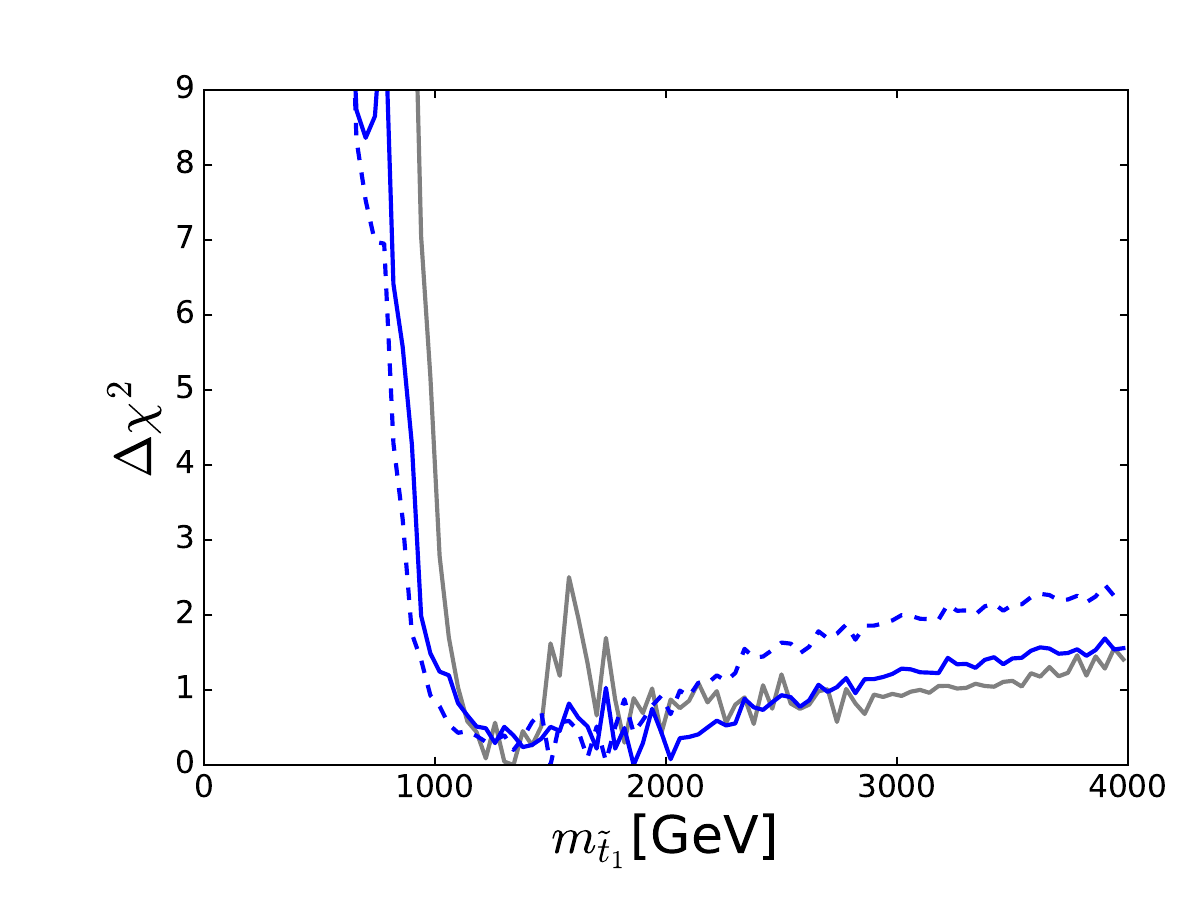}}
\resizebox{7.3cm}{!}{\includegraphics{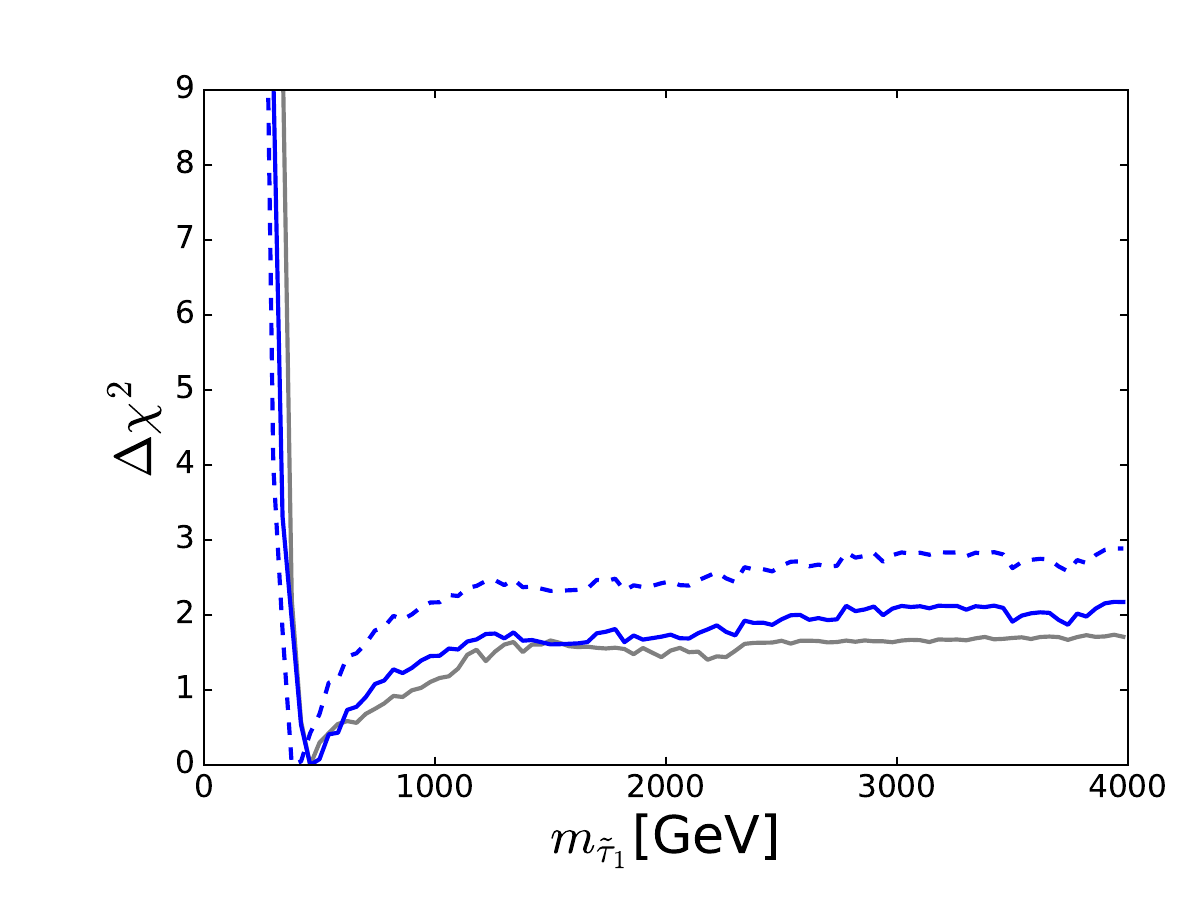}} \\
\end{center}
\vspace{-0.5cm}
\caption{\it 
The $\chi^2$ likelihood functions in the SUSY SU(5)
GUT model  (blue lines) for the gluino mass (top left panel), the left-handed squark mass (top right panel),
the right-handed down squark mass (centre left panel), the right-handed up squark mass (centre right panel),
the lighter stop squark mass (lower left panel) and the lighter stau slepton mass (lower right panel).
{The dashed blue lines show the result of omitting the LHC 13-TeV constraints, 
and the grey lines represent `fake' NUHM2 results obtained
by selecting a subset of the SU(5) sample with $m_5/m_{10} \in [0.9, 1.1]$.}
}
\label{fig:masschi2}
\end{figure*}

Concerning future $e^+e^-$ colliders,
one can see that the best-fit masses of the lightest neutralino and stau are
$\sim 500 \gev$, and some other 68\%CL ranges go down to $500 \gev$, offering the possibility of pair production at a
collider with $\sqrt{s} \sim 1 \tev$, as envisaged for the final stage
of the ILC~\cite{ILC-TDR,LCreport}. Going to higher centre-of-mass energies, e.g.,
$\sqrt{s} \lsim 3 \tev$~\cite{CLIC:2016zwp,LCreport} as
anticipated for CLIC, significant fractions of the 68\% CL ranges of electroweak sparticle masses
can be covered.

\begin{table*}[htb!]
\renewcommand{\arraystretch}{1.4}
\vspace{1em}
\begin{center}
\begin{tabular}{|c|c|c|c|c|c|c|c|} \hline
${\tilde \tau_1}$   &  ${\tilde \tau_2}$  & ${\tilde e_L}$ &  ${\tilde e_R}$ & ${\tilde \nu_\tau}$ & ${\tilde q_L}$ &  ${\tilde t_1}$ & ${\tilde t_2}$ \\ 
\hline         
470 & 660 & 630 & 678 & 570 & 2130 & 1840 & 2180 \\
\hline \hline
${\tilde b_1}$ & ${\tilde b_2}$ & ${\tilde u_R}$ & ${\tilde d_R}$ & ${\tilde g}$ & $M_{H,A}$ &  \mneu1    & $m_{\neu2,\cha1}$  \\ 
\hline         
1940 & 2090 & 2000 & 1980 & 2310 & 1620 & 460 & 860\\
\hline
\end{tabular}
\caption{\it Particle masses at the best-fit point in the SUSY SU(5) GUT model (in GeV units).
} 
\label{tab:spectrum}
\end{center}
\renewcommand{\arraystretch}{1.15}
\end{table*}

\begin{figure*}[htb!]
\begin{center}
\resizebox{10cm}{!}{\includegraphics{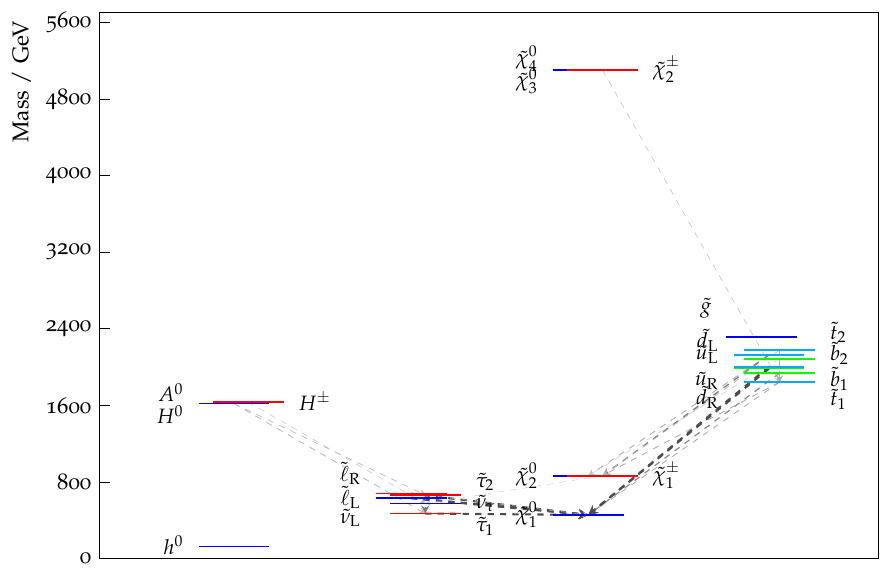}}
\vspace{-0.5cm}
\end{center}
\caption{\it 
  The spectrum at the best-fit point in the SUSY SU(5) GUT model.
  Decay branching ratios (BRs) exceeding 20\% are denoted by dashed lines, which are thicker for more important BRs.
}
\label{fig:bestfit}
\end{figure*}

\begin{figure*}[htb!]
\mbox{}\hspace{-3em}
\resizebox{19cm}{!}{\includegraphics{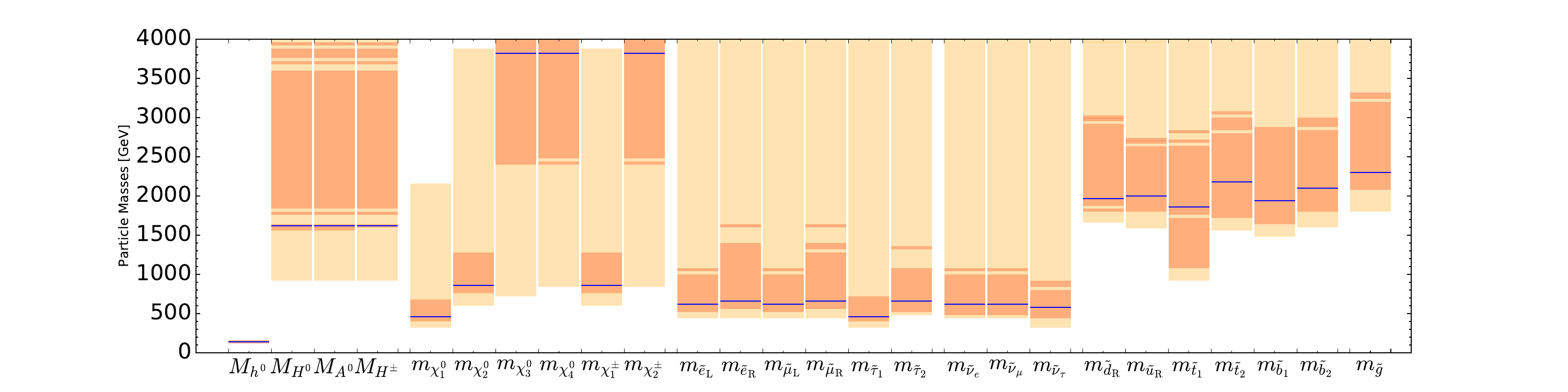}}
\vspace{-0.5cm}
\caption{\it The 1-dimensional 68 and 95\% CL ranges of masses we obtain for the current fit in the 
supersymmetric SU(5) model, shown in dark and light orange respectively. 
The best-fit point is represented by blue lines.
}
\label{fig:spectrum}
\end{figure*}

As already noted, a novel feature of the SUSY SU(5) GUT model with ($m_5 \ne m_{10}$)
is that the ${\tilde u_R}$ and ${\tilde c_R}$ may be much lighter than the other squarks. This leads to
the possibility of a ${\tilde u_R}/{\tilde c_R} - \neu1$ coannihilation strip where $m_{\tilde u_R}$ and $m_{\tilde c_R}
\sim 500 \gev$, which is visible as a second local minimum of $\chi^2$ with $\Delta \chi^2 > 4$ in the centre right
panel of Fig.~\ref{fig:masschi2}. 

We have checked specifically whether this strip is allowed by the available LHC constraints.
To this end, we verified using the {\tt Atom} simulation code that points
along this strip are consistent with the published constraints from the LHC 8-TeV data.
We have also checked that this strip is consistent with the preliminary simplified
model search for $\sq \sq + \sq \asq$ at
13 TeV reported by CMS. The left panel of Fig.~\ref{fig:smallmuR}
displays {as a solid/dashed blue line the one-dimensional $\chi^2$ function for $m_{\tilde u_R} - \mneu1$ 
including/omitting the 13-TeV data (the corresponding lines} for $m_{\tilde c_R} - \mneu1$
are} very similar), and the right panel of Fig.~\ref{fig:smallmuR} shows the region of the $(m_{\tilde u_R}, \mneu1)$
plane where $\Delta \chi^2 < {5.99}$, i.e., allowed at the 95\% CL.
We find that $\sigma (\sq \sq + \sq \asq)
< 0.1$~pb in this region, whereas the cross section upper limit as given
in~\cite{CMS:2016xva} is $\gtrsim 1$~pb. 
We conclude that this simplified model search does not affect the likelihood in this
${\tilde u_R}/{\tilde c_R} - \neu1$ coannihilation strip region.
{However, it will be explored further by future LHC data with increased luminosity.}

{Finally, we comment on the impact of the constraints from mono-jet 
searches~\cite{Aaboud:2016tnv, CMS:2016tns, CMS:2014yma}.
These searches are designed to be sensitive to the highly compressed mass region
by limiting the multiplicity of the high $p_T$ jets.
In the ${\tilde u_R}/{\tilde c_R} - \neu1$ coannihilation region, the mass difference is mildly compressed,
$m_{\tilde u_R/\tilde c_R} - m_{\neu1} \sim 40$ GeV, and the jets from ${\tilde u_R}/{\tilde c_R}$ decays are still resolvable from the background.
Such extra jets will spoil the characteristic of the mono-jet event and reduce the efficiency.
The degradation of the sensitivity for the mildly compressed region is clearly seen for example in Fig.~5 of \cite{Aaboud:2016tnv}. 
For this reason, the mono-jet searches lose sensitivity to the ${\tilde u_R}/{\tilde c_R} - \neu1$ coannihilation region,
compared to the jets + $\ETslash$ analysis~\cite{CMS:2016xva},
and we do not consider them in this paper.
}

\begin{figure*}[htb!]
\begin{center}
\resizebox{7.5cm}{!}{\includegraphics{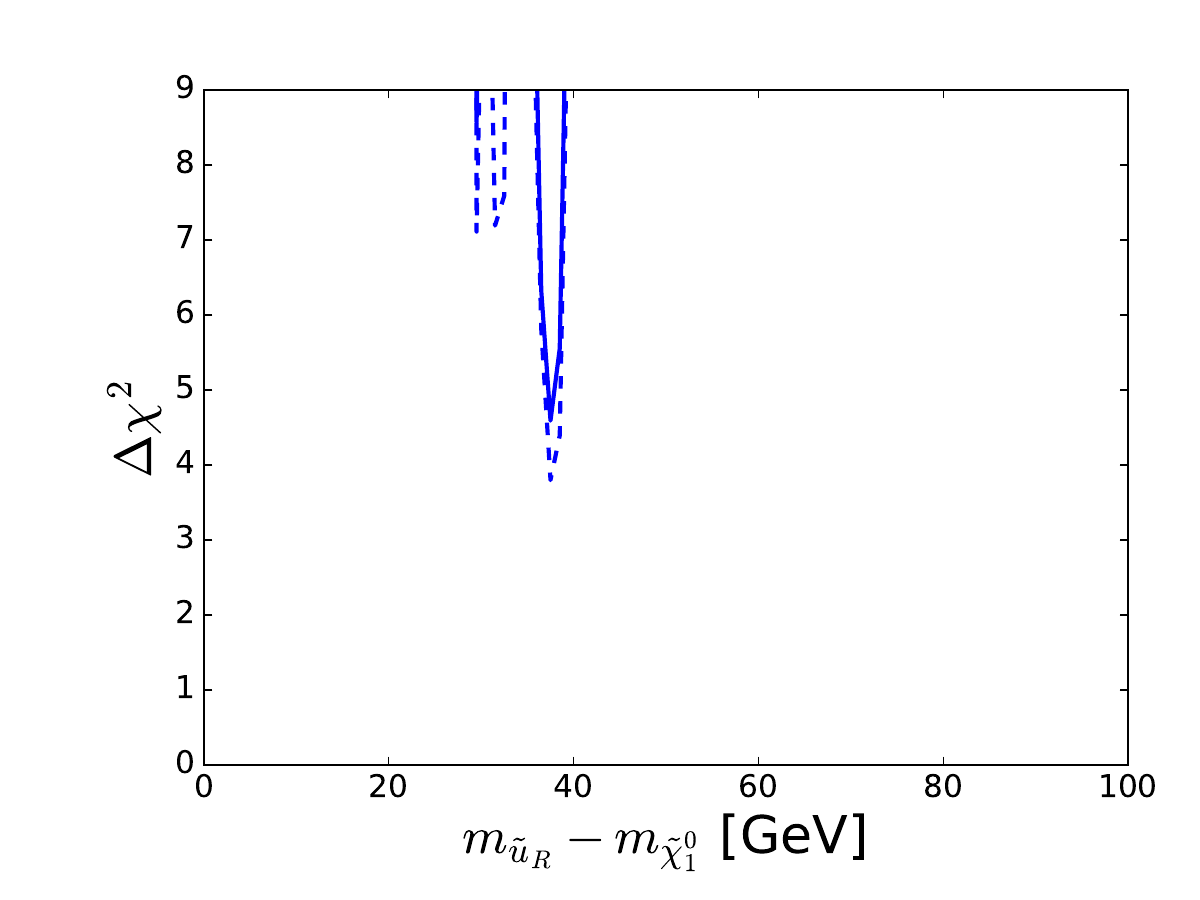}}
\resizebox{7.5cm}{!}{\includegraphics{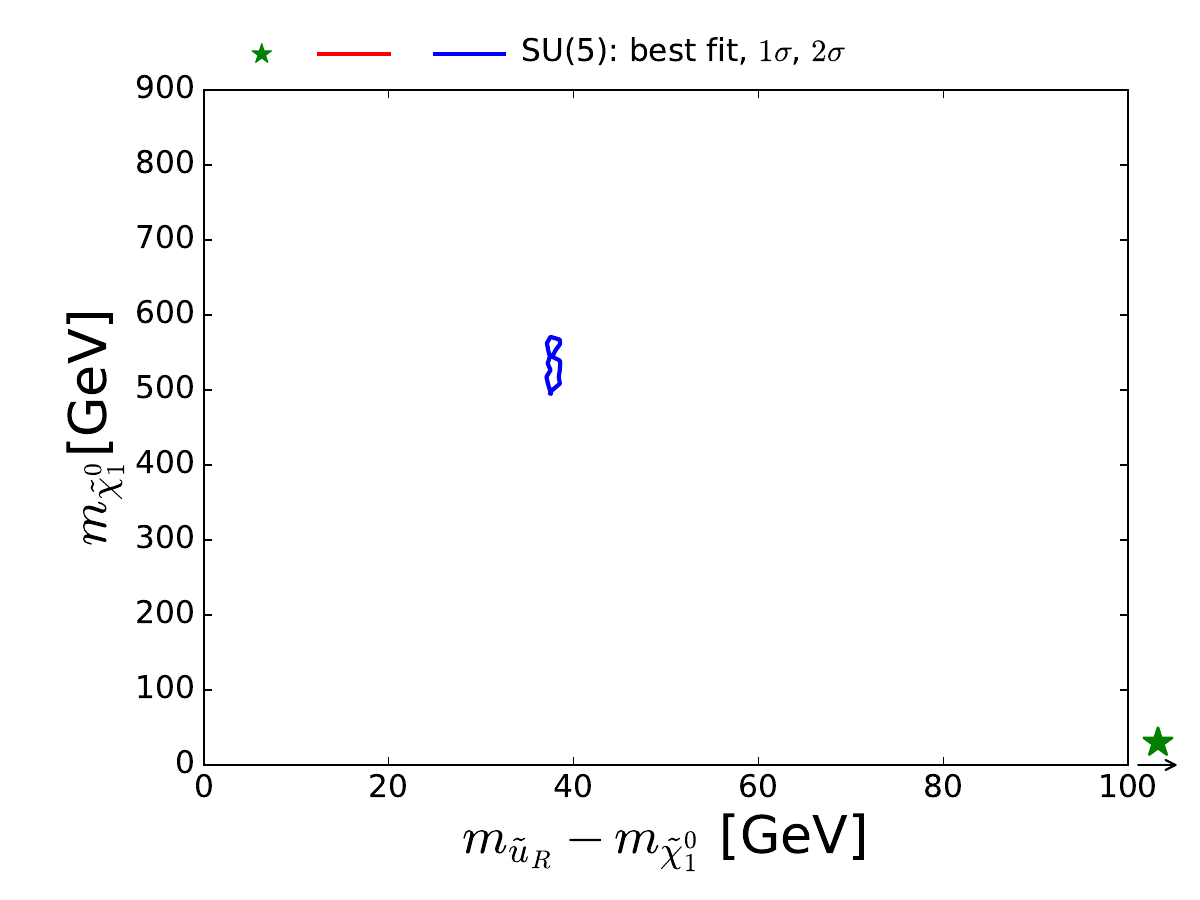}} \\
\end{center}
\vspace{-0.5cm}
\caption{\it 
Left panel: the $\chi^2$ likelihood function in the SUSY SU(5)
GUT model for $m_{\tilde u_R} - \mneu1$ in the ${\tilde u_R}/{\tilde c_R} - \neu1$ coannihilation strip region
({the solid/dashed line includes/omits the 13-TeV LHC data}).
Right panel:  the region of the $(m_{\tilde u_R}, \mneu1)$ plane where
$\Delta \chi^2 < {5.99}$.}
\label{fig:smallmuR}
\end{figure*}

{Another novel feature of the SUSY SU(5) GUT model is visible in
  \refta{tab:spectrum} and \reffi{fig:bestfit}.}
Having $m_5 \ne m_{10}$ allows the possibility of strong mixing between the ${\tilde \tau_R}$ in the $\mathbf{10}$
representation and the ${\tilde \tau_L}$ in the $\mathbf{\bar 5}$ representation. For example, at the
best-fit point the ${\tilde \tau_1}$ is an almost equal mixture of $\stau{L}$ and $\stau{R}$:
\begin{equation}
\stau1 \; = \; 0.70\, \stau{L} + 0.72\, \stau{R} \, .
\end{equation}
This large mixing explains the level repulsion $\Delta m \simeq 200 \gev$
between the ${\tilde \tau_1}$ and ${\tilde \tau_2}$ seen in Table~\ref{tab:spectrum},
which is much larger than the splitting $\Delta m \simeq 50 \gev$ between the
almost unmixed ${\tilde e_1} \sim {\tilde e_R}$ and ${\tilde e_2} \sim {\tilde e_L}$ 
that is also seen in Table~\ref{tab:spectrum}.

\smallskip
We show in Fig.~\ref{fig:gmtchi2} the contribution to the global
$\chi^2$ function of \gmt\ (in teal), as a function of $m_5$ (left panel), $m_{10}$ 
(middle panel) and $m_{1/2}$ (right panel). In each case, there is a
well-defined minimum that is lower than the plateau at large mass values by
$\Delta \chi^2 \gtrsim 2$. In contrast, the contributions to the global $\chi^2$ 
function of the other observables are relatively featureless over large ranges
of $m_5$, $m_{10}$ and $m_{1/2}$, with the exception of the contribution
from the LHC 13-TeV {data (mainly due to the} $\ETslash$ constraint), which rises sharply at low $m_{1/2}$,
as shown in the stacked red histogram in the right panel of Fig.~\ref{fig:gmtchi2}. Because we profile over the
other parameters, this does not have 
much impact on the dependence of $\chi^2$ on $m_5$ and $m_{10}$, as seen in the left and middle panels.
The well-defined minima seen in the \gmt\ contributions in the left and middle panels of Fig.~\ref{fig:gmtchi2}
occur at quite small values of $m_5$ and $m_{10}$, reflecting the fact that
\gmt\ is sensitive to the soft symmetry-breaking contributions to the masses
of both the ${\tilde \mu_L}$ and the ${\tilde \mu_R}$. These are $m_5$ and 
$m_{10}$, respectively, so maximizing the SUSY contribution to
\gmt\ and thereby minimizing the \gmt\ contribution to $\chi^2$
prefers small values of both $m_5$ and $m_{10}$. Similarly, the
SUSY contribution to \gmt\ is suppressed for large gaugino masses,
explaining the aversion to large $m_{1/2}$ seen in the right panel of 
Fig.~\ref{fig:gmtchi2}.

\begin{figure*}[htb!]
\begin{center}
\resizebox{5cm}{!}{\includegraphics{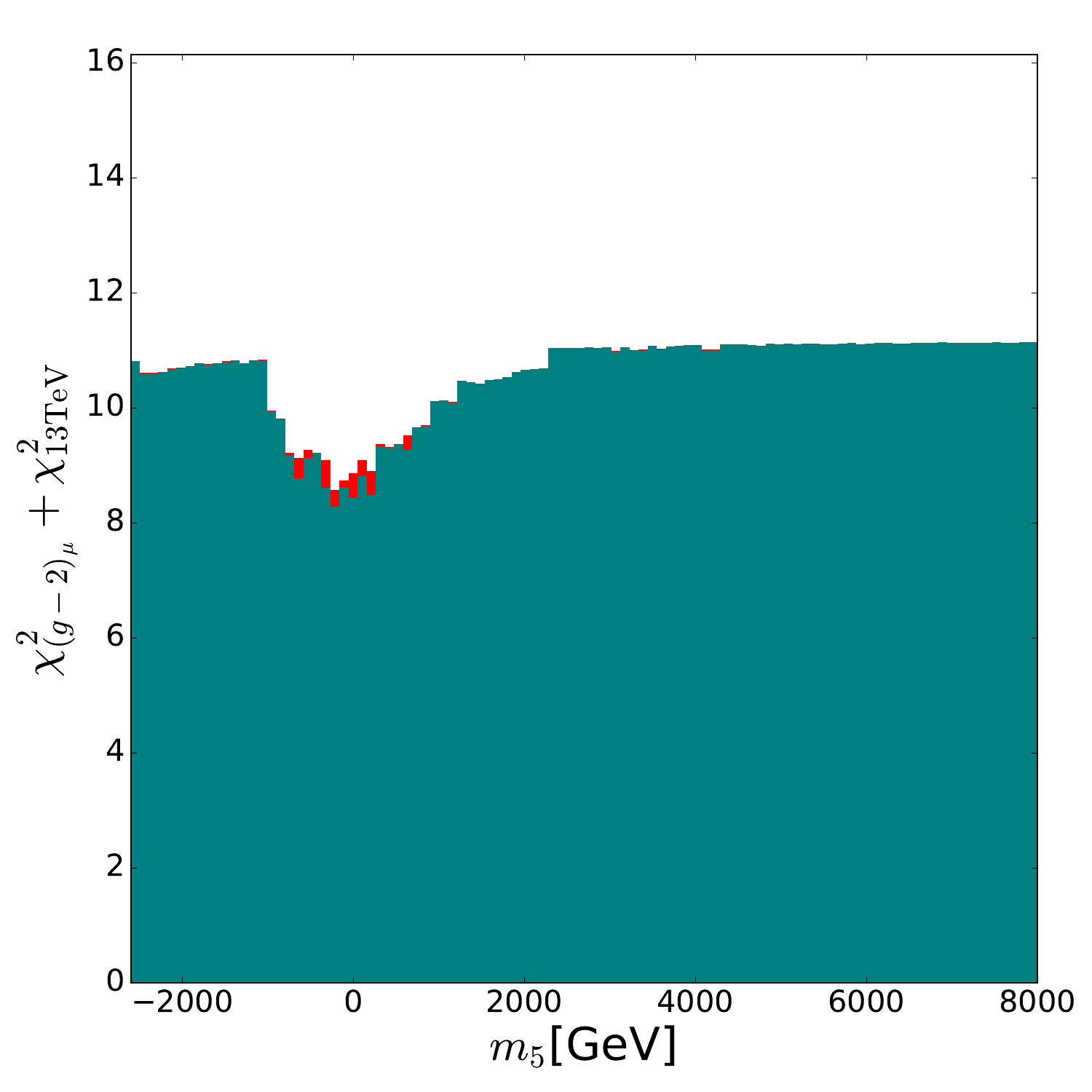}}
\resizebox{5cm}{!}{\includegraphics{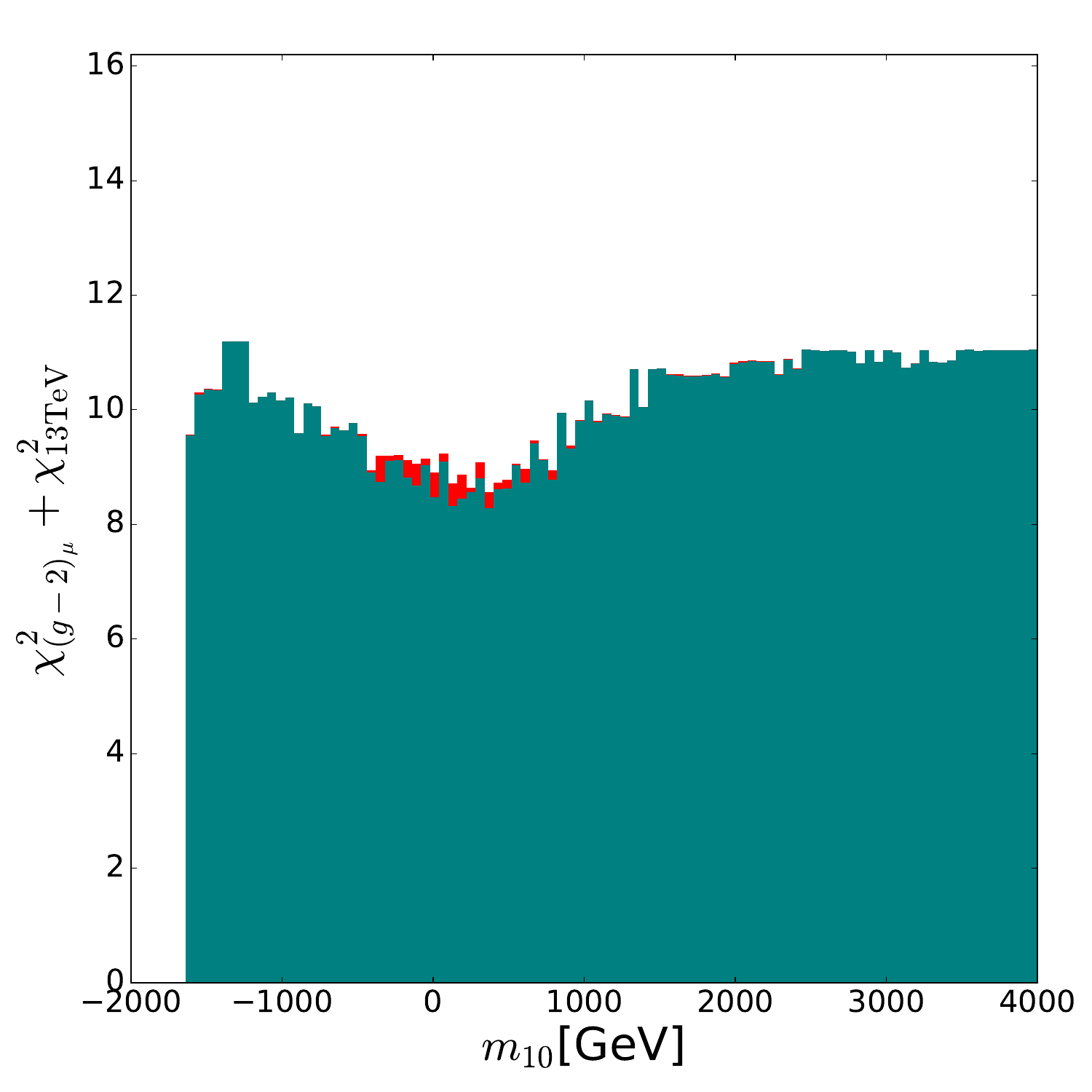}}
\resizebox{5cm}{!}{\includegraphics{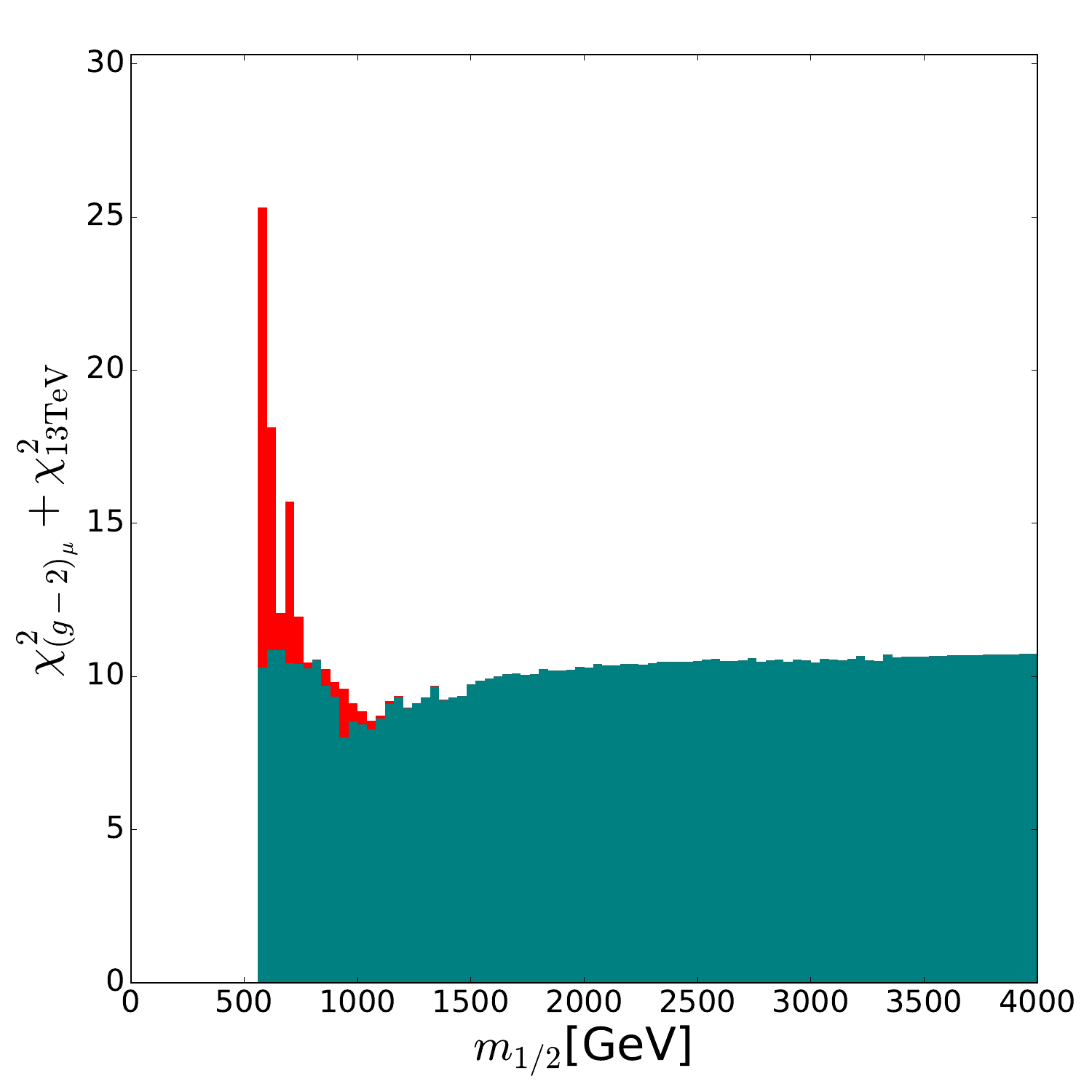}} \\
\vspace{-0.5cm}
\end{center}
\caption{\it 
The $\chi^2$ contributions of \gmt\ (teal) and {LHC 13-TeV data} (red) in the SUSY SU(5)
GUT model, as functions of $m_5$ (left panel), $m_{10}$ (middle panel) and
$m_{1/2}$ (right panel).
}
\label{fig:gmtchi2}
\end{figure*}

The principal contributions to the global $\chi^2$ function at the best-fit point for the SUSY SU(5) GUT model are
given in Table~\ref{tab:bestfitchi2}, and the corresponding pulls at the
best-fit point are
displayed graphically in Fig.~\ref{fig:pulls}.
Apart from \gmt, the other contributions deserving of comment include the
following. The large contribution from {\tt HiggsSignals} reflects the large number
of channels considered, and has negligible variation for most of the points in our sample.
We note that $A_{\rm FB} (b)$ makes a contribution that is not much smaller than that
of \gmt\ at the best-fit point, and that {\ALRe}
and $\sigma^0_{\rm had}$ also make relatively large contributions
to the global $\chi^2$ function. These observables reflect the residual tensions in the electroweak precision
observables at the $Z$ peak, which are present in the SM and the SUSY SU(5) GUT model
is unable to mitigate. 

\begin{table*}[htb!]
\begin{center}
\vspace{5mm}
\renewcommand{\arraystretch}{1.2}
\begin{tabular}{|c|c|c|c|c|c|} \hline
{\ALRe} &  $A_b$  & $A_{\rm FB}(\ell)$ & $A_{\rm FB}(b)$ &  $A_{\rm FB}(c)$ &  $A_{l}(P_\tau)$ \\ 
\hline         
3.40 & 0.35 & 0.78 & 6.79 & 0.82 & 0.08 \\
\hline
\hline
$R_b$ & \bsg\  & \btn\ & $\Omega_{\neu1} h^2$ &  {\ssi}   &   \bsdmm\  \\
\hline
0.26 & 0.00 & 0.18 & 0.00 & 0.00 & 2.09 \\
 \hline
 \hline
$\sweff$  & $M_W$ & $R_l$  & $R(K \to l \nu)$   &  \gmt\   & $M_h$     \\ 
\hline         
0.60 & 0.07  & 1.04 & 0.0 &  8.28  &  0.01  \\
\hline
 \end{tabular}

\vspace{0.5em}
\begin{tabular}{|c|c|c|c|c|c||c||}\hline
$\sigma_{\rm had}^0$ & ${\frac{\Delta M_{B_s}}{\Delta M_{B_d}}}$ & $\epsilon_K$ & $H/A \to \tau^+ \tau^-$ & {\tt HiggsSignals} & LHC $\ETslash$ & {\bf Total} \\ 
\hline         
2.54 & 1.78 & 1.94 & 0.00 & 67.95 & 0.3 & {\bf 100.34} \\
\hline
\end{tabular}
\caption{\it The principal $\chi^2$ contributions of observables at the
  best-fit point in the SUSY SU(5) GUT model, 
  together with the total $\chi^2$ function.}
\label{tab:bestfitchi2}
\end{center}
\end{table*}

\begin{figure}[htb!]
\begin{center}
\resizebox{5.5cm}{!}{\includegraphics{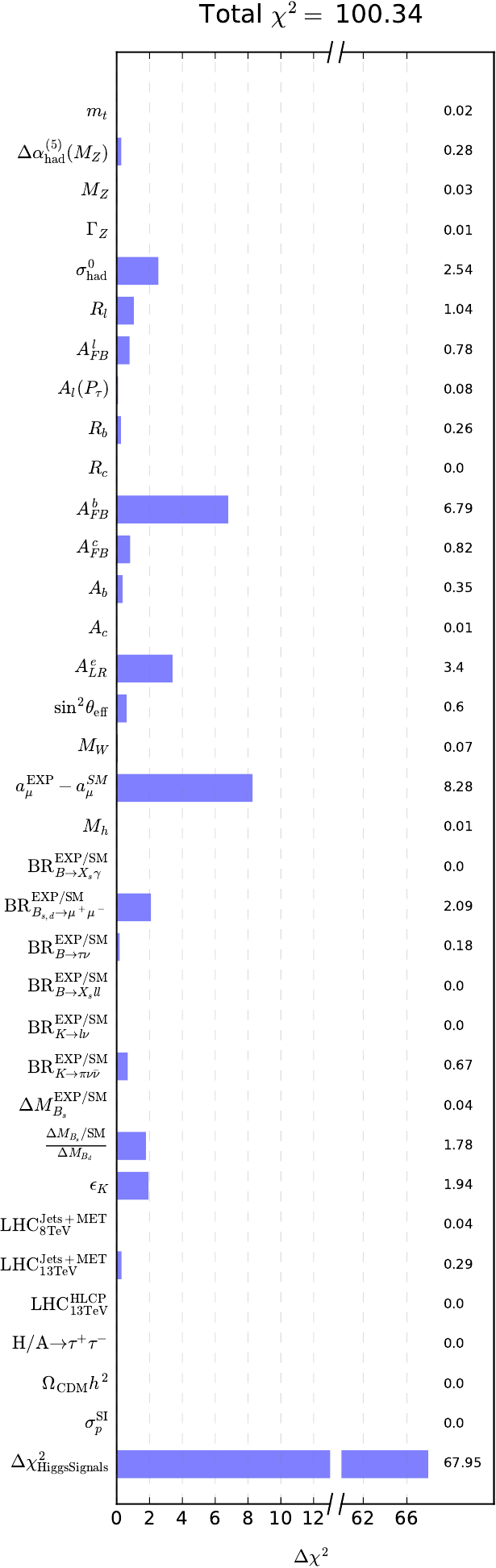}}
\end{center}
\vspace{-1.0cm}
\caption{\it 
The $\chi^2$ pulls for different observables at the best-fit point in
the SUSY SU(5) model.
}
\label{fig:pulls}
\end{figure}

{In order to compare the quality of the SU(5) fit to the results of
previous MasterCode analyses of competing models~\cite{mc11}, we follow
the prescription used there of subtracting from the total $\chi^2$ given
in Table~\ref{tab:bestfitchi2} and Fig.~\ref{fig:pulls}, namely 100.34,
the $\chi^2$ contributions originating from
{\tt HiggsSignals}~\cite{HiggsSignals}, which dominate the global $\chi^2$ function
and would bias the analysis. Fig.~\ref{fig:pulls} lists 36 separate contributions to the total $\chi^2$ function.
The first 3 ($m_t, M_Z$, and $\Delta \alpha_{\rm had}^{(5)} (M_Z)$) are treated as nuisance parameters
and the two LHC MET constraints at 8 and 13 TeV are applied as a single constraint.
Omitting the {\tt HiggsSignals} constraints in our determination 
of the number of degrees of freedom leaves 30 constraints,
with 7 parameters for the SU(5) model and hence 23 degrees of freedom.The $\chi^2$ contributions from the
relevant constraints sum to $32.39$, corresponding to a $\chi^2$ probability of 9\%.
This can be compared with the $\chi^2$ probability values of $11,12,11$ and 31\%
found in~\cite{mc11} for the CMSSM, NUHM1, NUHM2 and pMSSM10, respectively,
using LHC Run 1 constraints.
However, as in~\cite{mc11}, we stress that these $\chi^2$ probabilities
are only approximate since, for example, they neglect correlations
between the observables. A more complete treatment 
using toys, {as done in the last reference of~\cite{otherCMSSM}}, is beyond the scope of this work.}

There are a couple of important corollaries to this observation, one concerning
$m_{\tilde t_1}$. It is sensitive to $A_0$ as well as the soft 
SUSY-breaking contributions to the ${\tilde t_L}$ and ${\tilde t_R}$
mass parameters
(which are both given by {$m_{10}$} in the SUSY SU(5) GUT model).
Since $A_0$ is relatively poorly determined, 
the $\chi^2$ minimum for $m_{\tilde t_1}$ is relatively shallow, as seen in the
lower left panel of Fig.~\ref{fig:masschi2}.

The second observation concerns the sign of $\mu$. All our analysis has been
for $\mu > 0$, which is the sign capable of mitigating the discrepancy between
the experimental value of \gmt\ and the SM prediction. For $\mu < 0$,
the large-mass plateau would have a similar height as in Fig.~\ref{fig:gmtchi2},
but the $\chi^2$ function would rise monotonically at low values of $m_5$, 
$m_{10}$ and $m_{1/2}$, instead of featuring a dip. Thus, the $\mu < 0$
possibility would be disfavoured by $\Delta \chi^2 \gtrsim 2$, and the global
minimum would lie at large masses and be ill defined.

The $\chi^2$ distributions for some more observables are shown in Fig.~\ref{fig:moremasschi2},
We see that the minima for \mneu1\ (upper left panel) and \mcha1\ (upper right panel) are
quite well defined, mirroring the structure in the $\chi^2$ function for $m_{\tilde \tau_1}$ shown
in the lower right panel of Fig.~\ref{fig:masschi2}. The preference for a (very) small 
${\tilde \tau_1} - \neu1$ mass difference is seen in the lower left panel of Fig.~\ref{fig:moremasschi2},
and reflects the fact, commented on in connection with many previous figures, that the
best-fit point and much of the 68\% CL region lies in the ${\tilde \tau_1} - \neu1$
coannihilation region.
On the other hand, a small $\mste - \mneu1$ mass
difference is disfavoured, as seen in the lower right plot of
\reffi{fig:moremasschi2}, reflecting the fact that stop coannihiliation does not
play a significant role.

\begin{figure*}[htb!]
\begin{center}
\resizebox{7.5cm}{!}{\includegraphics{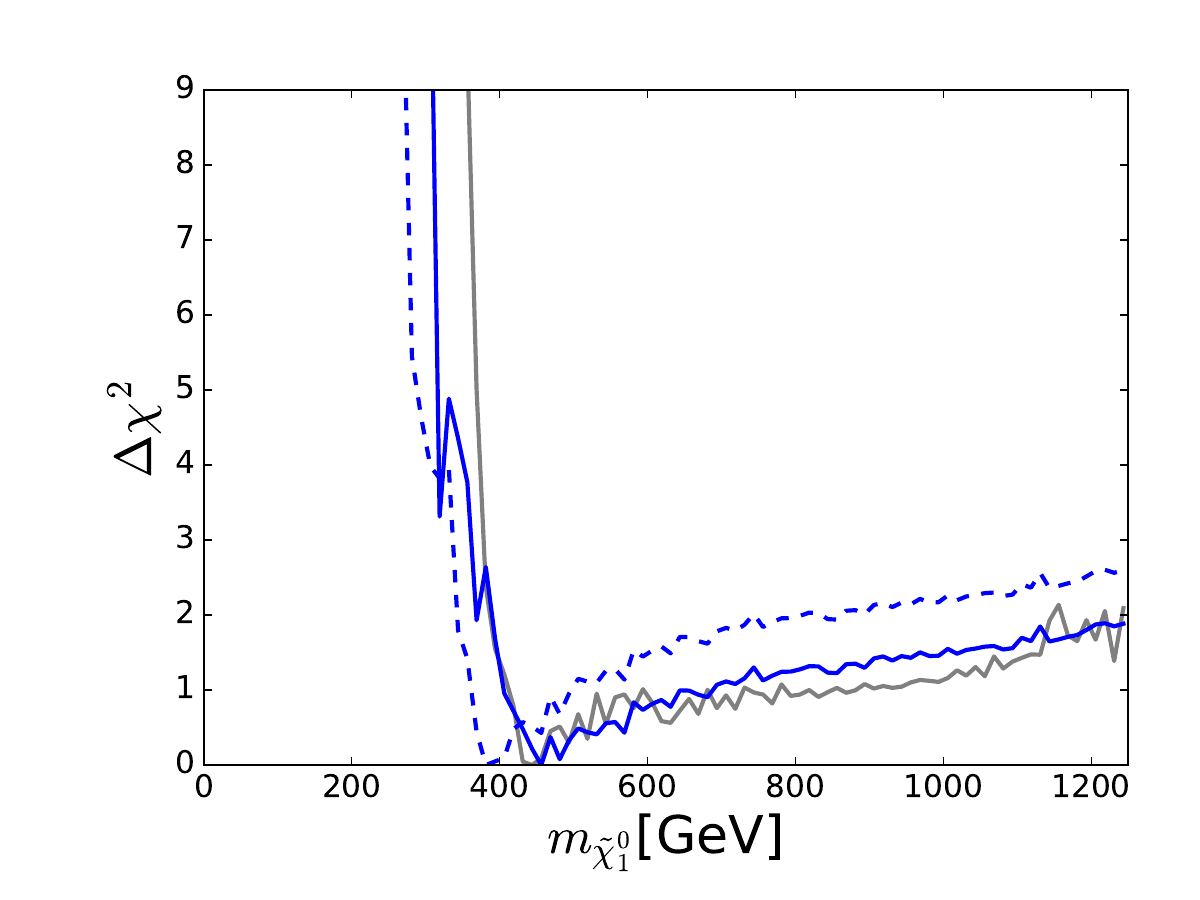}}
\resizebox{7.5cm}{!}{\includegraphics{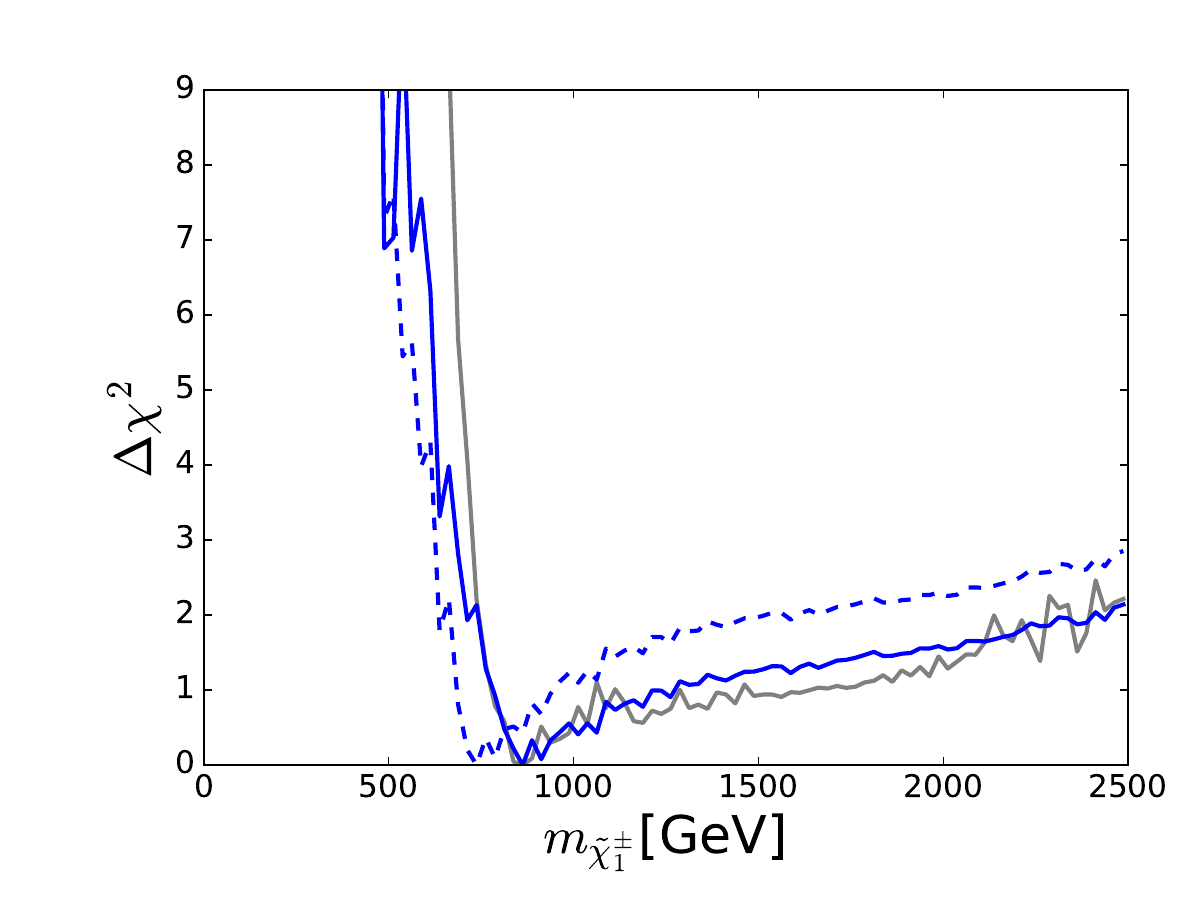}} \\
\resizebox{7.5cm}{!}{\includegraphics{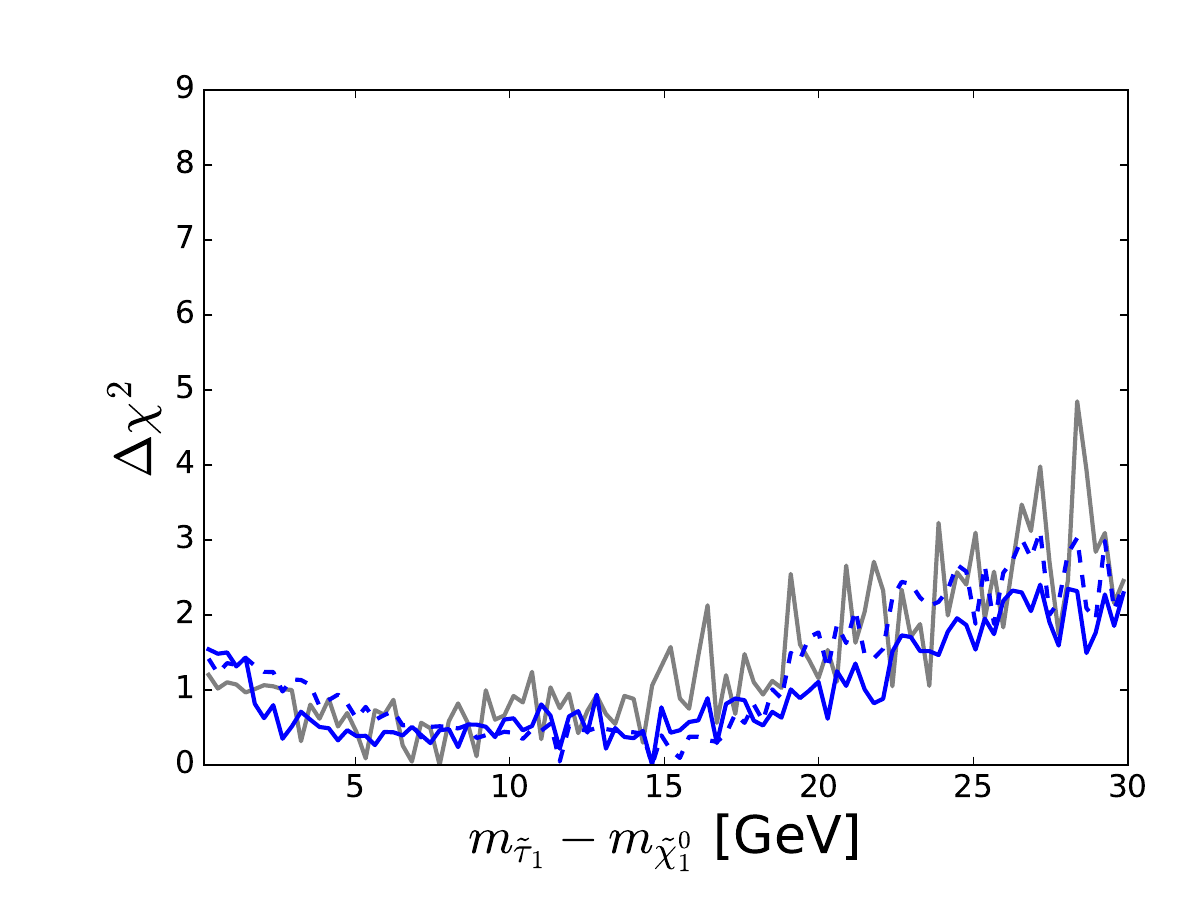}}
\resizebox{7.5cm}{!}{\includegraphics{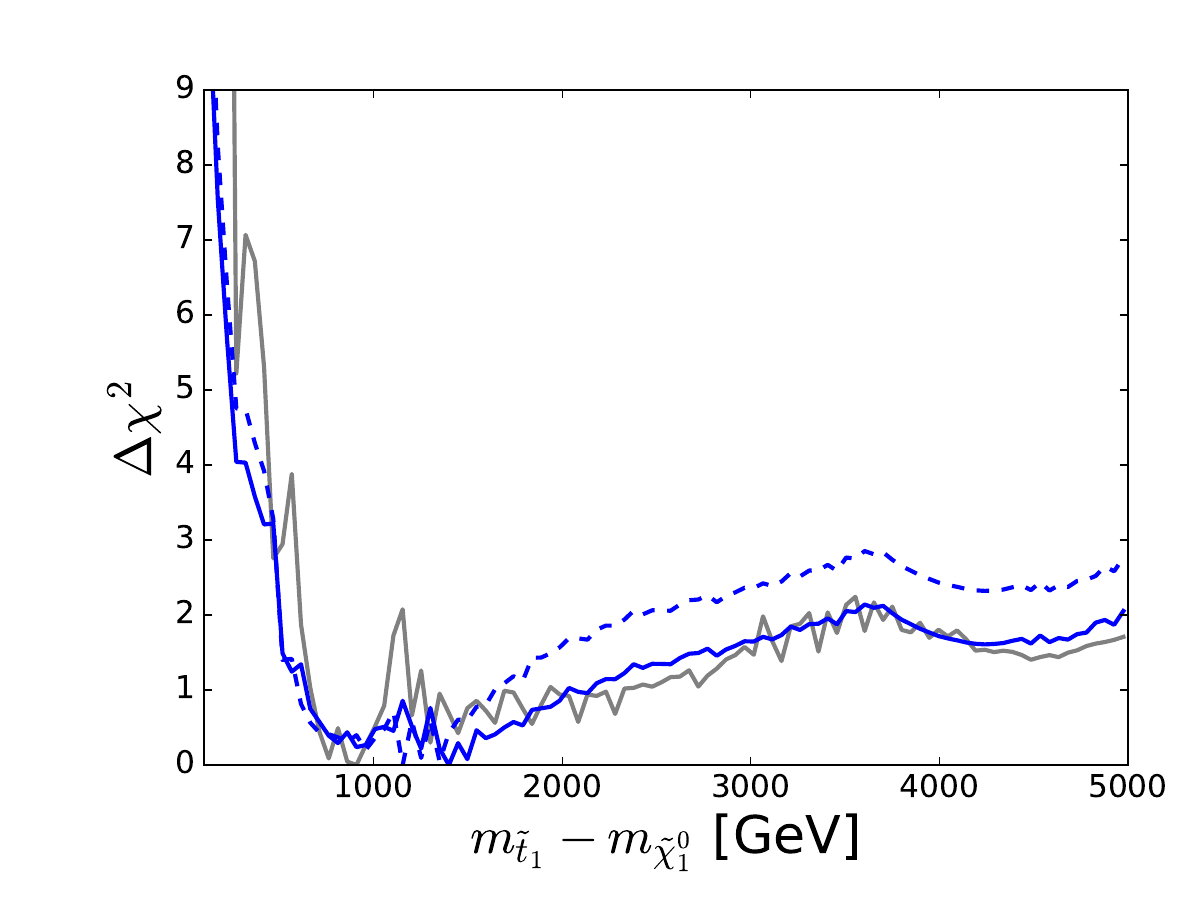}}
\end{center}
\vspace{-0.5cm}
\caption{\it 
The $\chi^2$ likelihood functions in the SUSY SU(5)
GUT model for the $\neu1$ mass (upper left panel), the $\cha1$ mass (upper right panel),
the ${\tilde \tau_1} -\neu1$ mass difference (lower left panel) and the ${\tilde t_1} - \neu1$ 
mass difference (lower right panel). {The dashed blue lines shows the result of omitting the LHC 13-TeV constraints, 
and the grey lines represent `fake' NUHM2 results obtained
by selecting a subset of the SU(5) sample with $m_5/m_{10} \in [0.9, 1.1]$.}
}
\label{fig:moremasschi2}
\end{figure*}

The $\cha1 - \neu1$ coannihilation region is prominent
in the previous figures, and also contains parameter sets that are preferred at the 68\% CL.
Hence a small $\cha1 - \neu1$ mass difference is also allowed at the $\Delta \chi^2 \gtrsim 1$
level, as seen in the left
panel of Fig.~\ref{fig:chachi2}, although the best-fit point has $\mcha1 - \mneu1 \sim 470 \gev$. 
However, values of the $\cha1$ lifetime that are allowed at the 95\% CL
are all too short to provide a long-lived particle signal, as seen in the right panel of Fig.~\ref{fig:chachi2}.
\footnote{For conditions to have a long-lived $\cha1$ with a bino-like LSP: see, e.g., \cite{Rolbiecki:2015gsa}.}

\begin{figure*}[htb!]
\begin{center}
\resizebox{7.5cm}{!}{\includegraphics{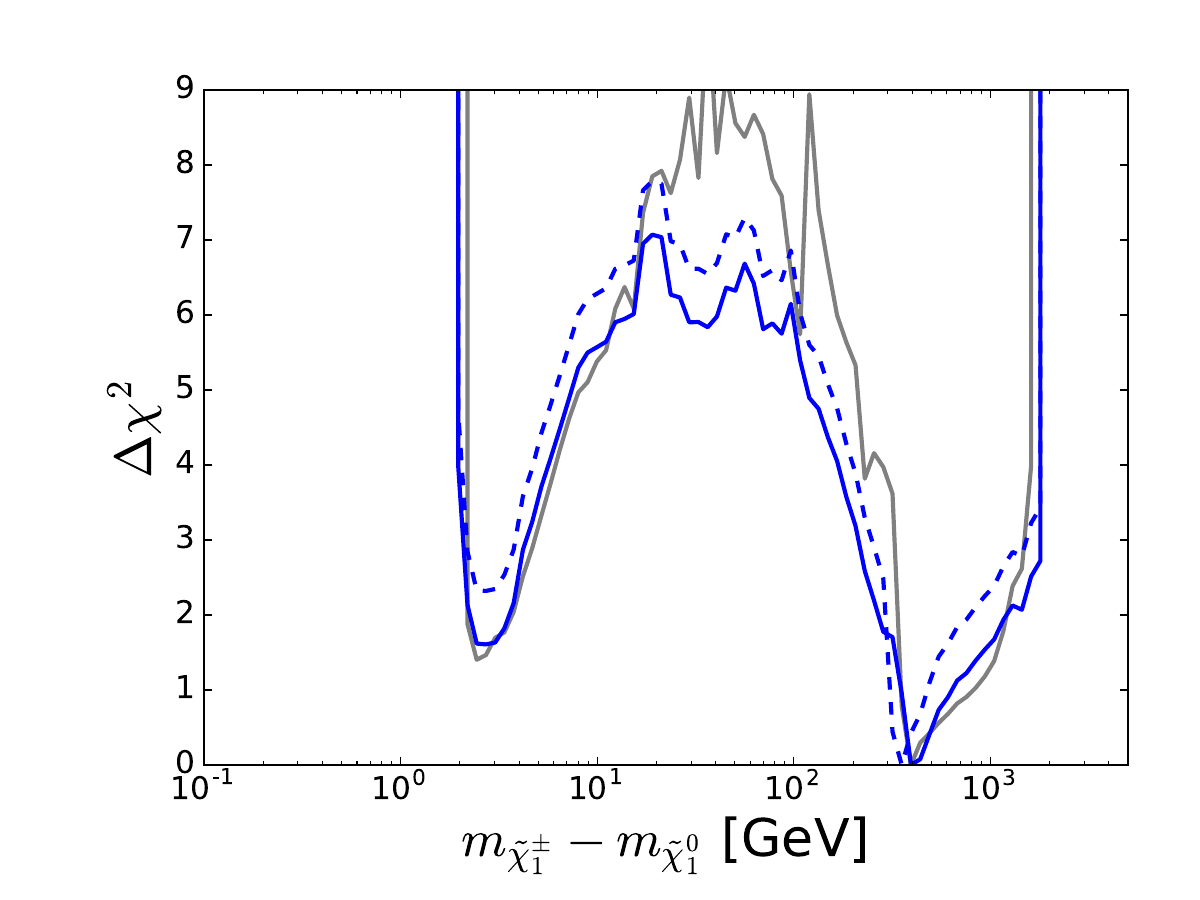}}
\resizebox{7.5cm}{!}{\includegraphics{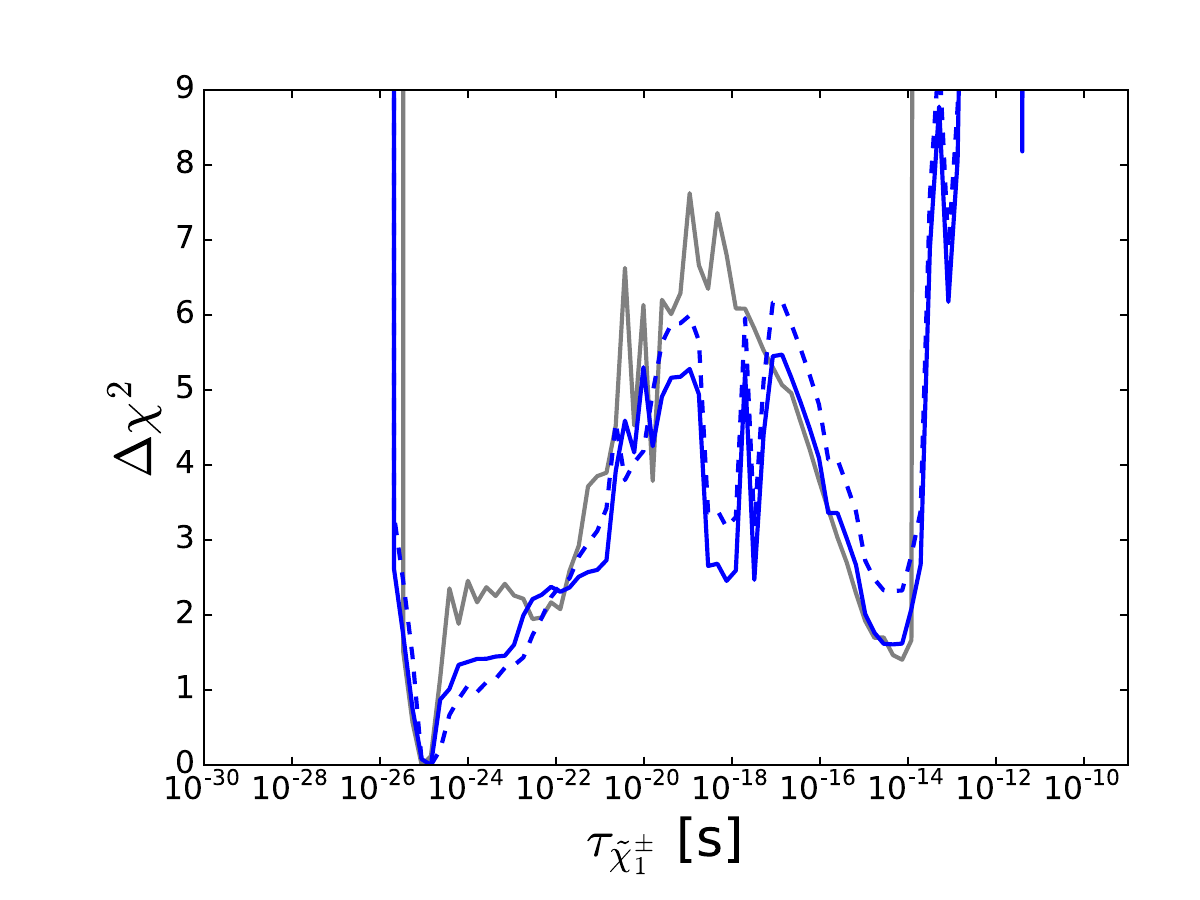}} \\
\end{center}
\vspace{-0.5cm}
\caption{\it 
The $\chi^2$ likelihood functions in the SUSY SU(5)
GUT model for the $\cha1 - \neu1$ mass (left panel) and the $\cha1$ lifetime (right panel).
{The dashed blue lines shows the result of omitting the LHC 13-TeV constraints, 
and the grey lines represent `fake' NUHM2 results obtained
by selecting a subset of the SU(5) sample with $m_5/m_{10} \in [0.9, 1.1]$.}
}
\label{fig:chachi2}
\end{figure*}

\medskip
We now discuss the 
one-dimensional likelihood functions for electroweak precision
observables and observables in the flavour sector shown in Fig.~\ref{fig:otherchi2},
together with the current experimental measurements and their uncertainties shown as dotted grey lines.
The upper left panel displays \gmt,
and we see that the global minimum occurs for $\Delta \gmt \simeq 0.4 \times 10^{-9}$,
with $\Delta \chi^2 \lesssim - 2$ compared to the case $\Delta \gmt = 0$.
{We see again that} the SUSY SU(5) GUT model is able to mitigate slightly
the discrepancy between the SM and the measurement of \gmt,
{although it} does not provide a substantial improvement over the SM
  prediction.

\begin{figure*}[htb!]
\begin{center}
\resizebox{7.5cm}{!}{\includegraphics{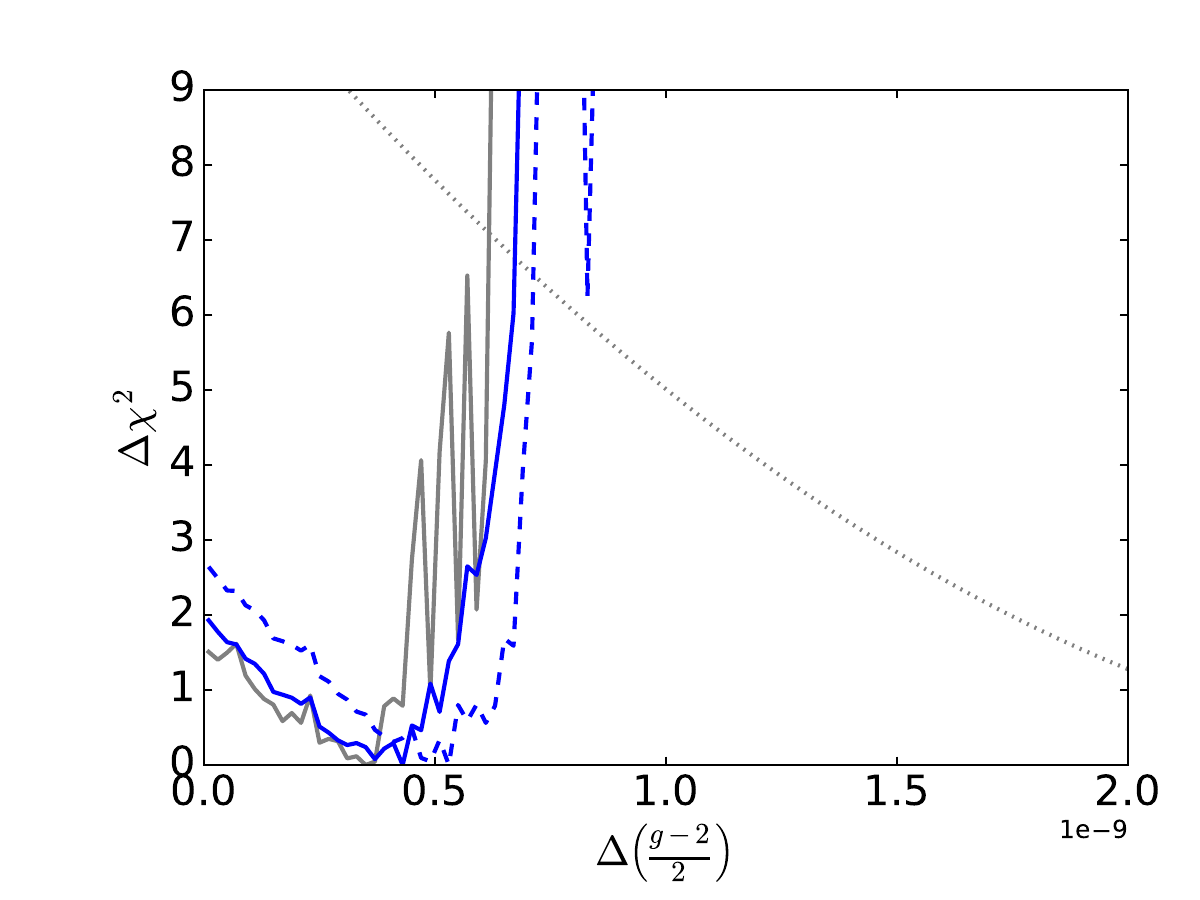}}
\resizebox{7.5cm}{!}{\includegraphics{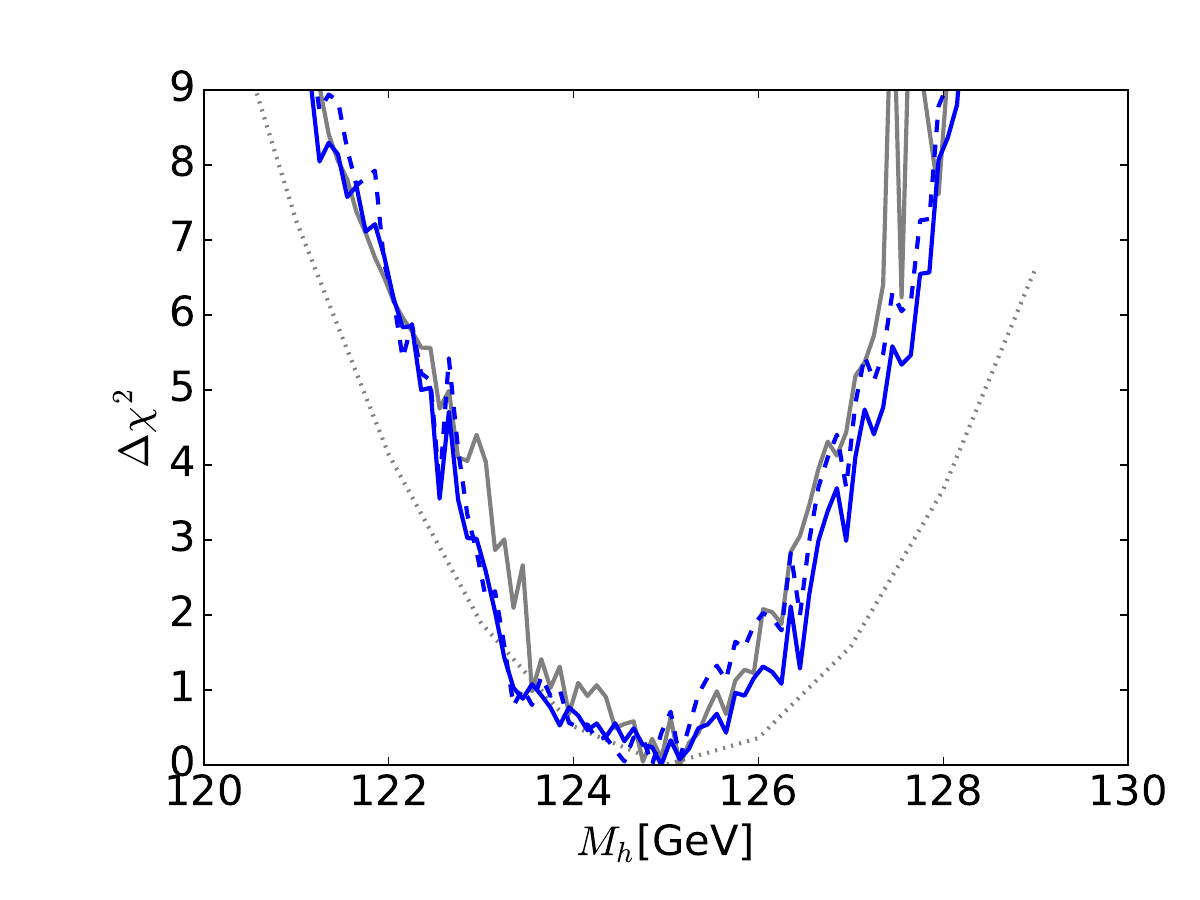}} \\[-1.5em]
\resizebox{7.5cm}{!}{\includegraphics{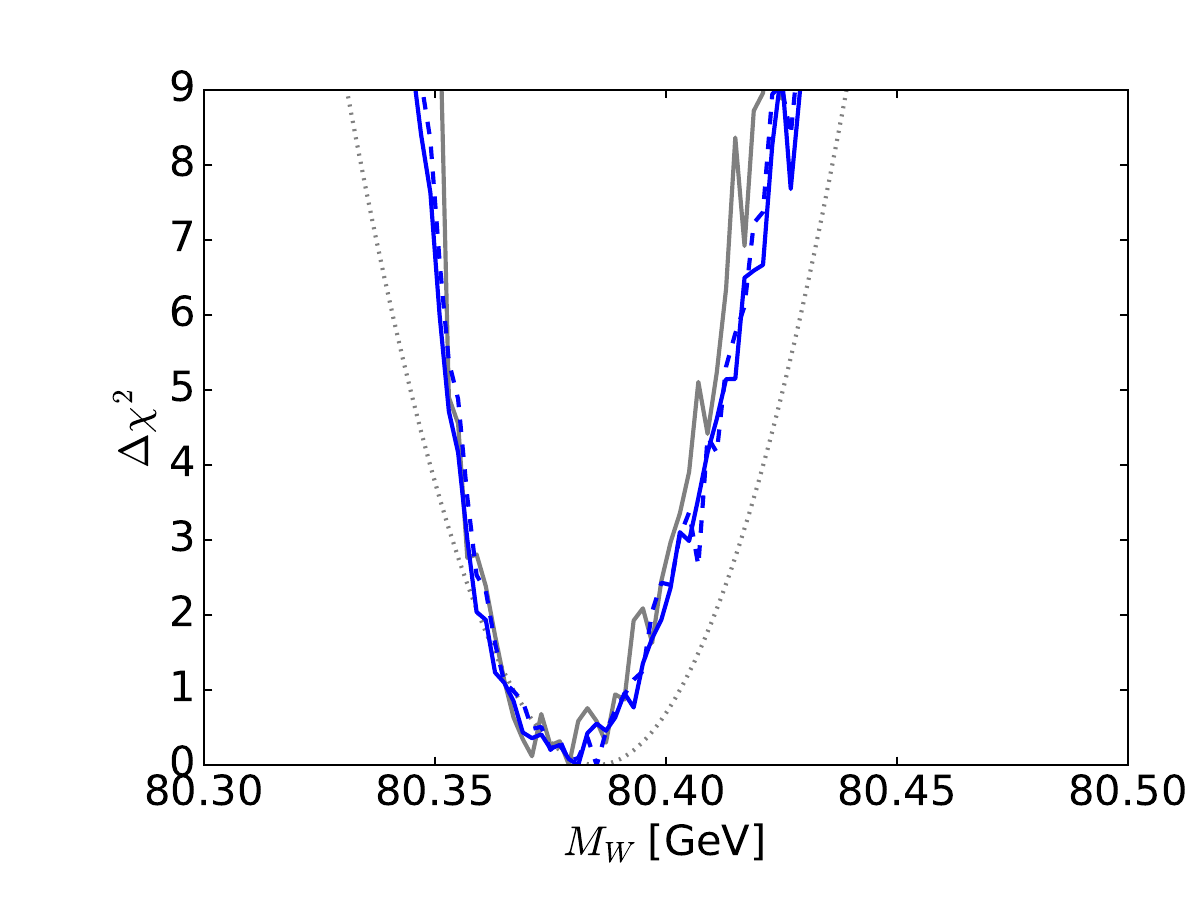}}
\resizebox{7.5cm}{!}{\includegraphics{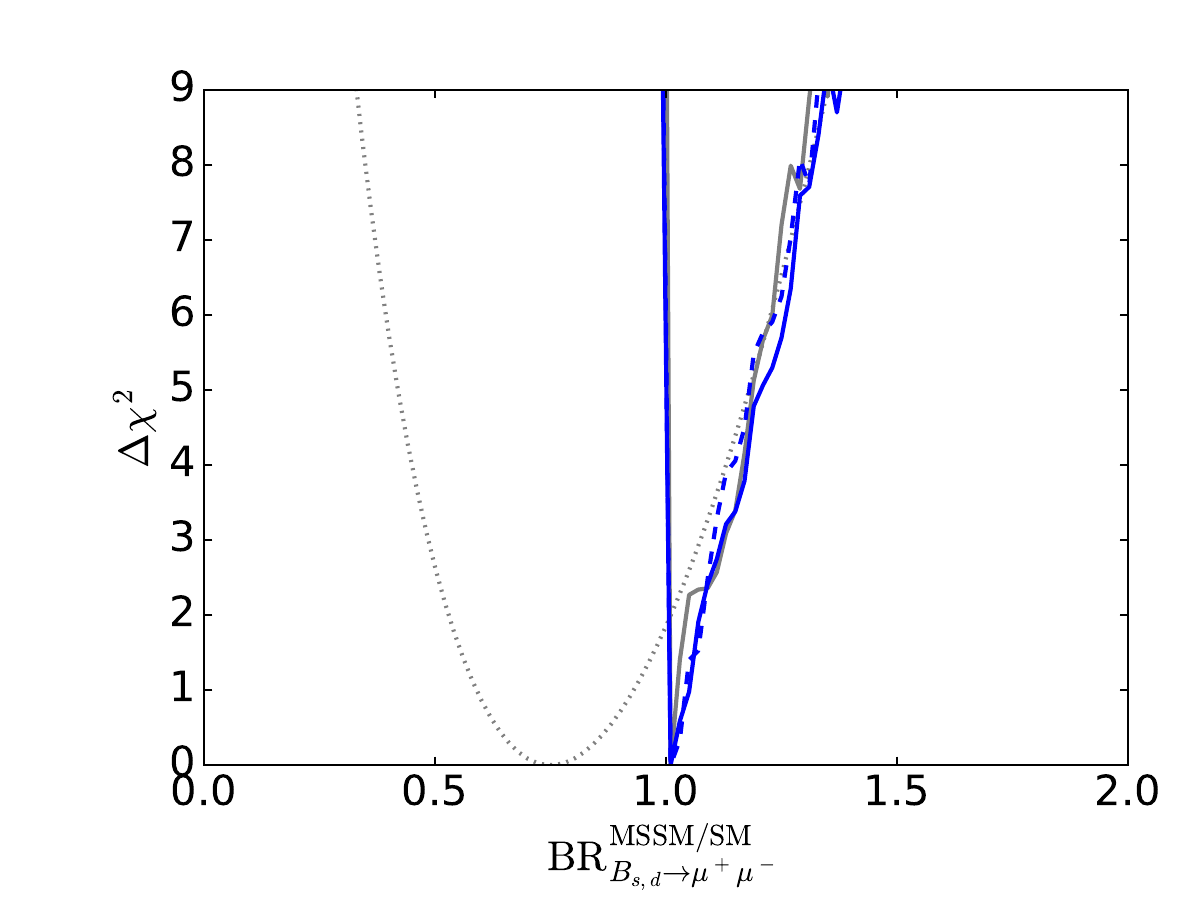}}
\end{center}
\vspace{-1.0cm}
\caption{\it 
The $\chi^2$ likelihood functions in the SUSY SU(5)
GUT model for $\gmt/2$ (upper left panel), $\Mh$ (upper right panel), $\MW$ (lower left panel),
and \bsdmm\ (lower right panel).
{The dashed blue lines shows the result of omitting the LHC 13-TeV constraints, 
and the solid grey lines represent `fake' NUHM2 results obtained
by selecting a subset of the SU(5) sample with $m_5/m_{10} \in [0.9, 1.1]$.,
and the dotted grey lines represent the current experimental measurements
with their uncertainties.}
}
\label{fig:otherchi2}
\end{figure*}

As for $\Mh$, as shown in the upper right panel of Fig.~\ref{fig:otherchi2}
the $\chi^2$ function is minimized close to the nominal experimental
value, and is quite symmetric, showing no indication of any tension in
the SUSY SU(5) GUT model fit. Likewise, the best-fit value
of $\MW$ (lower left panel of Fig.~\ref{fig:otherchi2}) is highly compatible 
with the experimental measurement, and that for \bsdmm\ (lower right panel)
is very close to the SM prediction, and hence also compatible
with the experimental measurement. We note that, whereas values of 
\bsdmm\ that are slightly larger than the SM value are possible,
smaller values are strongly disfavoured in the SUSY SU(5) GUT model.


\section{Higgs Branching Ratios}

We present in Fig.~\ref{fig:Higgs} the one-dimensional
likelihood functions for the ratios of supersymmetric SU(5) and SM predictions
for the BRs of $h \to \gamma \gamma$ (left panel), $h \to Z Z^*$ (middle panel)~\footnote{The
likelihood function for $h \to W W^*$ is very similar to that for $h \to Z Z^*$, because of
custodial symmetry.} and $h \to gg$ decays (right panel). We see that in
each case the
preferred region in the fit corresponds to a prediction in the 
SU(5) model that deviates from the SM case by at most a few \%, 
{whereas
the present experimental uncertainties in the different coupling
modifiers squared {(employing some theory assumptions)} are typically 
${\cal O}(30)$\%}~\cite{Khachatryan:2016vau}, {and a precision of
\order{5-10\%} (with the same theory assumptions) can be reached by
the end of the LHC programme.
On the other hand, future $e^+e^-$ colliders such as the ILC, CLIC or
FCC-ee anticipate a precision at the percent level for couplings to
fermions and at the permille level for couplings to massive gauge
bosons~\cite{LCreport,Gomez-Ceballos:2013zzn}. This offers the
possiblity that deviations from the SM in the SUSY SU(5) GUT model can
be measured in the future.}

\begin{figure*}[htb!]
\begin{center}
\resizebox{5cm}{!}{\includegraphics{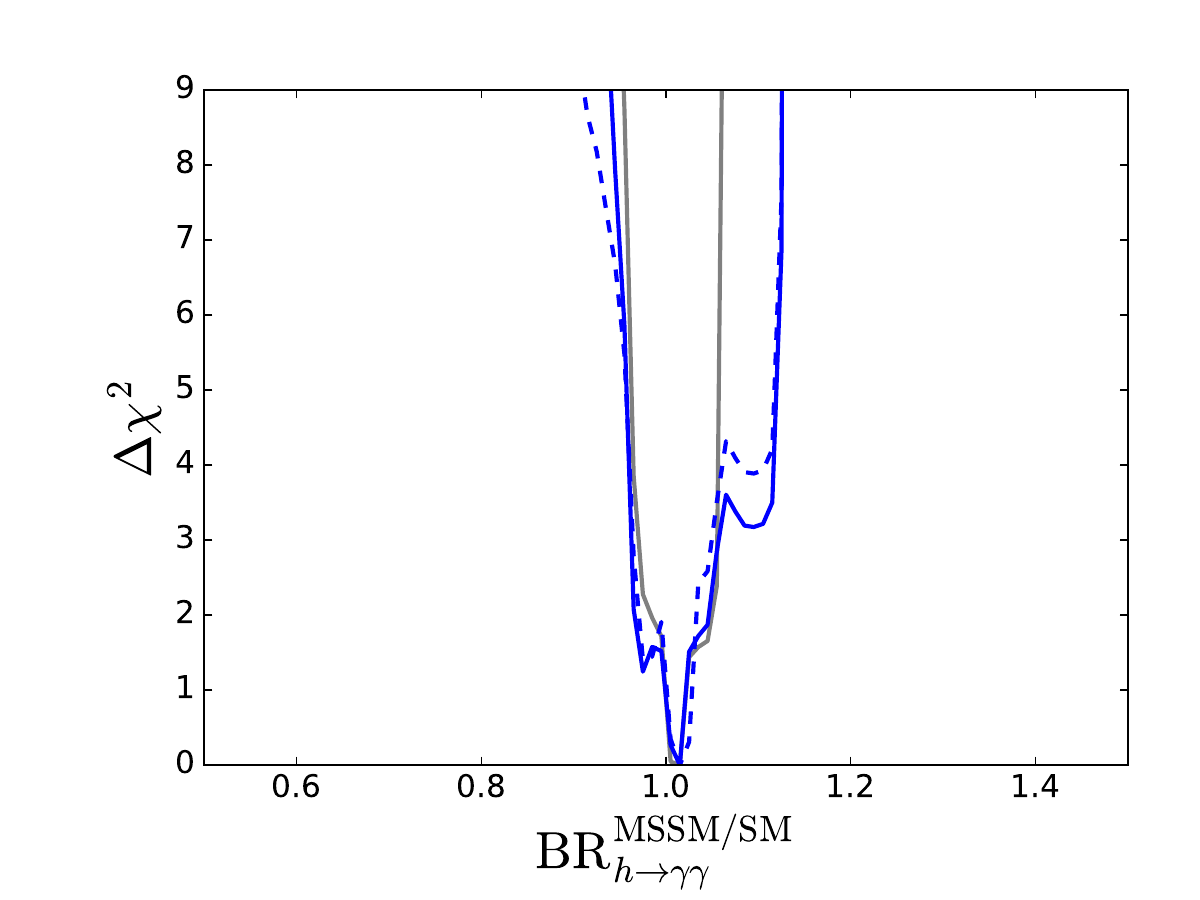}}
\resizebox{5cm}{!}{\includegraphics{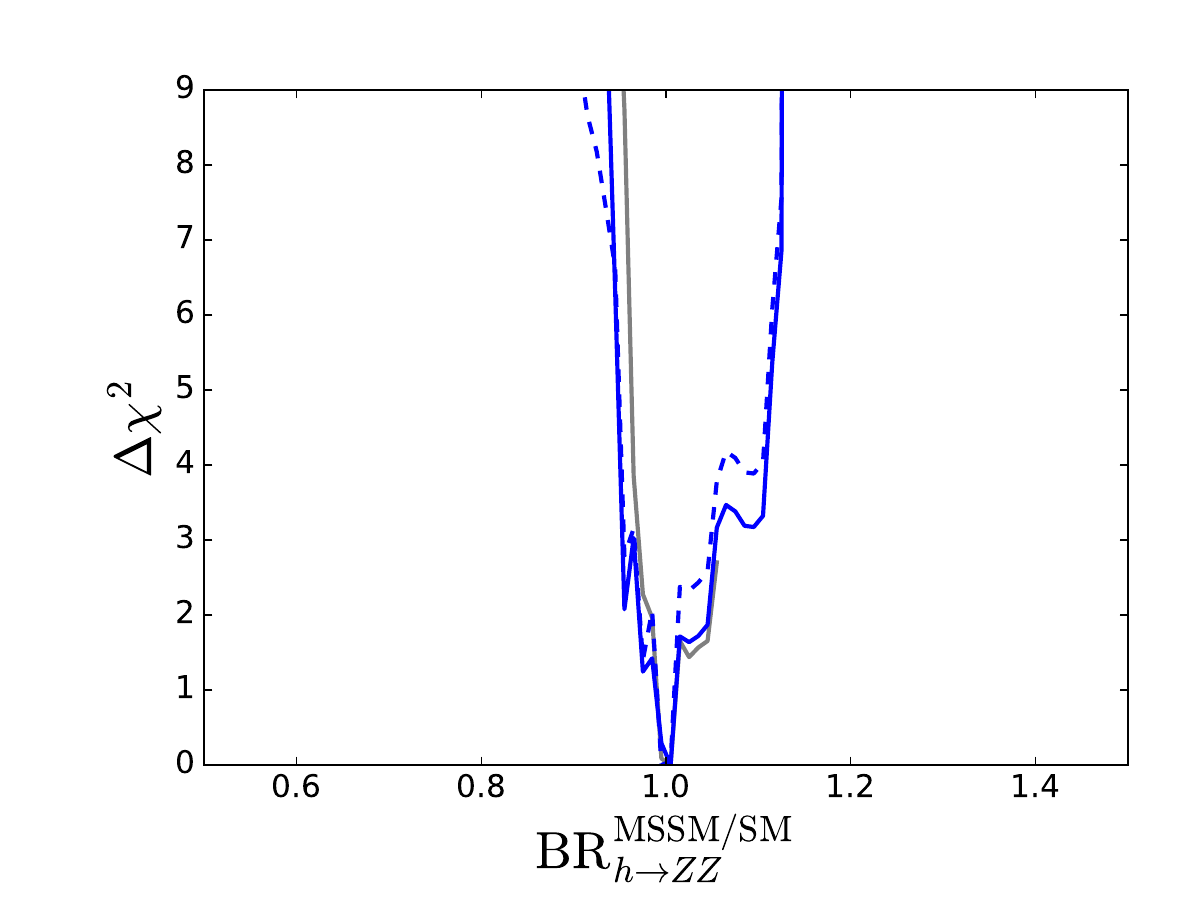}}
\resizebox{5cm}{!}{\includegraphics{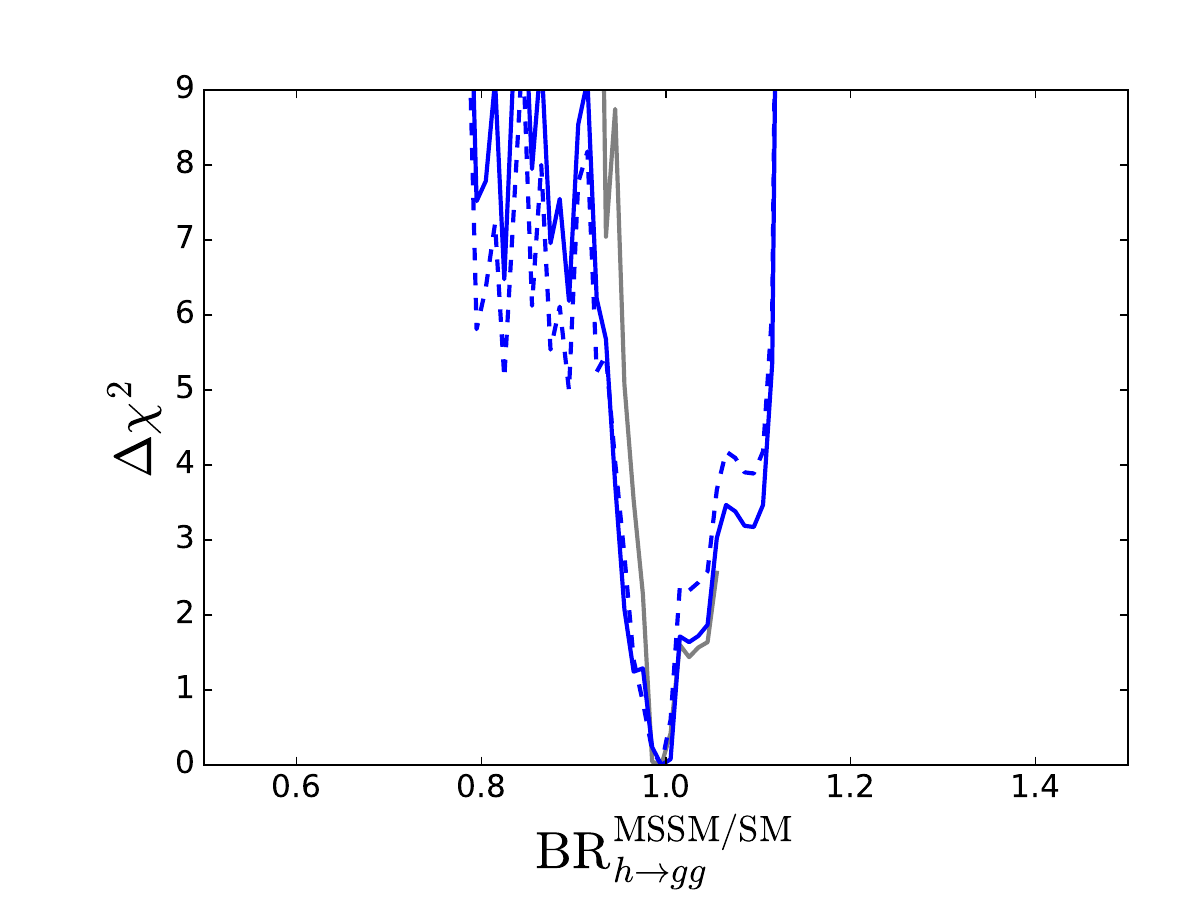}} \\
\end{center}
\caption{\it 
The $\chi^2$ likelihood functions for the ratios of the SUSY SU(5) and
SM predictions for the BRs of $h \to \gamma \gamma$ (left panel), $h \to
Z Z^*$ (middle panel) and  $h \to gg$ decays (right panel).
{The dashed blue lines shows the result of omitting the LHC 13-TeV
  constraints,  
and the grey lines represent `fake' NUHM2 results obtained
by selecting a subset of the SU(5) sample with $m_5/m_{10} \in [0.9, 1.1]$.}}
\label{fig:Higgs}
\end{figure*}


\section{Comparison with Previous Results}

{In previous papers we have studied the CMSSM, NUHM1 and NUHM2
using the LHC 8-TeV results
and earlier DM scattering constraints. None
of these models are directly comparable to the supersymmetric SU(5) model
studied here, which has 4 different soft SUSY-beaking scalar mass parameters,
$m_5, m_{10}, m_{H_u}$ and $m_{H_d}$. The most similar is the NUHM2,
which has the 3 parameters $m_0 = m_5 = m_{10}, m_{H_u}$ and $m_{H_d}$.
Here we compare the supersymmetric SU(5) 
results found in this paper using LHC 13-TeV data with `fake' NUHM2 results obtained
by selecting a subset of this SU(5) sample with $m_5/m_{10} \in [0.9, 1.1]$ (which
were also displayed as grey lines in Fig.~\ref{fig:masschi2}) and with previous
NUHM2 results~\cite{mc10}.

Fig.~\ref{fig:comparison} compares the one-dimensional $\chi^2$ likelihood functions
for $\mgl$ (upper left), $m_{\tilde q_R}$ (upper right), $m_{\tilde t_1}$ (lower left)
and $m_{\tilde \tau_1}$ (lower right) found in the SU(5) model including LHC 13-TeV constraints
(solid blue lines) with the restricted fake NUHM2 sample
(solid grey lines) and, for comparison, results from our previous NUHM2 analysis 
that used only the LHC 7- and 8-TeV constraints (dashed
grey lines)~\cite{mc10}. We see here and in Fig.~\ref{fig:masschi2}
that the restricted `fake' NUHM2 sample exhibits, in general, best-fit 
masses that are similar to those found in the full SU(5) sample. The
most noticeable differences 
are that lower masses are disfavoured in the restricted sample
relative those in the full SU(5) model, indicating that the latter has
some limited ability to relax the NUHM2 lower bounds on sparticle
masses, e.g., at the 95\% CL. The previous 
NUHM2 analysis~\cite{mc10} also yielded similar best-fit masses but, as
could be expected, gave 95\% CL lower limits on sparticle masses that
were further relaxed. 
{Similar features can also be observed in
\reffis{fig:moremasschi2} - \ref{fig:Higgs},
where we have also included the `fake' NUHM2 subsample.}

Restricting further our SU(5) to mimic the NUHM1, let alone the CMSSM, is not
useful because of the increased sampling uncertainties in such restricted samples.
However, we showed in~\cite{mc10} that our NUHM2 LHC 7- and 8-TeV results for the
exhibited sparticle masses were broadly similar to those for the NUHM1 and the CMSSM~\cite{mc9},
{and we expect the impacts of the LHC 13-TeV data on these models to be comparable to that in the NUHM2.}

{Finally, we ask whether or not there
is a significant improvement in the SU(5) fit compared to that in the NUHM2 subsample, thanks to the additional
parameter ($m_5$ and $m_{10}$ replacing $m_0$).
The NUHM2 subsample has a total $\chi^2 = 100.8$, which is reduced to 32.8 when we remove
the contributions from {\tt HiggsSignals}, as discussed earlier.
It should be noted that the NUHM2 subsample is statistically significantly smaller than that of the SU(5) sample.
The quoted NUHM2 $\chi^2$ represents only an upper bound on the $\chi^2$ of the best-fit point that would be found in a more complete sample of the NUHM2. 
Since the NUHM2 model has one less parameter than the SU(5) model, it has 24 degrees of freedom, and its $\chi^2$
probability is 11\%. {According to the Wilks test\cite{wilks}, the probability that the data are represented better by the
SU(5) model than by the NUHM2 subsample is 50\%, while the F-test\cite{Ftest} yields a 40\% probability.
Therefore we conclude that there is no evidence that the extra parameter of SU(5) provides a significant improvement.}

}

\begin{figure*}[htb!]
\begin{center}
\resizebox{7.5cm}{!}{\includegraphics{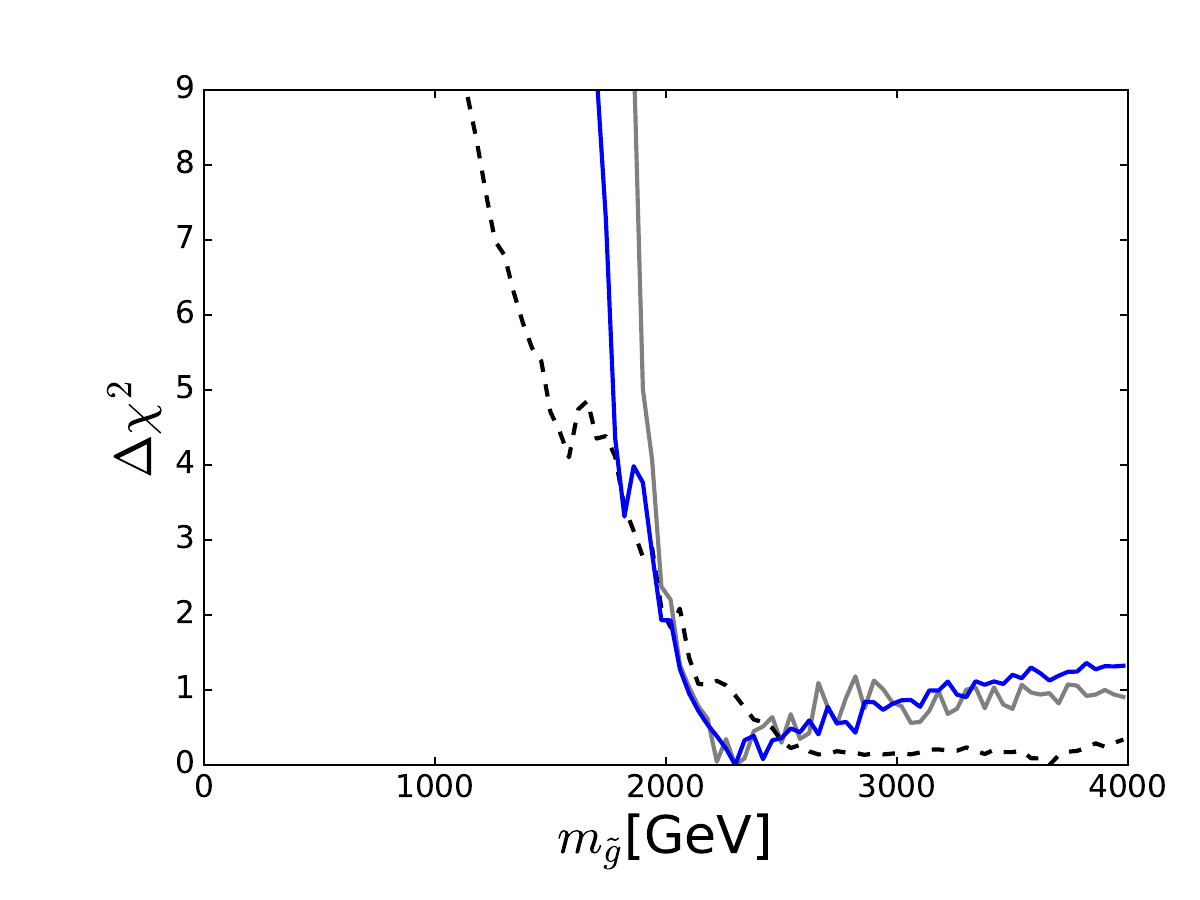}}
\resizebox{7.5cm}{!}{\includegraphics{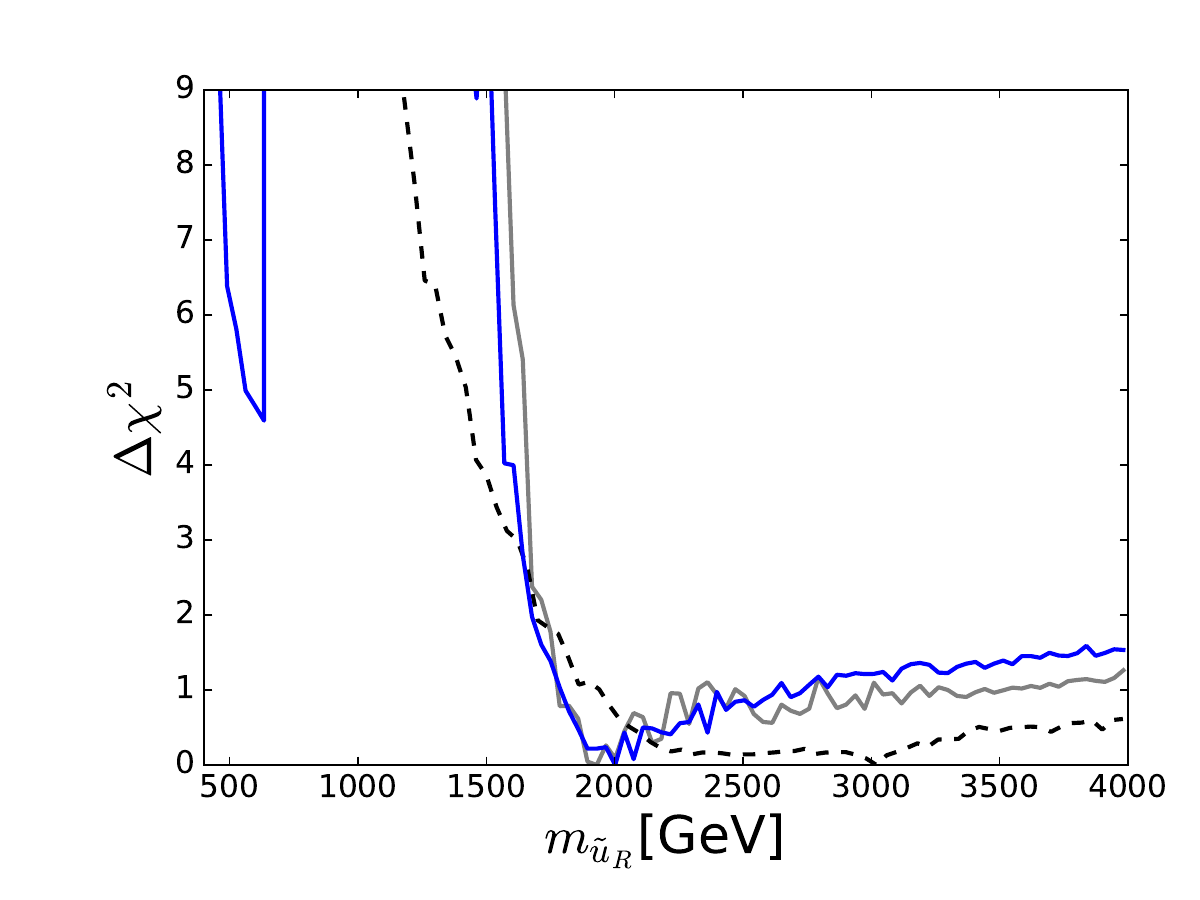}} \\
\resizebox{7.5cm}{!}{\includegraphics{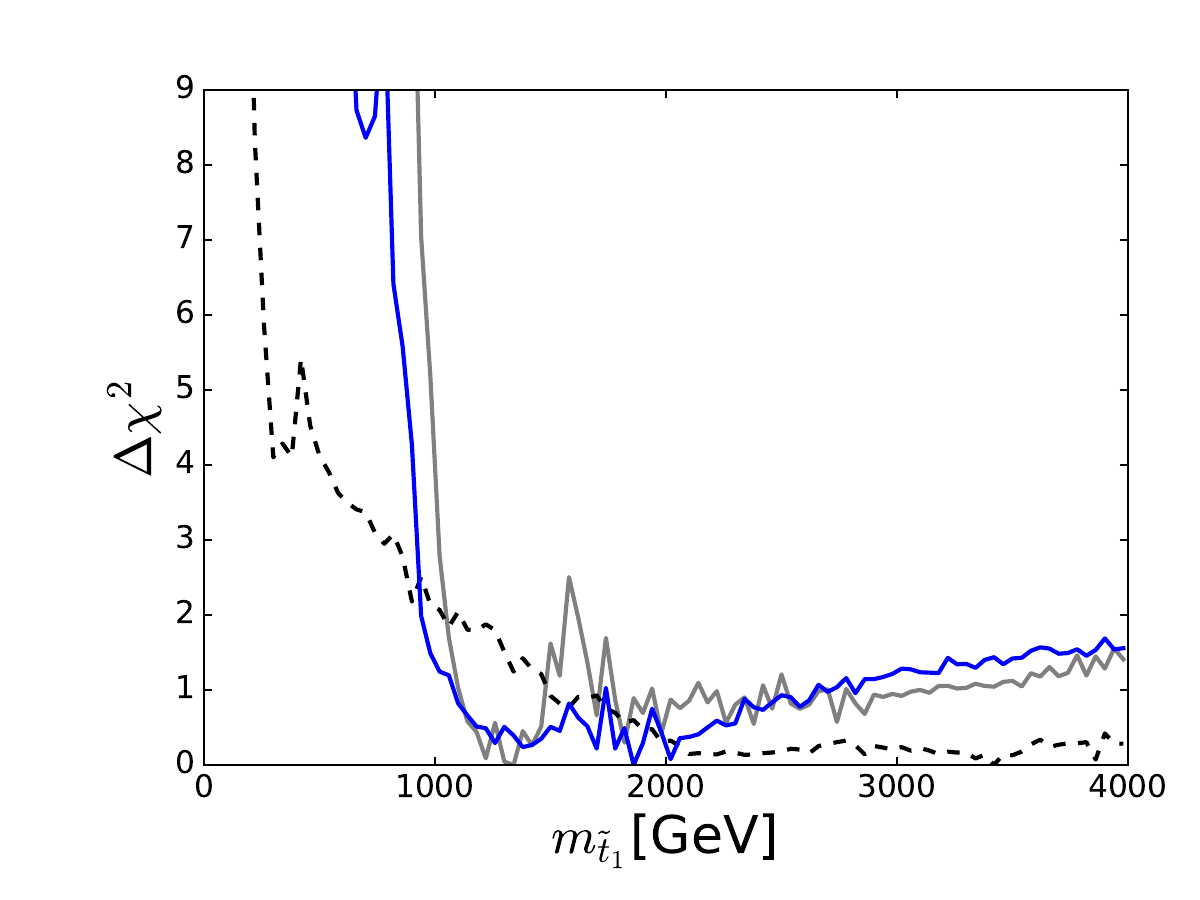}}
\resizebox{7.5cm}{!}{\includegraphics{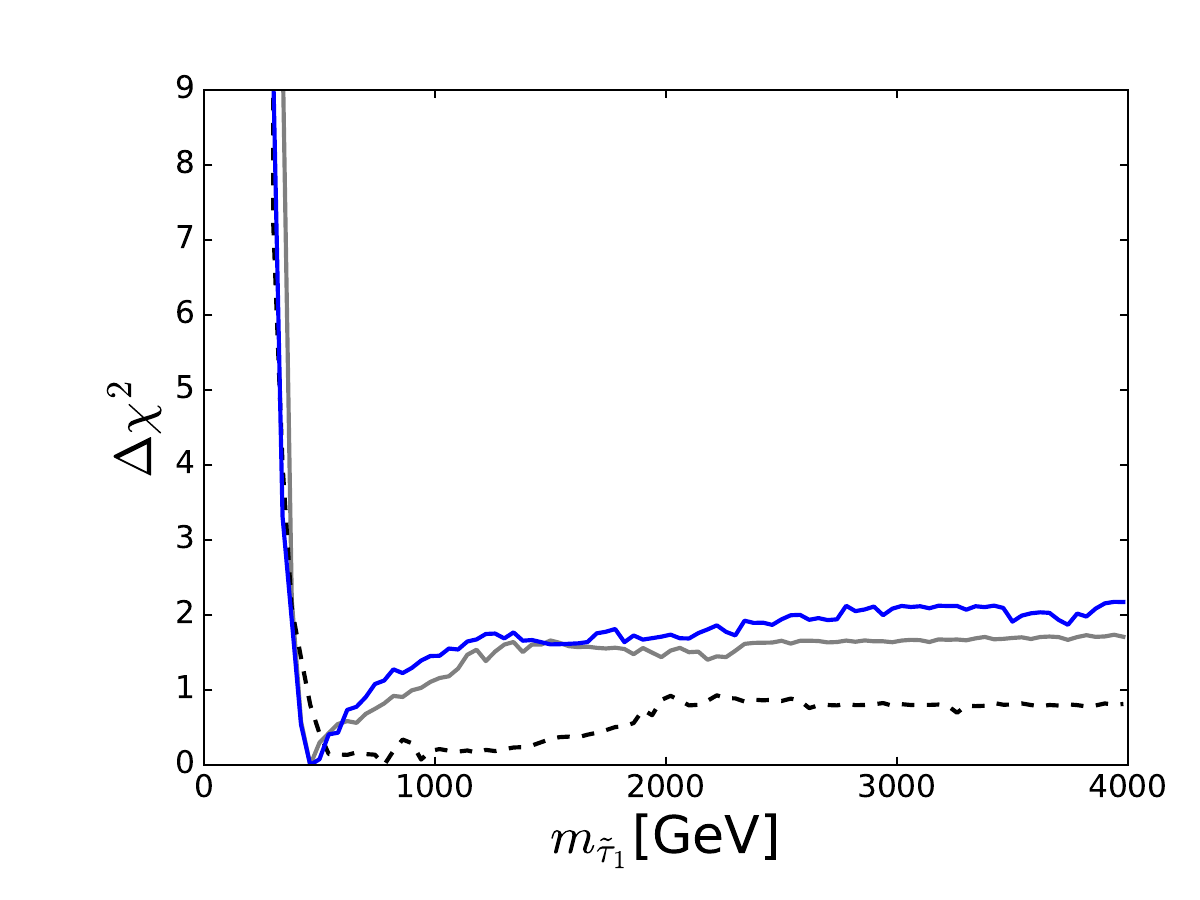}}
\end{center}
\vspace{-0.5cm}
\caption{\it 
The one-dimensional $\chi^2$ likelihood functions for the full SU(5) sample (solid blue lines)
and in the restriction of the SUSY SU(5) GUT model
sample to $m_5/m_{10} \in [0.9, 1.1]$ (solid grey lines) compared to those in our previous NUHM2
analysis~\protect\cite{mc10} (dashed grey lines)
for $\mgl$ (upper left panel), $\msq$ (upper right panel), $m_{\tilde t_1}$ (lower left panel),
and $m_{\tilde \tau_1}$ (lower right panel).
}
\label{fig:comparison}
\end{figure*}


\section{The Possibility of a Long-Lived \boldmath{$\staue$}}

The possibility of a very small ${\tilde \tau_1} - \neu1$ mass difference opens up
the possibility that the ${\tilde \tau_1}$ might have a long lifetime, as discussed
in the contexts of the CMSSM, NUHM1 and NUHM2 in \cite{MC12-DM}. 
This would occur if
$m_{\tilde \tau_1} - \mneu1 < m_\tau$. As seen in the lower left panel of Fig.~\ref{fig:moremasschi2},
the best-fit point  has a  mass difference $\sim 20 \gev$,
outside this range, but $m_{\tilde \tau_1} - \mneu1 < m_\tau$
is allowed with $\Delta \chi^2 \sim 1$. 
In \reffi{fig:lifetime} we analyze the lifetime of the \stau1.
We see in the upper {left} panel of Fig.~\ref{fig:lifetime}
that there is essentially no $\chi^2$ penalty for $10^{-9}$~s~$ \lesssim \tau_{\tilde \tau_1} \lesssim 10^{-2}$~s,
with lifetimes $\sim 10^{-10}$~s and $\lesssim 10^3$~s allowed with $\Delta \chi^2 \lesssim 1$.
Distinguishing a separated-vertex signature at the LHC would be challenging for smaller
values of $\tau_{\tilde \tau_1}$, and there would be significant disruption of the successful
conventional Big Bang nucleosynthesis calculations for $\tau_{\tilde \tau_1}
\gtrsim 10^3$~s \cite{bbn}.

\begin{figure*}[htb!]
\begin{center}
\resizebox{7.5cm}{!}{\includegraphics{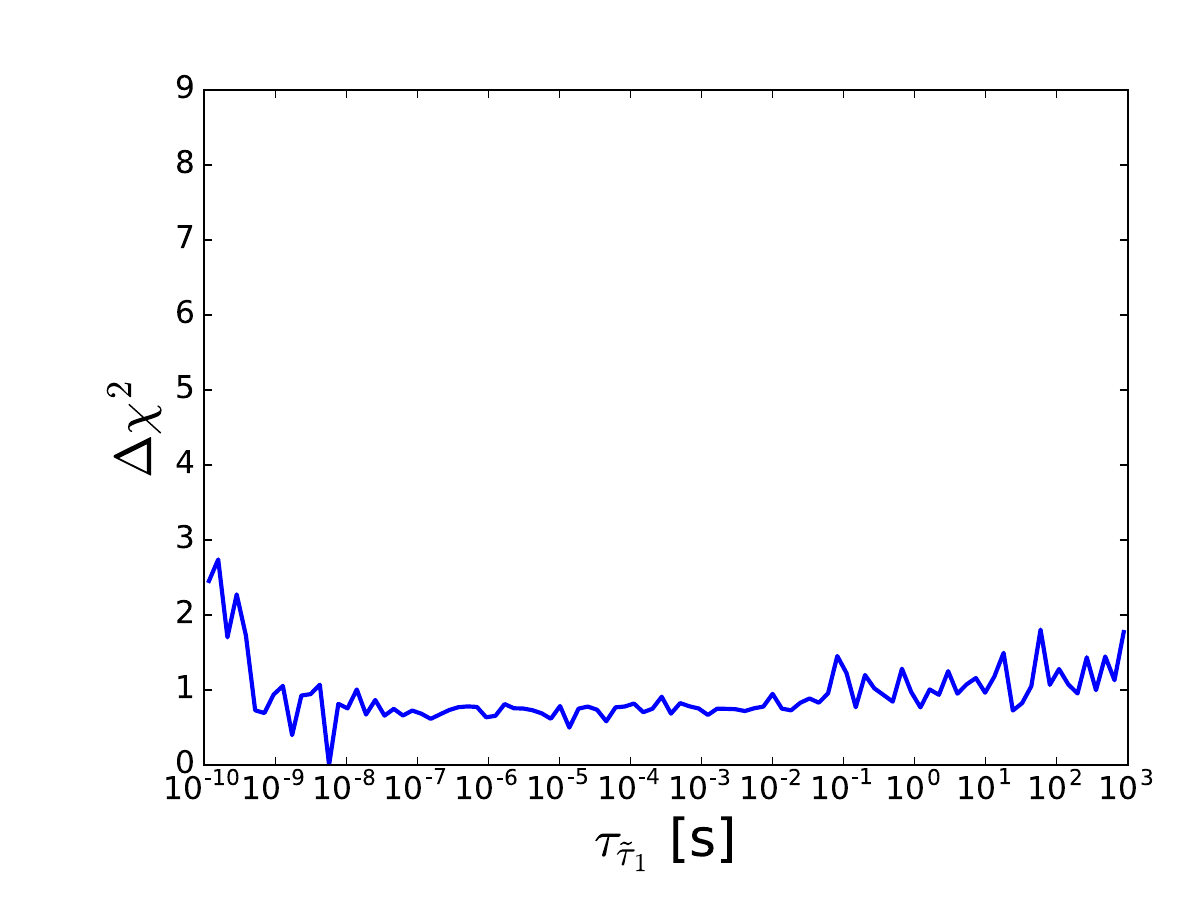}}
\resizebox{7.5cm}{!}{\includegraphics{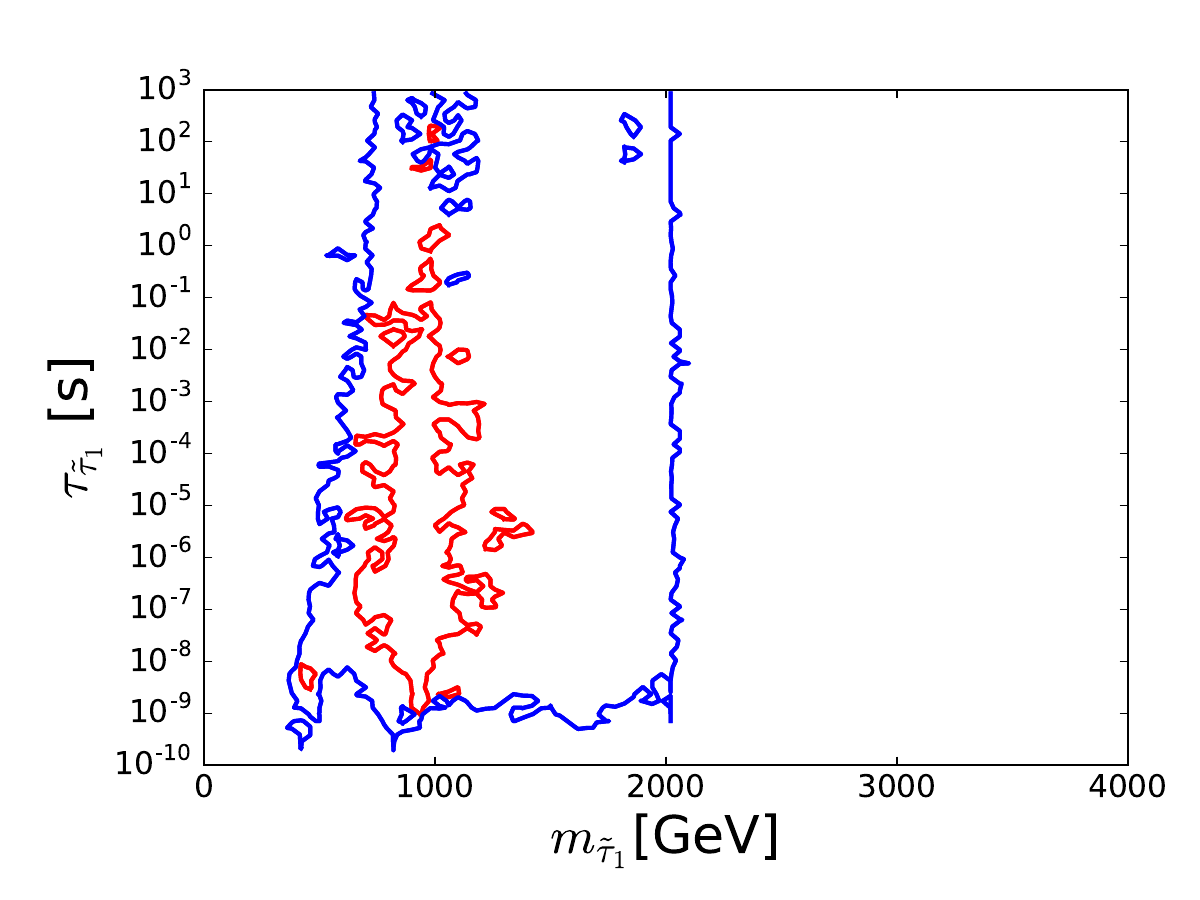}} \\
\resizebox{7.5cm}{!}{\includegraphics{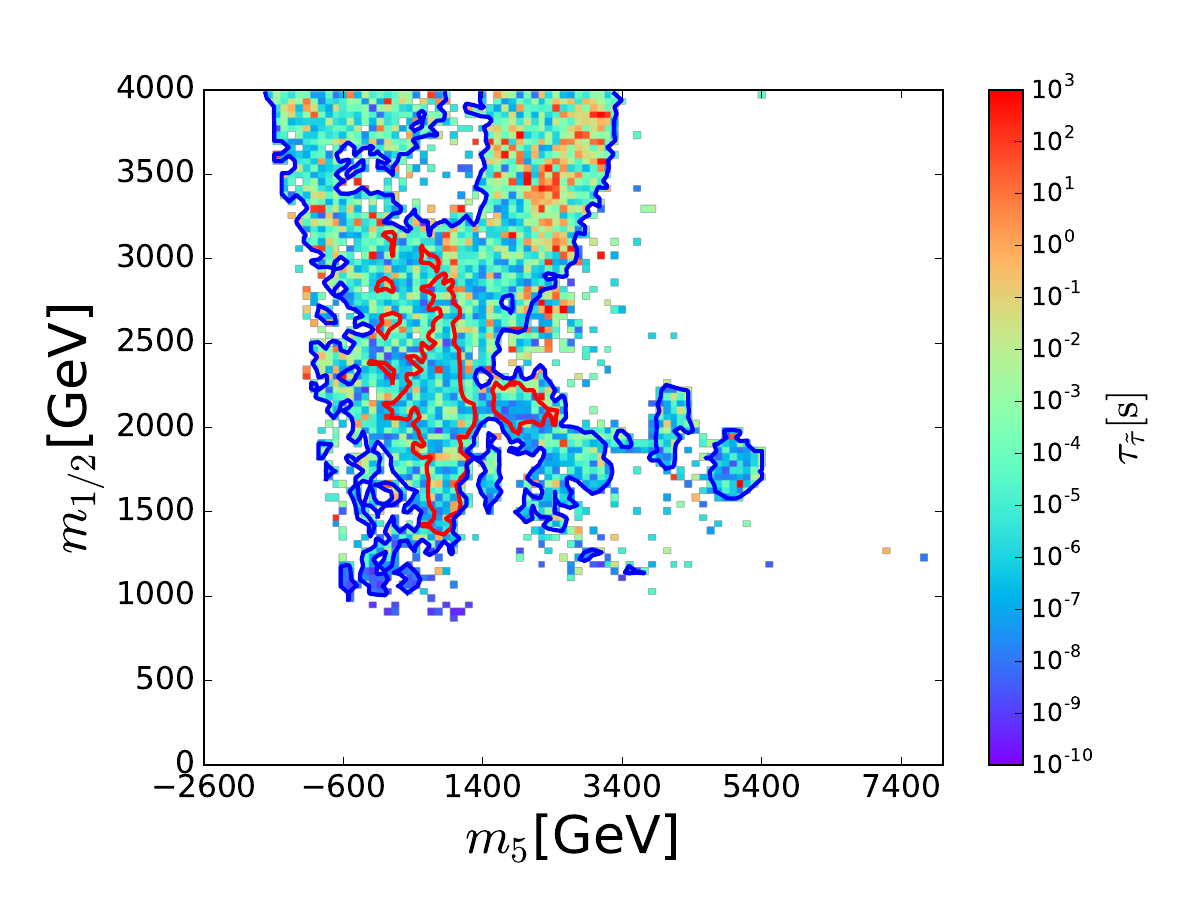}}
\resizebox{7.5cm}{!}{\includegraphics{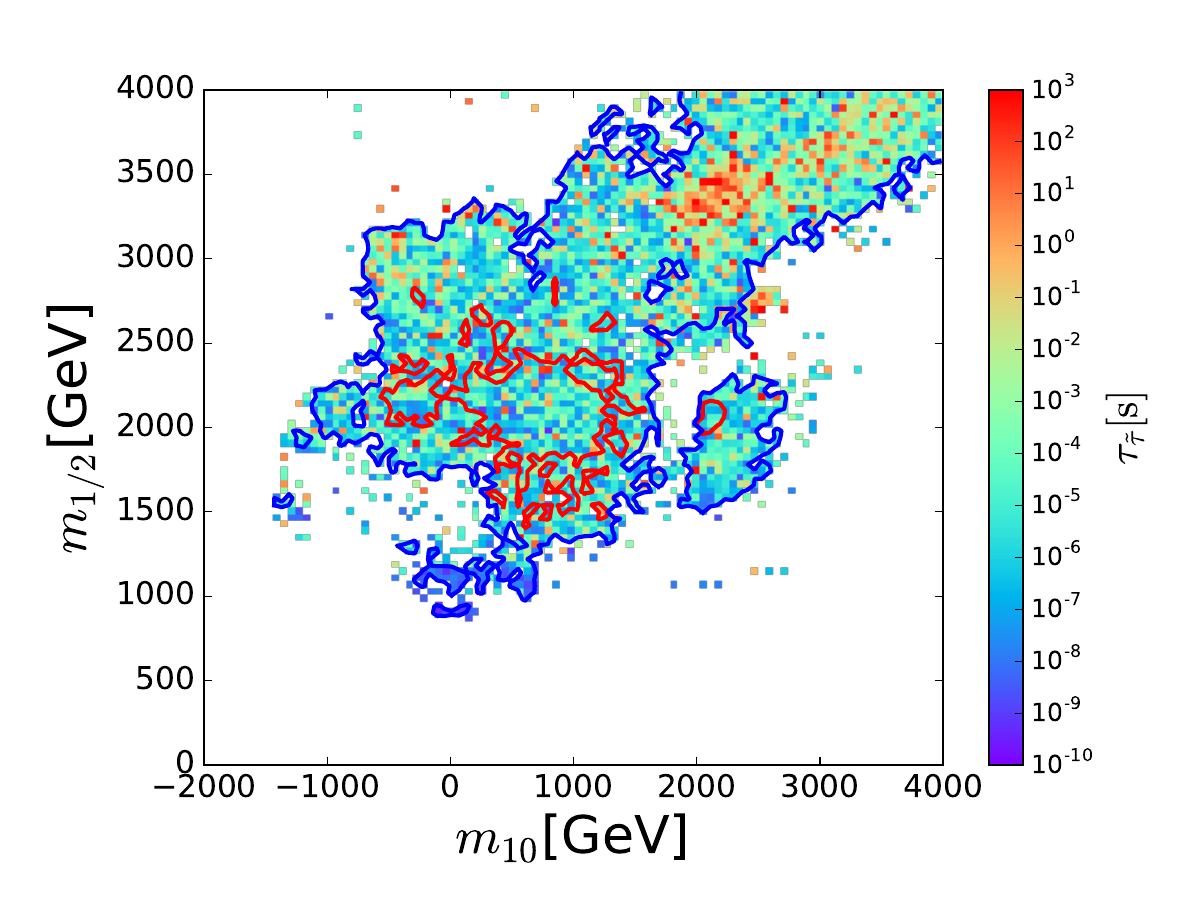}} \\
\end{center}
\vspace{-0.5cm}
\caption{\it 
Upper left panel: The global $\chi^2$ function in the SUSY SU(5) GUT model as a function of the $\tilde{\tau}_1$ lifetime.
Upper right panel: The $(m_{\tilde \tau_1}, \tau_{\tilde \tau_1})$ plane, shaded according to the values of $\tau_{\tilde \tau_1}$,
as indicated. Lower panels: The $(m_5, m_{1/2})$ and $(m_{10}, m_{1/2})$
planes, coloured according to the values of $\tau_{\tilde \tau_1}$. 
The 68\% and 95\% CL contours in these three planes are coloured red and blue, respectively. 
}
\label{fig:lifetime}
\end{figure*}

The upper right plot of \reffi{fig:lifetime} compares the \stau1\ lifetime with its mass.
The plane is characterized by a strip with $800 \gev \lesssim {\tilde \tau_1} \lesssim 1200 \gev$ allowed
at the 68\% CL, while the 95\% CL region is significantly wider, ranging from $m_{\tilde \tau_1} \sim 500 \gev$ to
$m_{\tilde \tau_1} \sim 2000 \gev$.

The lower panels of Fig.~\ref{fig:lifetime} display the regions of the $(m_5, m_{1/2})$ (left) and
$(m_{10}, m_{1/2})$ (right) planes in the SUSY SU(5) GUT model
where the lowest-$\chi^2$ points have 
$10^{-10} \, {\rm s} <  \tau_{\tilde \tau_1} < 10^3 \, {\rm s}$.
The colour-coding indicates the lifetimes of these points, as indicated in the legends. 
The contours for $\Delta \chi^2 < 2.30 (5.99)$ relative to the best-fit point in our
sample are shown as solid red and blue lines, respectively.
One can see that larger lifetimes occur all over the displayed parameter space, with
  a slight preference for larger $m_5$ or $m_{10}$ values.


\section{Direct Dark Matter Detection}

As already mentioned, the PandaX-II experiment \cite{pandax} has recently published results from its first 98.7 days of data,
which currently provide the most stringent upper limits on the spin-independent
DM scattering cross section on protons, \ssi. In parallel, the LUX Collaboration \cite{lux16}
has presented preliminary constraints on \ssi\ from 332 days of data. We have combined
these two constraints on \ssi\ into a single experimental likelihood function, which we have then
convoluted with an estimate of the theoretical uncertainty in the calculation of \ssi, as described
in~\cite{MC12-DM}, to constrain the SUSY SU(5) GUT parameter space. This
constraint has been used in obtaining the global fit whose results we have presented in the 
previous Sections. Here we discuss the future prospects for direct DM detection 
in light of our global fit.

Fig.~\ref{fig:DirectDM} displays our results for the SUSY SU(5) GUT model 
in the $(\mneu1, \ssi)$ plane. The combined PandaX-II/LUX constraint (black line) establishes a
95\% CL that reaches \ssi\ $\simeq 2 \times 10^{-46}$~cm$^2$ for $\mneu1 = 50 \gev$ and
$\simeq 10^{-45}$~cm$^2$ for $\mneu1 = 500 \gev$, providing the upper boundary
of the 95\% CL region in the $(\mneu1, \ssi)$ plane seen in Fig.~\ref{fig:DirectDM}. We see
that there are regions favoured at the 68\% CL that lie relatively close to this boundary,
whereas the main 68\% CL region and the best-fit point have smaller values of
\ssi .
We also note that the $H/A$ funnel  and $\cha1 - \neu1$ DM mechanisms 
favour values of \ssi\
that are relatively close to the PandaX-II/LUX boundary, whereas the ${\tilde \tau_1} - \neu1$ 
mechanism and its hybridization with the $H/A$ funnel favour smaller values of \ssi.
{The upcoming XENON1T~\cite{XENON1T} experiment will be able to probe the whole $\cha1$ coannihilation
region and a substantial part of the $H/A$ funnel region.}

\begin{figure*}[htb!]
\begin{center}
\resizebox{9cm}{!}{\includegraphics{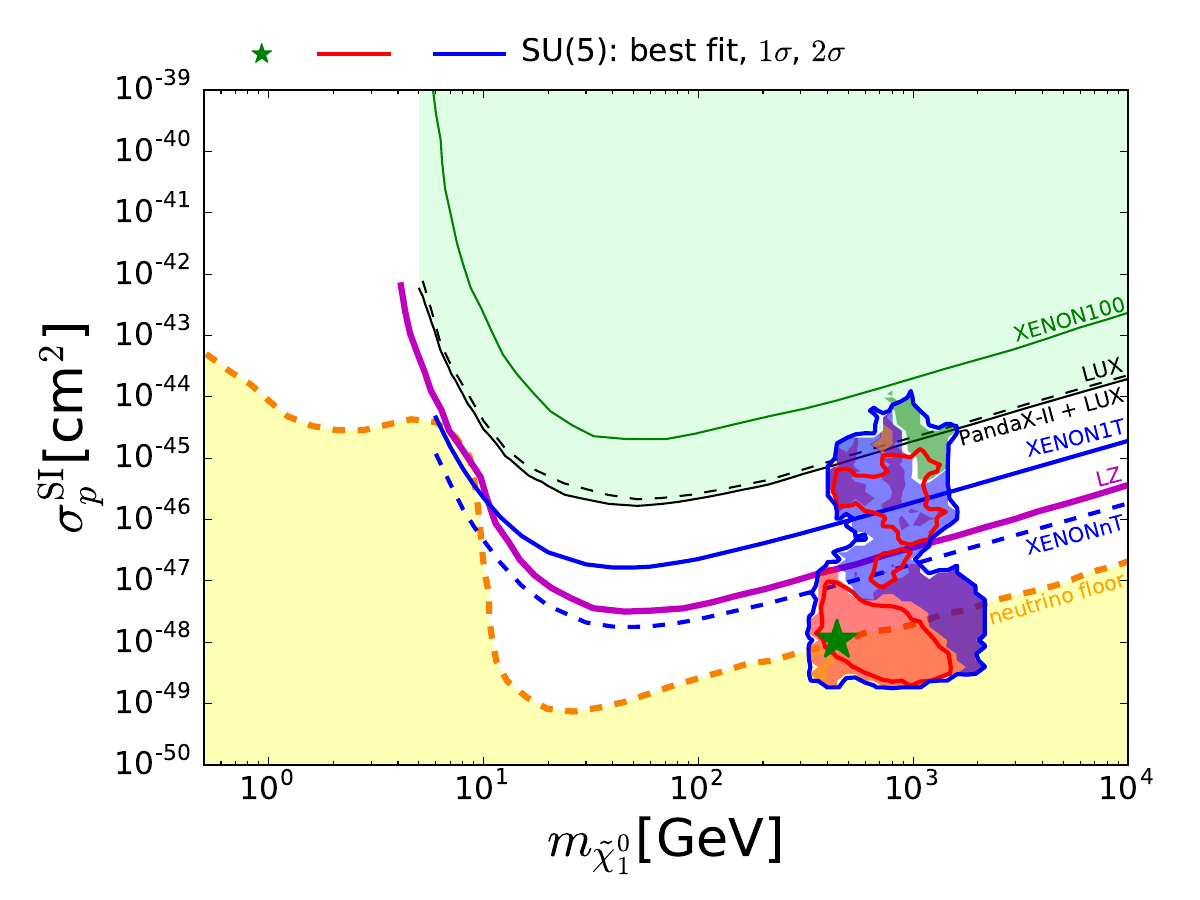}}
\vspace{0.75cm}
\resizebox{11cm}{!}{\includegraphics{dm_legend}} \\

\end{center}
\vspace{-1cm}
\caption{\it 
The $(\mneu1, \ssi)$ plane in the SUSY SU(5) GUT model.
The solid green line is the 95\% CL upper limit from the XENON100 experiment, and the dashed black solid line
is the new 95\% CL upper limit from the LUX experiment. The solid black line shows the 95\% CL exclusion contour for
our combination of the PandaX-II and LUX experiments,
the solid purple line shows the projected 95\% exclusion sensitivity of the LUX-Zeplin (LZ) experiment,
{the solid and dashed blue lines show the projected 95\% sensitivities of the XENON1T and XENONnT experiments,
respectively},
and the dashed orange line shows the astrophysical neutrino `floor', below which astrophysical neutrino 
backgrounds dominate (yellow region).
The other line colours and shadings within the 68\% and 95\% CL
regions are the same as in Fig.~\protect\ref{fig:m5m10m12}.
}
\label{fig:DirectDM}
\end{figure*}

We also display in Fig.~\ref{fig:DirectDM} the projected 95\% exclusion sensitivity of the 
future LUX-Zeplin (LZ) {and XENONnT experiments (solid purple and dashed blue lines 
respectively)~\cite{Malling:2011va,XENON1T}},
and the astrophysical neutrino `floor'
(dashed orange line) 
\cite{neutrino,cushman}, below which astrophysical neutrino backgrounds dominate (yellow region).
We see that much of the ${\tilde \tau_1} - \neu1$ coannihilation region and the region of its hybridization with
the $H/A$ funnel lie below the projected
sensitivities of the LZ {and XENONnT experiments}, and substantial portions of them also lie below the neutrino
`floor'. 
On the bright side, however, we recall that the ${\tilde \tau_1} - \neu1$
region, in particular, lies at relatively small values of $m_5, m_{10}$ and $m_{1/2}$, offering
greater prospects for detection at the LHC than, e.g., the $\cha1 - \neu1$ region, so there is
complementarity in the prospects of the LHC and direct DM experiments for probing
the SUSY SU(5) GUT model, as was noted previously for other SUSY
models~\cite{MC12-DM}. 


\section{Summary and Conclusions}

We have explored in this paper the experimental, phenomenological, astrophysical and cosmological
constraints on the minimal SUSY SU(5) GUT model. In this scenario the GUT-scale
universal soft SUSY-breaking scalar mass $m_0$ is replaced by independent masses
for the $\mathbf{10}$ and $\mathbf{\bar 5}$ sfermions. This flexibility introduces some features
that are novel compared to the GUT-universal CMSSM, NUHM1 and NUHM2.

{In general we observe that many best-fit values of the coloured
  particles are within the reach of the HL-LHC, but that the preferred
  regions clearly extend beyond the reach of the final stage of the LHC.
On the other hand,  the best-fit masses of some electroweakly-interacting particles are
$\sim 500 \gev$, offering the possibility of pair production at a
collider with $\sqrt{s} \sim 1 \tev$, as envisaged for the final stage
of the ILC. Going to higher centre-of-mass energies,
$\sqrt{s} \lsim 3 \tev$ as
anticipated for CLIC, significant fractions of the 68\% CL ranges of electroweak sparticle masses
can be covered.}

One novelty is the appearance of a ${\tilde u_R}/{\tilde c_R} - \neu1$ coannihilation region
that appears where $m_5^2$ is large and positive,
$m_{10}^2$ is small and negative, and $m_{H_u}^2$ and
$m_{H_d}^2$ are large and negative.
On the other hand, we find that ${\tilde t_1} - \neu1$
coannihilation is not important in the SUSY SU(5) GUT model, nor are
the focus-point region and rapid $\neu1 \neu1$ annihilation via direct-channel $h$
and $Z$ poles. We have checked that the ${\tilde u_R}/{\tilde c_R} - \neu1$ coannihilation
region is not yet excluded by searches for $\ETslash$ events at the LHC, because
the production rate is reduced compared to the case where all 8 squarks are mass degenerate and
the small ${\tilde u_R}/{\tilde c_R} - \neu1$ mass difference suppresses this signature.
{However, this region may be accessible with future LHC runs.}

{We have also highlighted the possibility that a ${\tilde \nu_\tau}$ NLSP might have
an important coannihilation role.}
Another novelty is the composition of the ${\tilde \tau_1}$ NLSP in a significant region of
the model parameter space. In the GUT-universal CMSSM, NUHM1 and NUHM2 models,
the universality of $m_0$ and the greater renormalization for SU(2) doublets impose a
substantial mass difference between the ${\tilde \tau_2}$ and the ${\tilde \tau_1}$, with the
latter being predominantly a ${\tilde \tau_R}$. However, in the SUSY SU(5) GUT
model with $m_5 \ne m_{10}$, the ${\tilde \tau_R}$ and ${\tilde \tau_L}$ may have similar
masses, and the off-diagonal entries in the ${\tilde \tau}$ mass matrix may cause large
mixing and repulsion between the ${\tilde \tau_1}$ and ${\tilde \tau_2}$ masses.

On the other hand, one experimental signature that is shared by the SUSY
SU(5) GUT model and GUT-universal models is the possible appearance of a long-lived
(metastable) ${\tilde \tau_1}$. This is a feature of a significant fraction (but not all) of
the ${\tilde \tau_1} -\neu1$ coannihilation region.

The prospects for direct DM detection are mixed: they are relatively good
in the $\cha1 - \neu1$ coannihilation region, {but} less promising in the rapid $H/A$
annihilation and hybrid regions, though potentially detectable in the  planned LUX-Zeplin
experiment. On the other hand, the ${\tilde \tau_1} -\neu1$ coannihilation region
probably lies beyond the reach of this experiment, as does part of the hybrid region.
Indeed, portions of these regions lie below the neutrino `floor'. On the other hand,
substantial parts of these regions are accessible to LHC searches 
for long-lived particles and $\ETslash$.

\section*{Acknowledgements}

The work of M.B., V.C., M.L. and 
D.M.-S. is supported by the European Research Council 
via Grant BSMFLEET 639068. 
The work of R.C. is supported in part by the National Science Foundation under Grant No. PHY- 1151640 
at the University of Illinois Chicago, and in part by Fermilab, operated by Fermi Research Alliance, LLC 
under Contract No. De-AC02- 07CH11359 with the United States Department of Energy. 
This work of M.J.D. is supported in part by the Australian Research Council. 
The work of J.E. is supported in part by STFC (UK) via the research grant ST/L000326/1, 
and the work of H.F. is also supported in part by STFC (UK). The work of S.H. is supported in part by 
CICYT (grant FPA 2013-40715-P) and also by the Spanish MICINN Consolider-Ingenio 2010 Program 
under grant MultiDark CSD2009-00064. The work of K.A.O. is supported in part by DOE grant de-sc0011842 at the 
University of Minnesota. 
The work of K.S. is partially supported by the National Science Centre, Poland, under research grants
DEC-2014/15/B/ST2/02157 and DEC-2015/18/M/ST2/00054.
The work of G.W. is supported in part by the Collaborative Research Center 
SFB676 of the DFG, ``Particles, Strings and the early Universe'', and in part by the European Commission 
through the ``HiggsTools'' Initial Training Network PITN-GA-2012-316704.

\end{document}